\documentclass[conference]{IEEEtran}
\usepackage{amsthm}
\usepackage[utf8]{inputenc}
\usepackage[nopar]{lipsum}
\usepackage{amsmath}
\usepackage{amssymb}
\usepackage[noend]{algpseudocode}
\usepackage{algorithm}
\usepackage{todonotes}
\usepackage{xspace}
\usepackage{url}
\usepackage{soul}
\usepackage{subcaption}
\usepackage{hyperref}
\usepackage{comment}
\usepackage{xcolor}
\usepackage{array}
\usepackage{colortbl}
\usepackage{color}
\definecolor{mygray}{rgb}{0.9,0.9,0.9}
\sethlcolor{mygray}

\theoremstyle{plain}
\newtheorem{theorem}{Theorem}

\usepackage{wrapfig}

\newcommand{\SH}{\textsf{SH}\xspace}
\newcommand{\SubstringHeavy}{\textsf{SubstringHK}\xspace}
\newcommand{\TOP}{\text{top}-$K$\xspace}
\newcommand{\TOPlong}{\textsc{TOP}-$K$-\textsc{SUB}\xspace}

\newcommand{\FASTQ}{\textsc{ECOLI}\xspace}
\newcommand{\HUM}{\textsc{HUM}\xspace}

\newcommand{\USIlong}{\textsc{Useful String Indexing}\xspace}
\newcommand{\USI}{\textsc{USI}\xspace}
\newcommand{\sa}{\textsf{SA}\xspace}

\newcommand{\frag}{\textsf{frag}}
\newcommand{\psw}{\textsf{PSW}\xspace}

\newtheorem{problem}{Problem}

\newtheorem{example}{Example}

\usepackage{enumitem}
\setlist[enumerate]{leftmargin=6mm}
\setlist[itemize]{leftmargin=6mm}
\setitemize{itemsep=1pt,topsep=1pt,parsep=1pt,partopsep=1pt}

\def\dd{\mathinner{.\,.}}

\newcommand{\cO}{\mathcal{O}}

\newcommand{\IoT}{\textsc{IoT}\xspace}
\newcommand{\ADV}{\textsc{Adv}\xspace}
\newcommand{\ECOLI}{\textsc{ECOLI}\xspace}
\newcommand{\XML}{\textsc{XML}\xspace}

\newcommand{\ExactTopK}{\textsf{Exact-Top-}$K$\xspace}
\newcommand{\ET}{\textsf{ET}\xspace}
\newcommand{\ApproxTopK}{\textsf{Approximate-Top-}$K$\xspace}
\newcommand{\AT}{\textsf{AT}\xspace}
\newcommand{\UAT}{\textsf{UAT}\xspace}
\newcommand{\UET}{\textsf{UET}\xspace}

\newcommand{\BLA}{\textsf{BSL1}\xspace}
\newcommand{\BLB}{\textsf{BSL2}\xspace}
\newcommand{\BLC}{\textsf{BSL3}\xspace}
\newcommand{\BLD}{\textsf{BSL4}\xspace}

\newcommand{\TopKTrie}{\textsf{Top-$K$ Trie}\xspace}
\newcommand{\TT}{\textsf{TT}\xspace}

\newcommand{\occ}{\textsf{occ}}

\mathchardef\mhyphen="2D

\newcommand{\USItop}{\USI_{\textsc{TOP}\mhyphen K}\xspace}

\newcommand{\tA}{\ensuremath{\mathtt{A}}}
\newcommand{\tC}{\ensuremath{\mathtt{C}}}
\newcommand{\tG}{\ensuremath{\mathtt{G}}}
\newcommand{\tT}{\ensuremath{\mathtt{T}}}

\usepackage{titlesec}
\titlespacing\section{0pt}{9pt plus 0pt minus 0pt}{5pt plus 0pt minus 2pt}
\titlespacing\subsection{0pt}{4pt plus 0pt minus 0pt}{3pt plus 0pt minus 0pt}

\usepackage{authblk}
\begin{document}

%\author{Anonymous}
%\title{Utility-Oriented String Indexing via Top-$K$ Frequent Patterns}
\title{Indexing Strings with Utilities}

\author[1]{Giulia Bernardini}
\author[2]{Huiping Chen}
\author[3]{Alessio Conte}
\author[3]{Roberto Grossi}
\author[3]{Veronica Guerrini}
\author[4]{\\Grigorios Loukides}
\author[3]{Nadia Pisanti}
\author[5]{Solon P.\ Pissis}

{\small
\affil[1]{University of Trieste, Trieste, Italy ~$^2$University of Birmingham,  Birmingham, UK ~$^3$University of Pisa, Pisa, Italy}
\affil[4]{King's College London, London, UK ~$^5$CWI \& Vrije Universiteit, Amsterdam, The Netherlands}
}
% \author{Giulia Bernardini\thanks{University of Trieste, giulia.bernardini@units.it}
% \and 
% Huiping Chen\thanks{University of Birmingham, h.chen.13@bham.ac.uk}
% \and Alessio Conte\thanks{University of Pisa, \{firstname.lastname\}@unipi.it}
% \and Roberto Grossi\footnotemark[3]
% \and Veronica Guerrini\footnotemark[3]
% \and Grigorios Loukides\thanks{King's College London, gloukides@acm.org}
% \and 
% Nadia Pisanti\footnotemark[3]
% \and
% Solon P. Pissis\footnotemark[4]\thanks{Centrum Wiskunde Informatica, solon.pissis@cwi.nl}
% }
\date{}

\maketitle

\begin{abstract}
Applications in domains ranging from bioinformatics to advertising feature strings (sequences of letters over some alphabet) that come with numerical scores (\emph{utilities}). The utilities quantify the importance, interest, profit, or risk of the letters occurring at every position of a string. For instance, DNA fragments generated by sequencing machines come with a confidence score per position. 
Motivated by the ever-increasing rate of generating such data, as well as by their importance in several domains, we
introduce Useful String Indexing (\USI), a natural
generalization of the classic String Indexing problem. 
Given a string $S$ (the \emph{text}) of length $n$, \USI
asks for preprocessing $S$ into a compact data structure
supporting the following queries efficiently: given a shorter string $P$ (the \emph{pattern}), return the global utility $U(P)$ of $P$ in $S$, where $U$ is a function that maps any string $P$ to a utility score based on the utilities of the letters of every occurrence of $P$ in $S$. Our work also makes the following contributions: (1) We propose a novel and efficient data structure for \USI based on finding the top-$K$ frequent substrings of $S$.  (2) We propose a linear-space data structure that can be used to mine the top-$K$ frequent substrings of $S$ or to tune the parameters of the \USI data structure. (3) We propose a novel space-efficient algorithm for \emph{estimating} the set of the top-$K$ frequent substrings of $S$, thus improving the construction space of the data structure for \USI. (4) We show that popular space-efficient top-$K$ frequent item mining strategies employed by state-of-the-art algorithms do not smoothly translate from \emph{items} to \emph{substrings}. (5) Using billion-letter datasets, we experimentally demonstrate that: (i) our top-$K$ frequent substring mining algorithms are accurate and scalable, 
unlike two state-of-the-art methods; and (ii) our \USI data structures are up to $15$ times faster in querying than $4$ nontrivial baselines while occupying the same space with them. 
\end{abstract}

\section{Introduction}\label{sec:intro}

\begin{figure}
    \centering
       % \includegraphics[width=\columnwidth,trim={1cm 0.6cm 1cm 0}]{images/DIAGRAM10.pdf}
       % \smallskip
       \includegraphics[width=.87\columnwidth]{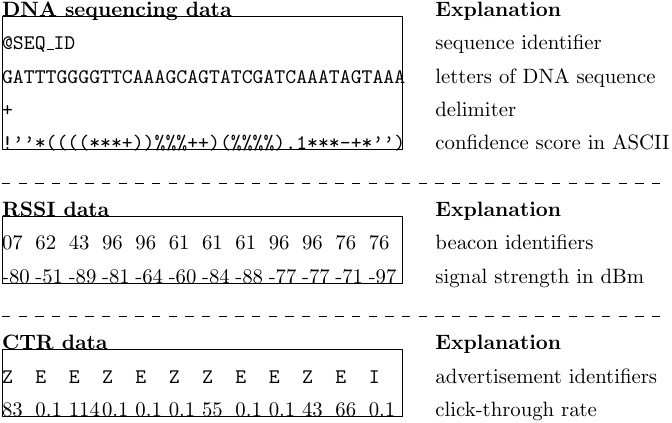}
        \caption{Illustration of real strings with associated utilities.}
        %:  (1) to (2) are construction steps, and (3) to (5) querying steps.}
        \label{fig:real_data}
   \end{figure}

Many application domains feature \emph{strings} (sequences of letters over some alphabet) associated with numerical scores (\emph{utilities}). The utilities quantify the importance, interest, profit, or risk of the letters occurring at every position of such strings~\cite{8845637,DBLP:conf/kdd/YinZC12,tkde15,infsci2020}; see Fig.~\ref{fig:real_data}. 
In bioinformatics, sequencing machines assign to each nucleotide a confidence score that represents the probability that this nucleotide has been correctly read by the machine and helps to identify sequencing errors~\cite{phred}. Thus, a DNA fragment is represented by a string where each letter is associated with a probability. 
In networks, each sensor is often assigned a Received Signal Strength Index (RSSI); i.e., a signal strength value that helps assessing network link quality~\cite{error_RSSI}. Thus, a sequence of sensor readings is represented by a string where each letter is associated with an RSSI. In advertising, each advertisement is often associated with a Click-Through Rate (CTR): an estimate of the probability that a user clicks on the advertisement, which helps advertisement pricing~\cite{marketing-book}. Thus, a sequence of advertisements is represented by a string where each letter is associated with a CTR.  In web analytics, each visited web page in a web server log is often associated with a score equal to the browsing time of the web page, which serves as a proxy of its importance~\cite{8845637}. Thus, a web server log is represented by a string where each letter is associated with such a score. In marketing, each product is often associated with a profit made by a sale~\cite{8845637}. Thus, a sequence of products is represented by a string where each letter is associated with a profit.  

%RG: very nice paragraph below!
Although strings with utilities are crucial to analyze in many application domains, existing research, which has been developed for more than $15$ years, focuses solely on the \emph{mining} of patterns from such strings (see~\cite{8845637} for a survey). However, \emph{querying}  such strings to find the utility of a query pattern
is equally important. For example, in bioinformatics, researchers are interested in evaluating the quality of a DNA pattern by computing its expected frequency in a collection of DNA strings with confidence scores~\cite{sdm24}. In advertising, evaluating the effectiveness of a series of advertisements is crucial for performing \emph{ad sequencing} (i.e., finding a good order of showing the advertisements to users), which increases consumers' interest or reinforces a message~\cite{robustadv} and is supported by Google Ads and YouTube. In web analytics, finding the total time spent visiting a sequence of web pages can improve website services, offer navigation recommendations, and improve web page design~\cite{webref1,8845637}. In marketing, computing the total profit made by selling some products in a certain order and/or comparing the profits made by selling products in different orders helps understanding consumers' behavior~\cite{DBLP:journals/isci/ZhangDGY21} as well as formulating commodity promotion and commodity procurement strategies~\cite{jerrytkdd21}.  In all these examples, the length of the strings is in the order of millions or billions, whereas the patterns are relatively short and occur a very large number of times~\cite{10.1093/gigascience/giy125,www18,dmkd05}.

In response, we introduce the \USIlong (\USI) problem, informally defined as follows (see Section~\ref{sec:preliminaries} for a formal definition): Let $S$ be a string of length $n$ (the \emph{text}), $w$ be a \emph{weight function} that assigns to each position of $S$ a utility, $u$ be a \emph{local}  utility function that aggregates the utilities of the positions of an occurrence of a string $P$ in $S$, and $\textsc{U}$ be a \emph{global}  utility function that aggregates the local utility of all occurrences of $P$ in $S$. The problem is to construct a data structure to answer the following queries: given a string $P$ (the \emph{pattern}) of length $m$, return the global utility $\textsc{U}(P)$ of $P$.  

\begin{example}
Consider the string $S$ below and the utilities of its positions assigned by $w$.  Consider also the following global utility function~\cite{8845637}:  $U(P)$ sums up the local utilities of all the
occurrences of $P$ in $S$, where the local utility of an occurrence of $P$ is the sum of the utilities of its letters. 
Let $P=\mathtt{TACCCC}$. $P$ occurs in $S$ at positions $1$ and $12$. \USI returns  
$U(P)=(1+ 3 + 2+ 0.7 + 1 + 1)+(1+ 1+ 1+ 0.9+ 1 + 1)=14.6$.
\begin{center}
\setlength{\tabcolsep}{0.05cm} 
\scalebox{0.74}
{
\begin{tabular}{c|*{20}{>{\centering\arraybackslash}p{0.025\textwidth}}}
%\hline
$i$ & $0$&$1$&$2$&$3$&$4$&$5$&$6$&$7$&$8$&$9$&$10$&$11$&$12$&$13$&$14$&$15$&$16$&$17$&$18$&$19$\\ \hline
$S$ & \tA & \cellcolor{mygray}\tT & \cellcolor{mygray}\tA & \cellcolor{mygray}\tC & \cellcolor{mygray}\tC &\cellcolor{mygray} \tC & \cellcolor{mygray}\tC & \tG & \tA & \tT & \tA & \tA & \cellcolor{mygray}\tT & \cellcolor{mygray}\tA & \cellcolor{mygray}\tC & \cellcolor{mygray}\tC & \cellcolor{mygray}\tC & \cellcolor{mygray}\tC & \tA & \tG \\ %\hline
$w$ &
.9 & 1 & 3 & 2 & .7 & 1 & 1 & .6 & .5 & .5 & .5 & .8 & 1 & 1 & 1 & .9 & 1 & 1 & .8 & 1 \\
%\hline 
\end{tabular}
}
\end{center}
\end{example}

We remark that indexing is arguably a more general problem than mining~\cite{sdm24}, as one can use \USI to: (1) query all patterns $P$ that are substrings of $S$, thus mining all patterns satisfying a global utility (or a length) constraint; (2) query any set of arbitrary patterns that are of interest in a specific setting.

\noindent{\bf Why is \USI Challenging?}
~\USI can be solved by constructing a classic text index over $S$, such as the suffix tree~\cite{DBLP:conf/focs/Farach97} or the suffix array~\cite{DBLP:journals/siamcomp/ManberM93}, finding the occurrences of the query pattern $P$ in $S$ and then computing and returning the global utility $U(P)$. The downside of this simple approach is that the computation of $U(P)$ requires \emph{aggregating the local utilities of all occurrences} of $P$, and this takes a large amount of time in practice for queries with reasonably many occurrences (e.g., in DNA sequencing data the occurrences are in the order of millions while $P$ is typically short~\cite{10.1093/gigascience/giy125}). This happens regardless of the way the global utility $U(P)$ is computed. Indeed, the simplest approach is to compute the local utility of each occurrence by applying the function $w$ to each position of an occurrence and then aggregating the results of all occurrences to obtain $U(P)$. This takes $\cO(m\cdot |\occ_S(P)|)$ time, where $m$ is the length of $P$ and $|\occ_S(P)|$ is the number of occurrences of $P$ in $S$. When the local utility function has the sliding-window property (see Section~\ref{sec:preliminaries}), a more efficient approach is to use \emph{prefix-sums}: we precompute the local utility of each prefix of $S$ and obtain the local utility of $P$ occurring at position $i$ as a function of the local utility of two prefixes of $S$, $S[0\dd i+m-1]$ and $S[0\dd i-1]$.  Then, we aggregate the results as in the simple approach. The precomputation takes $\cO(n)$ time and computing a single local utility of $P$ takes $\cO(1)$ time and thus $\cO(m+|\occ_S(P)|)$ time in total for $U(P)$.  Unfortunately, \emph{the query time is a function of $|\occ_S(P)|$}, thus it still requires a long time for frequent patterns, as these patterns have a very large number of occurrences.

\begin{figure}
\vspace{-1mm}
    \centering
       % \includegraphics[width=\columnwidth,trim={1cm 0.6cm 1cm 0}]{images/DIAGRAM10.pdf}
       % \smallskip
       \includegraphics[width=.93\columnwidth,trim={1cm 0.6cm 1cm 0}]{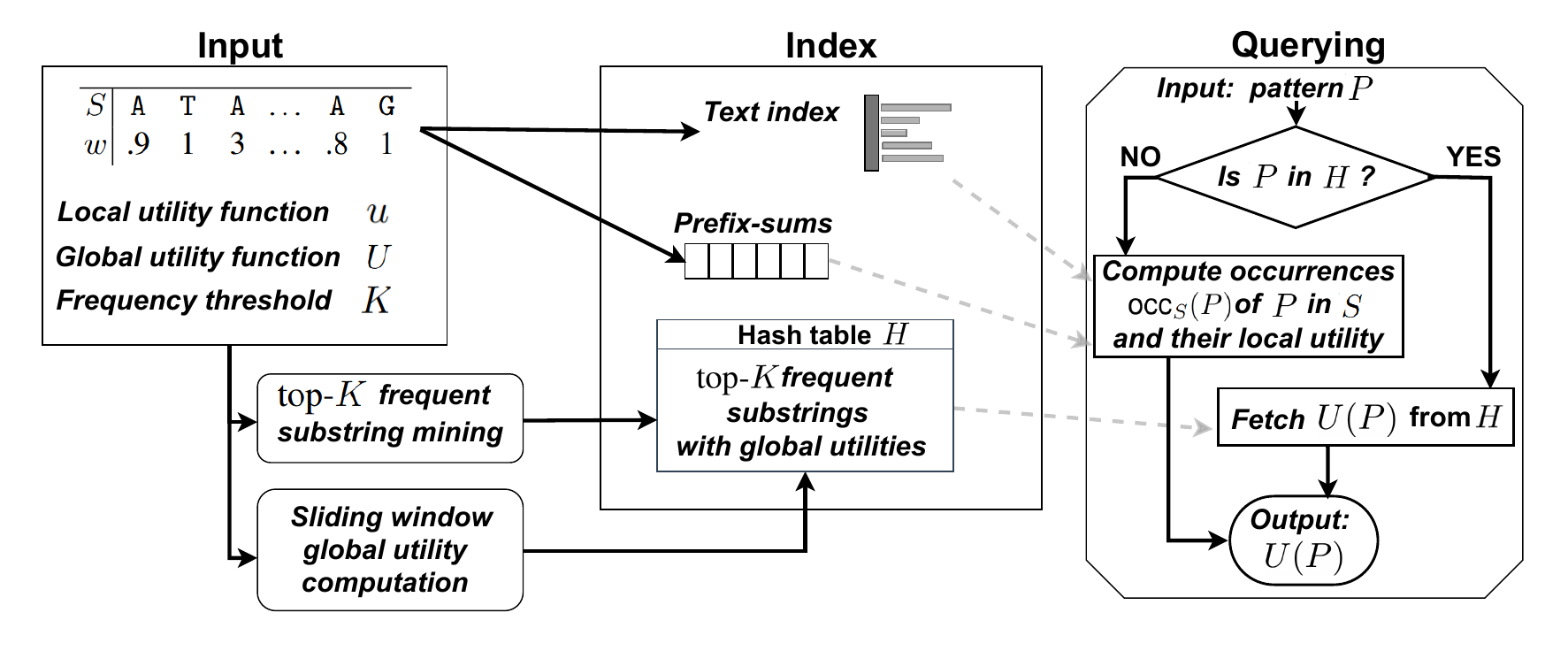}
        \caption{Overview of our approach for \USI.}
        %:  (1) to (2) are construction steps, and (3) to (5) querying steps.}
        \label{fig:diagram}
   \end{figure}

\vspace{+0.4mm}
\noindent{\bf Overview of our Approach.}~To address \USI efficiently, we propose to combine two different indexing schemes (see Fig.~\ref{fig:diagram}): one dedicated to queries with many occurrences; and a second index for the rest of the queries. Since every substring of $S$ could potentially be a query pattern $P$, we decompose the substrings of $S$ into the \TOP frequent substrings,  whose global utility is the most expensive to compute, and the rest of the substrings. For the frequent substrings, we use an explicit representation of size $\cO(K)$: we precompute their global utilities using a sliding-window approach and store them in a hash table. Each hash table key is the fingerprint~\cite{karp1987efficient} of a frequent substring of $S$ and the value is the global utility of the substring. Since we can read the global utility from the hash table in $\cO(1)$ time,
the query time for a frequent substring is $\cO(m)$: the time to compute the fingerprint of $P$.
The infrequent substrings are not represented explicitly: we index them by means of a classic text index over $S$~\cite{DBLP:conf/focs/Farach97} of size $\cO(n)$; and the computation of global utilities is offloaded to the query part which employs the prefix-sums approach. Let $\tau_K$ be the minimum support of the \TOP frequent substrings; $K$ can be efficiently determined by a data structure that we propose. Since computing a single local utility using prefix-sums takes $\cO(1)$ time, the query time for an infrequent substring is $\cO(m+\tau_K)$: the time to search $P$ in the text index plus the time to compute the local utilities of at most $\tau_K$ occurrences. Thus, the query time \emph{for any} $P$ is bounded by $\cO(m+\tau_K)$ and the size of the data structure is bounded by $\cO(n+K)$. 

\vspace{-1mm}
\begin{example}
    A bioinformatics researcher routinely evaluates the quality of DNA patterns of length $8$~\cite{kMer_count} occurring in a genomic dataset~\cite{fullgenom} with total size $n\approx 2.9\cdot 10^9$.  
    They indexed the dataset using a classic index (suffix array~\cite{DBLP:journals/siamcomp/ManberM93}) and computed the global utility of 5,000 DNA patterns of length $8$, randomly selected from the top-$(n/50)$ frequent substrings, based on the prefix-sums approach discussed earlier. The least frequent of these patterns occurred $104,262$ times. The average query time was $0.1\cdot10^{-3}$ seconds. 
     Using our approach with $K:=n/100\approx 2.9\cdot 10^7$ instead,
     the average query time was $0.7\cdot 10^{-6}$ seconds (i.e., it was almost three orders of magnitude faster). 
    The size of the suffix array index was $85.31$ GBs, and the size of our index was $86.38$ GBs.\label{example2}
\end{example}

\begin{table*}[t]
\caption{(a) Top-$4$ substrings with respect to global utility, among those having  length at least $3$, and their global utility ranks and scores (i.e., sum of their CTRs). (b) Top-$4$ frequent patterns with length at least $3$, and their global utility ranks and scores (i.e., sum of their CTRs). (c) Cluster categories and the letters they correspond to in Tables~\ref{tab:adv:1} and~\ref{tab:adv:2}.}\label{tab:adv1}
\begin{subtable}[h]{0.49\columnwidth}\centering
\resizebox{\columnwidth}{!}{%
\begin{tabular}{|c||c|c|c|c|}
\hline {\large\bf{Substring}} & {\large$\mathtt{aba}$} & {\large$\mathtt{ccc}$} & {\large$\mathtt{aaa}$} & {\large$\mathtt{ded}$} \\ \hline   
 {\large{\bf Rank}} & {\normalsize 1} & {\normalsize 2} & {\normalsize 3} & {\normalsize 4} \\\hline
  {\large{\bf Utility} $U$}  &  \normalsize{4075.6} & \normalsize{3998.3} & \normalsize{3229.5} & \normalsize{2885.8} \\ \hline 
\end{tabular}
}\caption{}\label{tab:adv:1}
\end{subtable}\hspace{+1mm}
\begin{subtable}[h]{0.49\columnwidth}\centering
\centering
\resizebox{\columnwidth}{!}{%
\begin{tabular}{|c||c|c|c|c|}
\hline {\large\bf{Substring}}  & {\large$\mathtt{ddf}$} & {\large $\mathtt{ffe}$} & {\large $\mathtt{fef}$} & {\large $\mathtt{gba}$} \\ \hline  
{\large{\bf Rank}} & {\normalsize 21} & {\normalsize 47} & {\normalsize 50} &  {\normalsize 46} \\ \hline  
{\large{\bf Utility} $U$} & \normalsize{1658.9} & \normalsize{1224.2}  & \normalsize{1222.3} & \normalsize{1226.5} \\ \hline 
\end{tabular}
}\caption{}\label{tab:adv:2}
\end{subtable}\hspace{+1mm}
\begin{subtable}[h]{.499\textwidth}\centering
\centering
\resizebox{\columnwidth}{!}{%
\begin{tabular}{|c||c|c|c|c|c|c|c|}
\hline {\large {\bf Letter}} & {\large$\mathtt{a}$} & {\large$\mathtt{b}$} & {\large$\mathtt{c}$} & {\large$\mathtt{d}$} & {\large$\mathtt{e}$} & {\large$\mathtt{f}$} & {\large$\mathtt{g}$} \\ \hline 
{\large {\bf Keyword}} & {\large Credit} & {\large Money} & {\large Mortgages} & {\large Credit} & {\large Financial} & {\large Non-vehicle} & {\large Web/software} \\
~ & {\large reporting} &   {\large making} & ~ & {\large  cards}  & {\large trading}  &  {\large insurance} & {\large  hosting}\\\hline 
\end{tabular}}\caption{}\label{tab:adv:3}
\end{subtable}
\end{table*}

The most crucial steps in our approach are to (i) efficiently compute the \TOP frequent substrings of $S$ and (ii) efficiently construct the hash table. Surprisingly, computing the \TOP frequent substrings has not been considered explicitly in the literature, unlike finding \TOP frequent \emph{items}. As we show, existing approaches for computing \TOP frequent items~\cite{DBLP:journals/scp/MisraG82,DBLP:conf/icdt/MetwallyAA05} cannot be modified to effectively compute \TOP frequent \emph{substrings}. Therefore, for step (i), we propose two algorithms
to compute the \TOP frequent substrings. The first algorithm computes the \emph{exact} set $T_K$ of \TOP frequent substrings in $\cO(n+K)$ time by combining efficient indexes of $S$ and sorting. The second algorithm \emph{estimates} $T_K$ but is more space-efficient: it uses space $\cO(n/s+K)$ and requires $\tilde{\cO}(n+sK)$ time, for a user-defined parameter $s$, which trades time efficiency and accuracy for space; the $\tilde{\cO}$ notation suppresses $\text{polylog}(n)$ factors. By combining the output of either of the two algorithms, an index of $S$, and a sliding-window approach, we can perform step (ii) efficiently. In particular, we bound the time complexity for step (ii) by $\cO(nL_K)$, where $L_K$ is the number of \emph{distinct lengths} of the $K$ reported substrings. We will show that $L_K$ is actually small in practice for suitable choices of $K$ (or $\tau$).

\vspace{+1mm}
\noindent{\bf Contributions.}~We introduce the \USI problem for indexing a string $S$ with utilities and make the following contributions. 

\noindent{\bf 1.}~We propose a novel, efficient data structure for \USI based on finding the \TOP frequent substrings of $S$ (Section~\ref{sec:USI_DS}).

\noindent{\bf 2.}~We propose a linear-space data structure that can be used to mine the \TOP frequent substrings of $S$ or to tune the parameter $K$ or $\tau$ in order to estimate the query time, size, and construction time of the \USI data structure (Section~\ref{sec:ESA}). 

\noindent{\bf 3.}~We propose a novel, space-efficient algorithm for \emph{estimating} the set of \TOP frequent substrings of $S$, thus improving the construction space of the \USI data structure (Section~\ref{sec:SSA}). 

\noindent{\bf 4.}~We consider modifying frequent item mining algorithms to estimate the set of \TOP frequent substrings. In particular, we demonstrate theoretically that popular space-efficient \TOP frequent item mining strategies~\cite{CormodeM12,DBLP:journals/scp/MisraG82,DBLP:conf/icdt/MetwallyAA05}, employed by state-of-the-art algorithms~\cite{HeavyKeeper2018, DBLP:conf/wea/DinklageFP24}, do not smoothly translate from \emph{items} to \emph{substrings} and thus lead to very inaccurate estimations of the \TOP frequent substrings (Section~\ref{sec:whynot}).

\noindent{\bf 5.}~We perform an extensive experimental study to assess the efficiency and effectiveness of all the proposed methods using 5 datasets of sizes up to $4.6$ billion letters (Section~\ref{sec:experiments}). We first show that our algorithm for estimating the set of \TOP frequent substrings is remarkably effective and $8.6$ times more space-efficient on average than the exact method. On the contrary, approaches based on state-of-the-art algorithms~\cite{HeavyKeeper2018, DBLP:conf/wea/DinklageFP24} for estimating \TOP frequent substrings are much less effective and/or less efficient. For instance, on the genomic dataset of Example~\ref{example2} and over a wide range of $K$ values, our algorithm had an average accuracy of $97.5\%$ and took at most $8.7$ hours. On the contrary, an approach based on a state-of-the-art algorithm~\cite{HeavyKeeper2018} had an average accuracy of $64.6\%$ and did not finish within $5$ days for the largest $K$. We then show that our data structure for USI offers on average $3.1$ times (and up to $15$ times) more efficient query answering than $4$ nontrivial baselines, which employ advanced string processing tools, while occupying a similar amount of memory than them. 

We organize the remainder of the paper as follows. 
In Section~\ref{sec:case_study}, we present a case study using a real advertising dataset (with CTR data) to showcase the applicability of our methods. In Section~\ref{sec:preliminaries}, we present the necessary definitions and notation. We discuss related work in Section~\ref{sec:related_work}. We conclude the paper in Section~\ref{sec:discussion} with some future directions.

\section{Case Study}\label{sec:case_study}

Consider: (1) An advertising company whose string $S$ (the text) is comprised of advertisements with a Click-Through Rate (CTR) assigned to each position to model its utility.
(2) Marketers who are interested in determining whether their own patterns of advertisements are sufficiently effective, according to the past experience of the advertising company, which is captured by $S$.   This operation helps marketers perform ad sequencing, as mentioned in Section~\ref{sec:intro}. The effectiveness  of a marketer's pattern $P$ can be measured by summing up the CTRs of all advertisements in an occurrence of $P$ in $S$, to obtain the local utility of that occurrence, and then 
summing up the local utilities of all occurrences of $P$ in $S$, to obtain the global utility $U(P)$.  
The effectiveness of $P$ can be determined efficiently by indexing $S$ using our \USI data structure and querying it using $P$. To demonstrate the applicability of our data structure, we used an advertising dataset from~\cite{ad_dataset} that we preprocessed by clustering its keyword phrases (advertisements) into $14$  categories, using an automated tool~\cite{clustering_keywords} and manual post-processing~\cite{adv_after}. 
We then replaced each advertisement by its category, but we retained the CTR of the advertisement. This led to a text, \ADV,  of length 218,987 over an alphabet of size 14. We used each substring of \ADV with length in $[3,200]$ as a marketer's pattern $P$, and the \emph{total query time} for all 187,883 such patterns was $3.4$ seconds. This highlights the \emph{efficiency} of our data structure. 

We also considered a setting where the advertising company wants to find the most useful substrings of length at least $3$ from $S$ based on the same function $U$. We constructed our \USI data structure, used each substring of $S$ with length at least $3$ as a query, and sorted the substrings in terms of decreasing global utility.
The top-$4$ substrings in terms of global utility and their global utilities are in Table~\ref{tab:adv:1}. Interestingly, \emph{these are different from the top-$4$ frequent substrings} in \ADV (see Table~\ref{tab:adv:2}) and have much larger utilities (e.g., the most frequent substring in Table~\ref{tab:adv:2}  has the $21$st largest global utility among those mined by our method). Furthermore, 
they are comprised of more semantically similar advertisements (e.g., credit reporting and money making vs. credit cards and non-vehicle insurance); see Table~\ref{tab:adv:3}. This highlights the \emph{usefulness} of our data structure. 

\section{Preliminaries and Problem Definitions}\label{sec:preliminaries}

\noindent 
{\bf Strings.}~An \emph{alphabet} $\Sigma$ is a finite set of elements called \emph{letters}. 
This can be any finite set; e.g., a set of integers (or reals).
A \emph{string} $S=S[0\dd n-1]$ of \emph{length} $|S|=n$ is a sequence of $n$ letters from $\Sigma$, where $S[i]$ denotes the $i$th letter of the sequence. 
We refer to each $i\in [0,n)$ as a \emph{position} of $S$.
We consider throughout that $\Sigma=[0,\sigma)$ is an integer alphabet of size $\sigma$, such that $\sigma=n^{\cO(1)}$; i.e., $\sigma$ is polynomial in $n$.
For instance, $\Sigma$ can be the set of integers from $0$ to $9$ and $S=0124966$ is a string over this integer alphabet $\Sigma=[0,9]$.

A substring $R$ of $S$ may occur multiple times in $S$. The set of its \emph{occurrences} in $S$ is denoted by $\occ_S(R)$, and its \textit{frequency} by $|\occ_S(R)|$; we may omit the subscript $S$ when it is clear from the context. An occurrence of $R$ in $S$ starting at position $i$ is referred to as a \emph{fragment} of $S$ and is denoted by $\frag_S(i,|R|)=S[i\dd i+|R|-1]$. Thus, different fragments may correspond to different occurrences of the same substring. 
 A \emph{prefix} of $S$ is a substring of the form $S[0\dd j]$, and a  \emph{suffix} of $S$ is a substring of the form $S[i \dd n-1]$. It should thus be clear that any fragment of $S$ is a prefix of some suffix of $S$. 

Karp-Rabin (KR) fingerprints (or fingerprints for short) is a rolling hash method, introduced by Karp and Rabin~\cite{karp1987efficient}. It associates strings to integers in such a way that, with high probability, no collision occurs among the substrings of a given string. The KR fingerprints  for all the length-$k$ substrings of a string $S$, $k>0$, can be computed in $\cO(|S|)$ total time~\cite{karp1987efficient}. 

We consider the following basic string problem that determines which substrings of $S$ will be explicitly stored in the hash table index of our data structure for the \USI problem.

\begin{problem}[\TOPlong] %(\TOP)] 
Given a string $S$ of length $n$ and an integer $K>0$, return the $K$ most frequent substrings of $S$ (breaking ties arbitrarily).\label{problem:topK} 
\end{problem}

\noindent 
{\bf Utility Definitions.}~Let $S$ be a string of length $n$ and let $w\!:\![0,n)\rightarrow \mathbb{R}$ be a function that assigns to each position $i\in [0,n)$ of $S$ a real number $w[i]$, referred to as the \emph{utility}  of $S[i]$. 
We may refer to the pair $(S, w)$ as a \emph{weighted string}. 
For any fragment $\frag_S(i, |R|)$, a \emph{local utility function} 
$u(i,|R|)$ aggregates the utilities of all letters of the fragment (i.e., aggregates $w[k]$, for each $k\! \in\! [i, i\!+\!|R|\!-\!1]$). For any substring $R$ of $S$, a \emph{global utility function} 
$U(R)$ aggregates the value of the local utility of all the occurrences of the substring in $S$. 

We define a class $\mathbb{U}$ of global utility functions, such that for every $U \in \mathbb{U}$: (1) $U$ is  \emph{linear-time} computable (e.g., $\mathrm{sum}$, $\min$, $\max$, or $\mathrm{avg}$); \emph{and} (2) the local utility function of $U$ has the \emph{sliding-window} property (e.g., sum): for any three fragments of $S$, $S[i\dd j]$, its prefix $S[i\dd i']$, $i\leq i'$, and its suffix $S[i'+1\dd j]$, $i'+1\leq j$, the local utility of any of these three fragments can be obtained from the local utilities of the other two in $\cO(1)$ time. We consider the following problem.

\begin{problem}[\USIlong (\USI)]\label{problem:usi} Given a string $S$ of length $n$, a weight function $w$, and a global utility function $\textsc{U}$ from class $\mathbb{U}$,  construct a data structure that answers queries of the following type: given a string $P$ of length $m$, return the global utility $\textsc{U}(P)$ of $P$.   
\end{problem}

\noindent{\bf String Indexes.}~A \emph{trie} $\textsf{T}(\mathcal{S})$ is a rooted tree whose nodes represent the prefixes of strings in a set $\mathcal{S}$ of strings~\cite{DBLP:books/daglib/0020103} over alphabet $\Sigma$.
The edges of a tree are labeled by letters from $\Sigma$; the prefix corresponding to node $v$ is denoted by $\textsf{str}(v)$ and is given by the concatenation of the letters labeling the path (sequence of edges) from the root to $v$. The node $v$ is called the \emph{locus} of $\textsf{str}(v)$. 
%The parent-child relationship in $\textsf{T}(\mathcal{S})$ is defined as follows: the root node is the locus of the empty string $\varepsilon$; and the parent $v$ of another node $x$ is the locus of $\textsf{str}(x)$ without the last letter. This letter is the label of the edge $(v,x)$. 
The order on $\Sigma$ induces an order on the edges outgoing from any node of $\textsf{T}(\mathcal{S})$. A node $v$ is \emph{branching} if it has at least two children and \emph{terminal} if $\textsf{str}(v) \in \mathcal{S}$. 

A \emph{compacted trie} is obtained from $\textsf{T}(\mathcal{S})$ by dissolving all nodes except the root, the branching nodes, and the terminal nodes. The dissolved nodes are called \emph{implicit} while the preserved nodes are called \emph{explicit}. The edges of the compacted trie are labeled by strings. The \emph{string depth} $\textsf{sd}(v)=|\textsf{str}(v)|$ of any node $v$ is the length of the string it represents, i.e., the total length of the strings labeling the path from the root to $v$. The \emph{frequency} $f(v)$ of node $v$ is the number of terminal nodes in the subtree rooted at $v$. The compacted trie takes $\cO(|\mathcal{S}|)$ space provided that edge labels are stored as pointers to fragments of strings in $\mathcal{S}$. Given the lexicographic order on $\mathcal{S}$ along with the lengths of the longest common prefixes between any two consecutive (in this order) elements of $\mathcal{S}$, one can compute the compacted trie of $\mathcal{S}$ in $\cO(|\mathcal{S}|)$ time~\cite{DBLP:conf/cpm/KasaiLAAP01}.  

The \emph{suffix tree} of a string $S$, denoted by $\textsf{ST}(S)$, is the compacted trie of the set of all suffixes of $S$. 
%; see Fig.~\ref{fig:esa1}. 
Each terminal node in $\textsf{ST}$ is labeled by the starting position in $S$ of the suffix it represents. 
$\textsf{ST}(S)$ can be constructed in $\cO(n)$ time for any string $S$ of length $n$ over $\Sigma=[0,n^{\cO(1)}]$~\cite{DBLP:conf/focs/Farach97}.
The \emph{suffix array} of $S$, denoted by $\textsf{SA}(S)$~\cite{DBLP:journals/siamcomp/ManberM93}, is the permutation of $[0,n)$ such that $\textsf{SA}[i]$ is the starting position of the $i$th lexicographically smallest suffix of $S$. 
%; see Figs.~\ref{fig:esa2} and~\ref{fig:esa3}. 
It can be constructed in $\cO(n)$ time for any string $S$ of length $n$ over $\Sigma=[0,n^{\cO(1)}]$~\cite{DBLP:conf/focs/Farach97}.
The $\textsf{LCP}(S)$ array~\cite{DBLP:journals/siamcomp/ManberM93} of $S$ stores the length of longest common prefixes of lexicographically adjacent suffixes. For $j>0$, $\textsf{LCP}[j]$ stores the length of the longest common prefix between the suffixes $\textsf{SA}[j-1]$ and $\textsf{SA}[j]$, and $\textsf{LCP}[0]=0$. 
%; see Fig.~\ref{fig:esa4}. 
Given the \textsf{SA} of $S$, we can compute the \textsf{LCP} array of $S$ in $\cO(n)$ time~\cite{DBLP:conf/cpm/KasaiLAAP01}.

\section{Data Structure for Useful String Indexing}\label{sec:USI_DS}

In this section, we describe an efficient data structure for \USI (Problem~\ref{problem:usi}). It is constructed for a weighted string $(S,w)$ of length $n$ and any global utility function $U$ from class $\mathbb{U}$, and it is parameterized by an integer $K\in [1,n^2]$. This parameter is defined by the user, and it trades query time for space, as we will explain in Section~\ref{sec:ESA}. Recall that a query consists of a string $P$ of length $m$, and we aim to return its global utility $U(P)$ fast. Our data structure relies on a precomputed set $T_K$ of \TOP frequent substrings, whose efficient computation will be discussed in Section~\ref{sec:ESA}.
Let $\tau_K$ be the smallest frequency of any substring from $T_K$, and $L_K$ be the number of distinct lengths of the substrings in $T_K$. Our data structure, coined $\USItop$, achieves the bounds stated by Theorem~\ref{the:USI_efficient}.

\begin{theorem}\label{the:USI_efficient}
    \USI can be solved for any weighted string $(S,w)$ of length $n$, any global utility function $U\in\mathbb{U}$, and any parameter $K\in [1,n^2]$, with a data structure that can be constructed in $\cO(nL_K)$ time, has size $\cO(n+K)$, and answers queries in $\cO(m+\tau_K)$ time. The construction space on top of the space needed by $(S,w)$ is $\cO(n + K)$. 
\end{theorem}

Before describing $\USItop$, let us comment on the bounds it achieves depending on the value of $K$. Consider the two extreme values: when $K=1$, $T_K$ consists of the single most frequent substring of $S$ (thus $L_K=1$), whose frequency $\tau_K$ can be as large as $\Theta(n)$: in this case, both the construction time and the size of $\USItop$ are $\cO(n)$, but queries require $\Theta(m+n)$ time, which is impractical as $n$ can be huge.  When $K=n^2$, the output of \TOP consists of all the substrings of $S$, thus $L_K=n$ and $\tau_k=1$, implying fast $\cO(m)$-time queries but $\cO(n^2)$ construction time and size, which is also clearly impractical. Let us now consider $K=\Theta(n)$ (for instance, $K=\frac{n}{100}$). This is arguably the most interesting case in practice: indeed, in this case, the size and construction space of $\USItop$ are $\cO(n)$, and we can expect both $\tau_K$ and $L_K$ to be small.\footnote{For instance, in random strings, the length of the longest repeating substring is $\cO(\log n)$ w.h.p.~\cite{DBLP:journals/ejc/BollobasL18}, implying $L_K=\cO(\log n)$ and $\tau_K=\cO(1)$.} In Section~\ref{sec:experiments}, we confirm this intuition using several datasets with different characteristics.
\medskip

\noindent\textbf{High-Level Idea.} $\USItop$ encodes the global utility of the substrings of $S$ storing them in two different indexes depending on whether they are in $T_K$ or not. The data structure consists of two indexes (a hash table $H$ and the suffix tree $\textsf{ST}(S)$) and of an array $\psw$ of length $n$.
We assume the leaves of $\textsf{ST}(S)$ are stored as an array: by definition, this array is equal to $\textsf{SA}(S)$. The hash table $H$ stores the precomputed global utilities of the \TOP frequent substrings: a fingerprint of each such substring is added to $H$ as a key, and the value is its global utility. The array $\psw$ implements the prefix-sums strategy discussed in Section~\ref{sec:intro} and stores the \emph{local} utility of each prefix of $S$: $\psw[i]=u(0,i+1)$, for each $i\in [0,n-1]$. 

To answer a query for a pattern $P$ of length $m$, we first compute its fingerprint in $\cO(m)$ time and look it up as a key in $H$: if we find it, the query can be answered in $\cO(1)$ time simply returning the associated value. If $P$ is not found in $H$ (thus its frequency is bounded by $\tau_K$), then its global utility is computed on the fly using $\psw$ and $\textsf{ST}(S)$. We next provide the details of how $\USItop$ is constructed and how queries are answered within the bounds claimed in Theorem~\ref{the:USI_efficient}.
\medskip

\noindent\textbf{Construction.} There are three phases in the construction process: (i) compute the \TOP frequent substrings; (ii) compute the global utility of these substrings and add them into the hash table $H$; and (iii) construct $\textsf{ST}(S)$ and \psw.

\emph{Phase (i).}~We compute the set $T_K$ of \TOP frequent substrings using a data structure that will be described in Section~\ref{sec:ESA}. This data structure returns these substrings as a set of triplets $\langle \textsf{lcp}, \textsf{lb}, \textsf{rb} \rangle$, where $\textsf{lcp}$ is the length of the substring and $\textsf{SA}[\textsf{lb}\dd \textsf{rb}]$ is the interval of $\textsf{SA}(S)$
%\footnote{$\textsf{SA}(S)$ coincides with the array of leaves of $\textsf{ST}(S)$, thus it does not need to be constructed separately.} 
containing all the occurrences of the substring. Equipped with this information, we can thus populate the hash table $H$ as follows. 

\emph{Phase (ii).}~We first group the substrings of $T_K$ according to their length. To do this, we radix sort the tuples from Phase (i) according to their $\textsf{lcp}$ value $\ell$. We obtain $L_K$ groups of tuples, each for a distinct value $\ell$. For group $\ell$, we use an auxiliary bit vector  $B_\ell$ with $n$ entries: its $i$th entry is $1$ if and only if a substring of length $\ell$ in the current group occurs at position $S[i]$ (this information is stored in the $\sa$ interval $[\textsf{lb}, \textsf{rb}]$ of the tuples from the group), and $0$ otherwise.  
We then slide a window of size $\ell$ over $S$. For each starting position $i\in[0,n-\ell]$ of the window, we compute the KR fingerprint of $S[i \dd i + \ell -1]$ and its local utility $u(i,\ell)$ in $\cO(1)$ time~\cite{karp1987efficient}. We then 
check if $B_\ell[i]=1$. If this is the case, we use the fingerprint as a key in $H$ and we aggregate $u(i,\ell)$ with the current value according to function $U$ (if no such key is found in $H$, we create a new entry and initialize its value with $u(i,\ell)$). If $B_\ell[i]=0$, we do nothing and slide the window to the next position. When the window reaches the end of $S$, we have computed the global utilities of all the \TOP frequent substrings of length $\ell$ and stored them in $H$ in $\cO(n)$ time. We then proceed accordingly to process the next group.

\emph{Phase (iii).}~We construct $\textsf{ST}(S)$~\cite{DBLP:conf/focs/Farach97} and $\psw$. For the latter, we use a single scan through $S$ and $w$, exploiting the sliding-window property of the local utility function.\footnote{Recall the property from Section~\ref{sec:preliminaries}. E.g., if the local utility function $u$ is the sum, $\psw[i]=\psw[i-1]+w[i]$, for all $i\in[1,n-1]$.}

\emph{Analysis.}~Let us start with the construction time. For Phase (i), we apply the algorithm in Section~\ref{sec:ESA}, which requires $\cO(n+K)$ time. For Phase (ii), observe that for any fixed $\ell$, the total number of occurrences of all substrings of length $\ell$ is bounded by $n$. This is because no two distinct substrings of the same length can occur at the same position in $S$.
For each $\ell$, we store the occurrences of the \TOP frequent substrings of length $\ell$ in a bit vector with $n$ entries.
Using a sliding window $S[i \dd i + \ell -1]$, for all $i\in[0,n-\ell]$, we compute each KR fingerprint in $\cO(1)$ time~\cite{karp1987efficient} and each local utility in $\cO(1)$ time, and by looking up the fingerprints in the hash table $H$, we aggregate the global utility of all length-$\ell$ substring in $\cO(n)$ overall time. The total time for processing all $L_K$ lengths is thus $\cO(nL_K)$. Phase (iii) requires $\cO(n)$ time. The total construction time is thus bounded by $\cO(nL_K+K)=\cO(nL_K)$ as $K=\cO(nL_K)$.

The construction space is bounded by the $\cO(n)$ space required to construct $\textsf{ST}(S)$ and $\psw$, in addition to the space $\cO(n+K)$ required to construct and store $T_K$ (see Section~\ref{sec:ESA}). The total space occupied by $\USItop$ (i.e., its final size) is $\cO(n+K)$, as we have one entry of $H$ for each \TOP frequent substring, and both $\textsf{ST}(S)$ and $\psw$ occupy $\cO(n)$ space.

\medskip

\noindent\textbf{Query.} Consider a query pattern $P$ of length $m$. We first compute its KR fingerprint in $\cO(m)$ time~\cite{karp1987efficient} and search it as a key in $H$ to check if $U(P)$ has been precomputed. If so, then the answer to the query is the value stored in $H$, returned in $\cO(1)$ time. Consider now the case where the fingerprint is not found in $H$. To compute $U(P)$, we first locate its set $\occ_S(P)$ of occurrences in $S$ using $\textsf{ST}(S)$ in $\cO(m+|\occ_S(P)|)$ time~\cite{DBLP:conf/focs/Farach97}. We then retrieve the local utility $u(i,m)$ of each occurrence $i\in\occ_S(P)$ in $\cO(1)$ time from $\psw[i+m-1]$ and 
$\psw[i-1]$ exploiting the sliding-window property of $u$. Aggregating all such values, we compute and return $U(P)$. Observe that, since we defined $\tau_K$ as the smallest support of any \TOP frequent substring, any $P$ that is not in $T_K$ occurs \emph{at most} $\tau_K$ times, therefore the querying takes $m+|\occ_S(P)|=\cO(m+\tau_K)$ time.

Combining the last bound with the bounds proved in the \emph{Analysis} section, we have proved Theorem~\ref{the:USI_efficient}.

\section{Top-$K$ Frequent Substrings \& $\USItop$ Tuning }\label{sec:ESA}

In this section, we present a \emph{linear-space} data structure for performing the following tasks that are necessary for \USI: 
\begin{itemize}
\item[i] Compute a representation of the \TOP frequent substrings of $S$ used in $\USItop$ as a set of triplets.  
\item[ii] Estimate the query and construction time of $\USItop$ before constructing it when a user has a value for $K$, and thus knows the size $\cO(n + K)$ of $\USItop$. 
\item[iii] Estimate the size and construction time of $\USItop$ before constructing it when a user has a value for $\tau$, and thus knows the query time $\cO(m+\tau)$ of $\USItop$. 
\end{itemize}

Let us explain why such a data structure is useful in light of Theorem~\ref{the:USI_efficient}.
Consider task (iii) that uses a value of $\tau$ to infer the number $K_{\tau}$ of $\tau$-frequent substrings of $S$, which determines the size and construction time of $\USItop$. 
Space-efficient hash tables usually come with some guarantees: e.g., if we are storing $K$ $w$-bit keys,  then the total space usage should be $(1 + \epsilon)wK$ bits for some small $\epsilon$~\cite{DBLP:journals/jacm/BenderCFKT23}. Thus by computing $K_{\tau}$, we can estimate the \emph{size of our hash table} and from thereon, since we also know $n$, the total size of $\USItop$~\cite{DBLP:journals/spe/Kurtz99}.

\noindent{\bf Construction.}~The data structure comprises of the suffix tree $\textsf{ST}(S)$ and of $3$ arrays of size at most $n$. %(inspect the table below). 
The first is an array $\mathcal{T}$ of triplets $
\langle v, f(v), q(v) \rangle$, sorted in decreasing order w.r.t. $f(v)$, where $v$ is an explicit node in $\textsf{ST}(S)$ with frequency $f(v)$ and there are $q(v)$ letters labeling the edge between $v$ and its parent. Each such letter represents a distinct substring of $S$ having the same frequency $f(v)$. The second and third arrays are parallel to $\mathcal{T}$: $\mathcal{Q}$ is such that $\mathcal{Q}[i]=\sum_{j=1}^{i}q(v_j)$ is equal to the total number of distinct substrings  represented by the first $i$ triplets of $\mathcal{T}$; $\mathcal{L}$ is such that $\mathcal{L}[i]$ is the total number of distinct lengths of the substrings represented by the same triplets. The data structure size is thus $\cO(n)$.

% \begin{table}[!ht]
% {\small
% \centering
% \resizebox{0.99\columnwidth}{!}{\large 
% $\begin{array}{l||c|c|c|c
% }
% \mathcal{T}&\langle v_1, f(v_1), q(v_1) \rangle & \langle v_2, f(v_2), q(v_2) \rangle & \ldots & 
% \langle v_r, f(v_r), q(v_r) \rangle\\\hline
% \mathcal{Q}&q(v_1) & q(v_1)+q(v_2) & \ldots & \sum_{j=1}^{r}q(v_j) \\\hline
% \mathcal{L}&l_1 & l_2 & \ldots & l_r\\
% \end{array}$
% }}
% \end{table}

To construct the data structure, we first construct $\textsf{ST}(S)$. Values $q(v)$ can be computed for all explicit nodes $v$ with a traversal of $\textsf{ST}(S)$ by subtracting the string depth $\textsf{sd}(p(v))$ of the parent $p(v)$ of $v$ from $\textsf{sd}(v)$. Values $f(v)$ can be computed with a bottom-up tree traversal. 
%too: see Section~\ref{sec:preliminaries}. 
As we traverse $\textsf{ST}(S)$, we extract all triplets $\langle v, f(v), q(v)\rangle$. We then radix sort them in decreasing order of their values $f(v)$, breaking ties so that a triplet $\langle v_i, f(v_i), q(v_i)\rangle$ precedes $\langle v_j, f(v_j), q(v_j)\rangle$ with $f(v_i)=f(v_j)$ if $\textsf{sd}(v_i)\leq\textsf{sd}(v_j)$: in other words, for equal frequency,
triplets representing shorter substrings precede the triplets representing longer ones. Values $\textsf{sd}(v_i)$ can be read directly from $\textsf{ST}(S)$. 
We store the sorted sequence in $\mathcal{T}$. 

To compute $\mathcal{Q}$, we scan $\mathcal{T}$ from left to right and progressively sum up the values $q(v)$ from the triplets. To compute $\mathcal{L}$, we consider the triplets of $\mathcal{T}$ one by one from left to right, maintaining a counter $c$ of the distinct lengths and the current maximal string depth $M$. When reading the leftmost triplet $\langle v_{1},f(v_{1}),q(v_1)\rangle$, we set $c=M=\mathcal{L}[1]=\textsf{sd}(v_1)$. When processing a triplet $\langle v_i,f(v_i),q(v_i)\rangle$, $i>1$, we first compare $\textsf{sd}(v_i)$ with the current value of $M$. If $\textsf{sd}(v_i)>M$, we set $M=\textsf{sd}(v_i)$, increase $c$ by $\textsf{sd}(v_i)-M$, and store the new value of $c$ into $\mathcal{L}[i]$. Otherwise, we move on to the next triplet.
This is correct because the only distinct lengths that have not been accounted for before the $i$th triplet are the ones longer than $M$.
Since $\textsf{ST}(S)$ has exactly $n$ leaves, it has fewer than $n$ explicit nodes, thus the length of $\mathcal{T}$, $\mathcal{Q}$, and $\mathcal{L}$ is bounded by $n$, and values $f(v)$ are also bounded by $n$. This implies that the triplets can be radix sorted in $\cO(n)$ time and $\mathcal{T}$, $\mathcal{Q}$, and $\mathcal{L}$ can be computed in $\cO(n)$ time. $\textsf{ST}(S)$ can also be constructed in $\cO(n)$ time~\cite{DBLP:conf/focs/Farach97}, thus the whole construction requires $\cO(n)$ time.
The space is also bounded analogously by $\cO(n)$.

\noindent{\bf Task (i).} We scan $\mathcal{T}$ from left to right, i.e., for decreasing values $f(v)$. For each triplet $\langle v, f(v), q(v)\rangle$,  we list all the substrings represented by the implicit nodes on the edge between the parent $p(v)$ of $v$ and $v$, terminating the scan of $\mathcal{T}$ when $K$ substrings have been listed. We represent each listed substring as a different kind of triplet: $\langle\textsf{lcp},\textsf{lb}, \textsf{rb}\rangle$. From any triplet $\langle v, f(v), q(v)\rangle$, we list $q(v)$ distinct output triplets. Consider the output triplet corresponding to the $\ell$th letter (i.e., implicit node) on the edge from $p(v)$ to $v$. The value $\textsf{lcp}=\textsf{sd}(p(v))+\ell$ is the substring length; $\textsf{lb}$ and $\textsf{rb}$ are the endpoints of the interval of leaves descending from $v$. All such intervals can be computed in $\cO(n)$ time with a traversal of $\textsf{ST}(S)$. Note that values $\textsf{lb}$ and $\textsf{rb}$ are the same for all the output triplets computed for implicit nodes on the same edge. 

This procedure is called \ExactTopK and requires $\cO(n+K)$ time.
We have arrived at Theorem~\ref{the:topSA}. 

\begin{theorem}\label{the:topSA}
For any string of length $n$ and any integer $K>0$, \textnormal{\ExactTopK} solves \TOPlong in $\cO(n+K)$ time.
\end{theorem}

The output triplets of \ExactTopK are given as input to the construction algorithm of $\USItop$. Each such triplet can be converted to an explicit \TOP frequent substring $S[\textsf{SA}[\textsf{lb}]\dd \textsf{SA}[\textsf{lb}]+\textsf{lcp}-1]$ of $S$ should one require the \TOP frequent substrings in an explicit form. 

\noindent{\bf T ask (ii).} Given any $K$ value, we seek to compute the minimum frequency $\tau_K$ of any \TOP frequent substring of $S$ and the number $L_K$ of their distinct lengths. 
This is because $\tau_K$ directly determines the query time of $\USItop$, and $L_K$ its construction time (see Theorem~\ref{the:USI_efficient}). To do this, we binary search for $K$ in $\mathcal{Q}$ to find the smallest index $i$ such that $\mathcal{Q}[i]\geq K$ (the values of $\mathcal{Q}$ are increasing from left to right). Let $\mathcal{T}[i]=\langle v_i, f(v_i), q(v_i)\rangle$: then by construction $\tau_K=f(v_i)$ and $L_K=\mathcal{L}[i]$. Since the length of $\mathcal{Q}$ is bounded by $n$, the whole process requires $\cO(\log n)$ time.

\noindent{\bf Task (iii).} Given any $\tau$ value, we seek to compute the number $K_{\tau}$ of $\tau$-frequent substrings of $S$ and the number $L_\tau$ of distinct lengths of all substrings with frequency at least $\tau$. This is because $K_{\tau}$ and $L_\tau$ directly determine the space occupied by $\USItop$ and its construction time (see  Theorem~\ref{the:USI_efficient}). 
To do this, we binary search for $\tau$ in the values $f(v)$ of the triplets of $\mathcal{T}$ ($\mathcal{T}$ is sorted in decreasing order of $f(v)$) to find the largest index $i$ such that $f(v_i)\geq \tau$. Then, by construction, we have $K_{\tau}=\mathcal{Q}[i]$ and $L_\tau=\mathcal{L}[i]$. Since the length of $\mathcal{T}$ is bounded by $n$, the whole process requires $\cO(\log n)$ time.
\medskip

In Section~\ref{sec:SSA}, we present an approximate, space-efficient algorithm for Task (i) alternative to the exact one above.

\section{Estimating \TOP in Small Space}\label{sec:SSA}

%In this section, we 
We present \ApproxTopK, an algorithm for \emph{estimating} the set of \TOP frequent substrings %in a string $S$ of length $n$, for any integer $K>0$, 
in small space. 

\medskip

\noindent{\bf High-Level Idea.}~\ApproxTopK employs sampling and indexing data structures that need small space. It uses a user-defined parameter $s\in [1,n]$, which trades time efficiency (and accuracy) for space, and executes $s$ rounds of sampling. In Round $i\in [0, s)$, it performs the following steps:
\begin{enumerate}
    \item Samples positions $i+r\cdot s$ of $S$, for each $
    r\in[0, \lceil n/s \rceil]$. 
    \item Constructs a \emph{sparse} index only for the suffixes starting at the sampled positions. 
    \item Finds the $K$ substrings which occur the most at the sampled positions (i.e., \TOP frequent in the sample). 
    \item Merges the set of substrings found in Step 3 with those found until Round $i-1$ (if any). 
\end{enumerate}
After all rounds of sampling, the algorithm returns the set of substrings that are constructed in Step 4. Note that this approach does not always produce the true set $T_K$ of \TOP frequent substrings of $S$ because, by design, the frequency of a substring in $T_K$ may be computed incorrectly if this substring is not part of the \TOP frequent substrings in at least one sample. However, the error in the frequencies is one-sided: the frequencies reported by \ApproxTopK lower bound the true frequencies of the output substrings, thus no frequency is over-estimated. In addition, the following bounds hold. 
\begin{theorem}\label{the:topSSA}
For any string $S$ of length $n$, any integer $K>0$,
and any parameter $s\in[1,n]$, Algorithm \ApproxTopK takes $\tilde{\cO}(n+sK)$ time and the extra space on top of the space needed by $S$ is $\cO(n/s+K)$.
\end{theorem}

Let us look at the two extremes: when $s=1$, we have one sample and so the algorithm is essentially the same as the one in Section~\ref{sec:ESA} that uses $\cO(n+K)$ extra space, $\cO(n+K)$ time, and is \emph{exact};
when $s=\Theta(n)$, we have $\Theta(n)$ samples, the algorithm takes $\cO(K)$ extra space, $\tilde{\cO}(nK)$ time, 
and the estimation will be most probably very bad. In practice, we set $s$ to a small function of $n$, such as $\cO(\log n)$. This results in sublinear extra space, a reasonable running time, and high accuracy, as we will show later in Section~\ref{sec:experiments}.

Note that, although \ApproxTopK works for any $K>0$, it makes sense to use it when $K<n$, as otherwise one can simply use \ExactTopK that takes $\cO(n+K)$ time using $\cO(n)$ extra space. %Some other choice for $s$ is thus $s:=n/K$, which gives $\tilde{\cO}(n)$ time and $\cO(K)$ extra space.

\noindent {\bf Details.}~As a preprocessing step, we construct on $S$ the in-place \emph{Longest Common Extension} (LCE) data structure of Prezza~\cite{DBLP:journals/talg/Prezza21} in $\cO(n)$ time and $\cO(1)$ extra space. This data structure answers LCE queries on $S$ in $\cO(\text{polylog}(n))$ time: given two integers $i,j\in[0,n)$, the LCE query asks for the length of the longest common prefix of $S[i\dd n-1]$ and $S[j\dd n-1]$. This data structure will efficiently implement the string comparison functions used in Steps $2$ to $4$. After each Round $i$, we store the set of the \TOP frequent substrings as a set of tuples $\langle j,\ell,f_{[0,i]}\rangle$: $S[j\dd j+\ell-1]$ is a witness occurrence for substring $U=S[j\dd j+\ell-1]$; and $f_{[0,i]}$ is the total estimated frequency of $U$ in Rounds $0,\ldots,i$. Step 1 is straightforward. The next steps are provided below. 

\noindent {\bf Step 2.}~At any round $i$, we construct an index consisting of the sparse suffix array $\textsf{SSA}_i$~\cite{ssapaper} and the sparse LCP array $\textsf{SLCP}_i$~\cite{ssapaper} for the sampled positions.
$\textsf{SSA}_i$ consists of the lexicographically sorted sequence of the suffixes of $S$ starting at the sampled positions. To construct it, we use in-place mergesort~\cite{DBLP:journals/njc/KatajainenPT96}, where any two substrings can be compared in $\cO(\text{polylog}(n))$ time by finding the length $\textsf{lce}$ of their longest common prefix with the LCE data structure of Prezza~\cite{DBLP:journals/talg/Prezza21}, and comparing the letters at position $\textsf{lce}+1$ in each substring. 
Therefore, lexicographically sorting the $\lceil n/s \rceil$ sampled suffixes via in-place mergesort requires $\tilde{\cO}(n/s)$ time and no extra space. To construct $\textsf{SLCP}_i$, we then compute the length of the longest common prefix of every two consecutive entries of $\textsf{SSA}_i$, again using $\cO(\text{polylog}(n))$-time \textsf{LCE} queries. This procedure thus requires $\tilde{\cO}(n/s)$ time per round.

\noindent {\bf Step 3.}~To compute the \TOP frequent substrings from $\textsf{SSA}_i$ and $\textsf{SLCP}_i$, we apply the algorithm of Abouelhoda et al.~\cite[Algorithm 4.4]{DBLP:journals/jda/AbouelhodaKO04}, which simulates a bottom-up traversal of the compacted trie of the suffixes in $\textsf{SSA}_i$ (note that, for using this algorithm, we do not need to construct such a trie). 
This traversal requires $\cO(n/s)$ time and produces one tuple $\langle \textsf{lcp},\textsf{lb},\textsf{rb},\textsf{childList} \rangle$ per explicit node $v$ of the trie: $\textsf{lcp}=|\textsf{str}(v)|$ is the string depth of node $v$, i.e., the length of the substring represented by node $v$ in the trie; $[\textsf{lb},\textsf{rb}]$ encodes all suffixes $\textsf{SSA}_i[\textsf{lb}\dd \textsf{rb}]$ in $\textsf{SSA}_i$ with $\textsf{str}(v)$ as prefix, thus $\textsf{rb}-\textsf{lb}+1$ is the frequency of $\textsf{str}(v)$ in the sample; and, finally, \textsf{childList} is a list storing the children of node $v$ in the trie. These $\cO(n/s)$ tuples have the same role as the tuples used in Task (i) of Section~\ref{sec:ESA}, and we can sort them in ascending order of $\textsf{rb}-\textsf{lb}+1$ (i.e., by frequency) in linear time using radix sort.
This is because any frequency is at most $n/s$, the maximal number of suffixes in Round $i$. 
Listing the \TOP frequent substrings of the $i$th round, similarly to Task (i) of Section~\ref{sec:ESA}, takes $\tilde{\cO}(n/s+K)$ time. 
Each substring $U$ in the list is represented by a tuple $\langle j,\ell,f_{[i,i]} \rangle$: $S[j\dd j+\ell-1]$ is a witness occurrence for $U$ and $f_{[i,i]}=\textsf{rb}-\textsf{lb}+1$ is its frequency in the $i$th sample.

\noindent\textbf{Step 4.}~We efficiently merge the list of the \TOP frequent substrings found in Rounds $0, \ldots, i-1$ with the list of the \TOP frequent substrings computed in Step 3 of Round $i$, and only keep the \TOP frequent substrings in the merged list. The merged list consists of the union of the substrings appearing in either list: the frequency of a substring in the merged list is given by the sum of its frequency in each of the two original lists. To produce the merged list, we thus need to efficiently find out which substrings appear in both lists to sum their frequencies.
To do so, we concatenate the list of $K$ tuples $\langle j,\ell,f_{[0,i-1]}\rangle$ produced by Step 4 of Round $i-1$ with the list of $K$ tuples $\langle j,\ell,f_{[i,i]} \rangle$ produced by Step 3 of Round $i$ and sort the resulting list in lexicographic order of the substrings represented by the tuples using in-place mergesort~\cite{DBLP:journals/njc/KatajainenPT96}, similarly to Step 2. We then scan this sorted list: substrings appearing in both lists will be represented by adjacent tuples, thus we can sum up their frequencies to produce the new estimated frequencies $f_{[0,i]}$ in total $\tilde{\cO}(K)$ time and no extra space. 
Once we have produced the merged list, we sort it again, this time in decreasing order of the frequencies $f_{[0,i]}$, using another round of mergesort.  
We finally output the first $K$ tuples $\langle j,\ell,f_{[0,i]}\rangle$ of this sorted list, representing the \TOP frequent substrings found until Round $i$. 
We have arrived at Theorem~\ref{the:topSSA}.

\vspace{+1mm}
\noindent\textbf{A Space-Efficient Construction Algorithm for $\USItop$.} \ApproxTopK can be employed in the construction of $\USItop$ instead of \ExactTopK to make it more space-efficient. However, the worst-case query time of this space-efficient version of $\USItop$ is no longer $\cO(m+\tau_K)$: indeed, since the output of  \ApproxTopK is not exact, we can no longer guarantee that $\tau_K$ bounds the frequency of any substring whose global utility is not stored in the hash table $H$. This implies that, in the worst case, the query time can be $\cO(n)$. However, in Section~\ref{sec:experiments}, we show that queries are much faster in practice and competitive to the exact counterpart.

As for the construction of this space-efficient data structure, we can use the sliding-window approach described in Section~\ref{sec:USI_DS}.
The total construction time is $\tilde{\cO}(nL_k+sK)$: $\tilde{\cO}(n+sK)$ time to run \ApproxTopK plus $\cO(nL_K)$ to construct the data structure with the sliding-window procedure. 
Although asymptotically the construction space of this space-efficient data structure is $\cO(n+K)$, thus the same as for $\USItop$, in practice it is 
determined by the space needed by \ApproxTopK, which is significantly lower than that of \ExactTopK (see Section~\ref{sec:experiments}).

\section{Why not Modifying a Frequent Item Mining Algorithm?}\label{sec:whynot}

Assume we have a stream of $N$ \emph{items}. 
%Misra and Gries~\cite{DBLP:journals/scp/MisraG82} extended the Boyer-Moore Majority algorithm~\cite{BoyerMoore1981} to find the items occurring more than $N/K$ times in a stream $S$ of $N$ items, where $K \geq 1$ is an input parameter. The algorithm maintains $K-1$ pairs $(\mathit{item}, \mathit{counter})$, suitably updating them while scanning $S$. 
Demaine et al.~\cite{DemaineESA} proved that for any $N$ and $K$, a one-pass deterministic algorithm storing at most $K$ items may fail to identify the top-$K$ most frequent items in certain sequences (this is related to~\cite{DBLP:journals/scp/MisraG82}).
%: exactly finding the top-$K$ most frequent items requires large memory, proportional to the number of distinct items, which can be $\cO(N)$ in the worst case. 
%Approximate streaming algorithms for the top-$K$ most frequent items follow two strategies: \emph{count-all} (e.g., CM sketch~\cite{CormodeM12}) and \emph{admit-all-count-some} (e.g.,  Misra-Gries~\cite{DBLP:journals/scp/MisraG82} or Space-Saving~\cite{DBLP:conf/icdt/MetwallyAA05}). 
Thus, approximate streaming algorithms for the top-$K$ most frequent items have been proposed~\cite{CormodeM12,DBLP:journals/scp/MisraG82,DBLP:conf/icdt/MetwallyAA05}. 

We consider a variation to the task of estimating the \TOP most frequent items. We treat the stream as our string $S$ of length $n=N$, where letters are streamed one by one, and we aim to identify the top-$K$ most frequent \emph{substrings} of $S$. Thus, we must consider not only single letters (items) $S[i]$, but whole substrings $S[i \dd i+\ell - 1]$, for any length $\ell > 1$. While there may be $\cO(N^2)$ distinct substrings, the problem can be solved exactly using $N$ counters by means of the suffix tree, which can be built online, one letter at a time~\cite{Ukkonen}: indeed, for $K \geq N$, the suffix tree provides an exact solution. 

We now consider whether a solution exists for $K < N$. For $K \leq |\Sigma|$, an exact solution is not possible, as implied by~\cite{DemaineESA}. In the remaining case $|\Sigma| < K < N$, we are not aware of an exact solution, but we can discuss two approximate solutions, which are from the literature and can fail in estimating the \TOP most frequent substrings when using $K$ counters: (1) \SubstringHeavy, an adaptation of \textsf{HeavyKeeper}~\cite{HeavyKeeper2018} 
to \emph{substrings} of a single string $S$ (\textsf{HeavyKeeper} is the state of the art for computing the \TOP frequent \emph{strings} in a database of several strings); and (2) \TopKTrie~\cite{DBLP:conf/wea/DinklageFP24}, which implements a variant of the Misra-Gries algorithm to
approximately find the \TOP frequent substrings of $S$. 
%\footnote{In~\cite{DBLP:conf/wea/DinklageFP24},  \TopKTrie is employed in an online manner to summarize the top-$K$ frequent consecutive patterns based on a combination of the Lempel-Ziv compression scheme and the Misra-Gries algorithm.}
We note that both solutions fail due to the extension from \textit{items}  to \textit{substrings} in $S$ of the Misra-Gries/Space-Saving scheme with $K$ counters. 

\smallskip
\noindent
{\bf {\small \SubstringHeavy}.}~The strategy in~\cite{HeavyKeeper2018} combines the count-all and admit-all-count-some methods for strings and relies on a CM sketch-like structure~\cite{CormodeM12} with exponential decay and on a summary \emph{ssummary} which tracks the frequency of $K$ strings for fast queries. We adapt it to substrings from $S$ with this rule: for any $i$, try to insert $S[i]$ into \emph{ssummary}, and then try to insert $S[i \dd i+\ell]$, $\ell\geq1$, only if $S[i \dd i+\ell - 1]$ is in \emph{ssummary}.
A string is successfully inserted into \emph{ssummary} if its estimated frequency, stored in the CM sketch table, is larger than the frequency of a string in \emph{ssummary}, or if the latter contains fewer than $K$ strings. Here, the frequency value of a string is the number of times it has been a candidate for insertion into \emph{ssummary}.
%
% In the generalization to substrings, any processed element is a substring: while scanning the stream $S$, for each position $i$, we consider a variable number $c_i>0$ of substrings starting at $i$ with length from 1 to $c_i$, depending on the previous summary insertions. These strings have a counter in \textit{ssummary} if their CM sketch-like value is large enough. 
% Differently from the original paper~\cite{HeavyKeeper2018}, where $c_i = 1$ as individual elements are considered, the difficulty is to establish (determinstically or randomly) the value of $c_i$, as we can only fit $K$ counters in \textit{ssummary}. 
%
To hash substrings we use KR fingerprints~\cite{karp1987efficient} and pay $\cO(1)$ time per substring; hence, for a total number $z$ of hashed substrings, this requires $\cO(z)$ time and $\cO(K)$ space. On average, $z$ is linear in $n$, since the probability of extending the next letter of the current length-$\ell$ substring is programmatically chosen to be $1/c^\ell$ for a constant $c > 1$, which implies expected $\cO(1)$ time per letter. 
\SubstringHeavy can fail for, say, $S=(\texttt{AB})^{n/2}$ 
%, namely $\texttt{AB}$ repeated $n/2$ times,
where $n/2 \geq K > 4$, $K$ is even, and $|\Sigma|=2$. %(other choices are possible). 
In this example, \SubstringHeavy fails to report half of the output. The details are deferred to~\cite{supplement}.

\smallskip
\noindent
{\bf {\small \TopKTrie}.}~The authors of~\cite{DBLP:conf/wea/DinklageFP24} introduced \TopKTrie, a novel trie data structure that approximately maintains the top-$K$ most frequent substrings in $S$ in $\cO(K)$ space, and reports them in $\cO(n+K)$ time. %Each trie node except the root has a counter, initially set to zero. 
Just as \SubstringHeavy, it can fail to report half of the output. The details are deferred to~\cite{supplement}.  

\section{Related Work}\label{sec:related_work} 

%The abundance of high-dimensional data, such as set-valued data or strings, which are associated with utilities, led to the development of a large number of works. 
%In particular, t
There are many works for \emph{mining}  
utility-oriented itemsets~\cite{DBLP:journals/tkde/AhmedTJL09},  
association rules~\cite{icdm2002}, 
and episodes~\cite{10.1145/2487575.2487654, 9960731,icdm13}. 
Unlike set-valued data, strings with utilities have duplicate elements \emph{and} the order of the elements is crucial. % (e.g., think of a long DNA sequence).  
Unlike event sequences, strings with utilities have no specific temporal information, and thus the notion of time window is irrelevant. Thus, mining strings  with utilities requires specialized algorithms~(e.g.,~\cite{sdm24}). %above algorithms are not applicable to strings.

Recently, two algorithms for mining utility-oriented substrings were proposed in~\cite{sdm24}. Both take $\cO(n\log n)$ time but one of them offers drastic space savings in practice when, in addition, a lower bound on the length of the output strings is provided as input.  There are also algorithms that are applied to a collection of short strings comprised of letters or itemsets~\cite{DBLP:conf/kdd/YinZC12,tkde15,infsci2020}. These algorithms mine subsequences. Our work differs from the aforementioned works in that it focuses on query answering and in that it uses different utility functions. %along two important dimensions. First, we focus on a query-answering (rather than on a mining) setting. Second, we use different utility functions than the existing algorithms.  

Our approach includes finding top-$K$ frequent substrings. Thus, it is related to the literature on
top-$K$ pattern mining. There are works for 
mining top-$K$ frequent patterns (e.g., frequent itemsets~\cite{vldbjfreq,dmkdfreq} and closed frequent itemsets~\cite{vldbjfreq}) or association rules~\cite{tkddfreq} from set-valued data, and works for mining top-$K$ patterns from sequential data (e.g., sequential patterns~\cite{tattifreq} and closed sequential patterns~\cite{kaisfreq}). These are not alternatives to our approach, as they mine different types of patterns than substrings. The most relevant to our work are algorithms for estimating the top-$K$ most frequent items in a stream~(e.g.,~\cite{HeavyKeeper2018,DBLP:conf/wea/DinklageFP24}). As discussed in Section~\ref{sec:whynot} and will be experimentally shown in Section~\ref{sec:experiments}, these algorithms cannot be suitably adapted to mine top-$K$ frequent substrings.

\section{Experimental Evaluation}\label{sec:experiments}

\subsection{Data and Environment}

\noindent {\bf Data.}~We used $5$ real datasets of sizes up to 4.6 billion letters; see Table~\ref{table:data} for their characteristics. 
These include the \ADV dataset from Section~\ref{sec:case_study}, in which every advertisement is associated with a real CTR value.
The strings \IoT~\cite{iotdataset} and  \FASTQ~\cite{ecolidataset} are also associated with real utilities. In \IoT, the utilities are RSSIs (Received Signal Strength Indicators representing signal strength values of sensors) normalized in $[0,1]$, and in \FASTQ, confidence scores~\cite{phred} in $[0,1]$; see Section~\ref{sec:intro}. 
In \XML~\cite{xmldataset} and \HUM~\cite{fullgenom}, there are no real utilities. Thus, we selected each utility $w[i]$, for all $i\in[0,n)$,  uniformly at random from $\{0.7, 0.75, \ldots, 1\}$ as in~\cite{sdm24}.  
%as it was too small for the purpose of our evaluation. 

\begin{figure*}[!ht]
    \centering
    \begin{subfigure}{0.19\textwidth}
        \includegraphics[width=\textwidth]{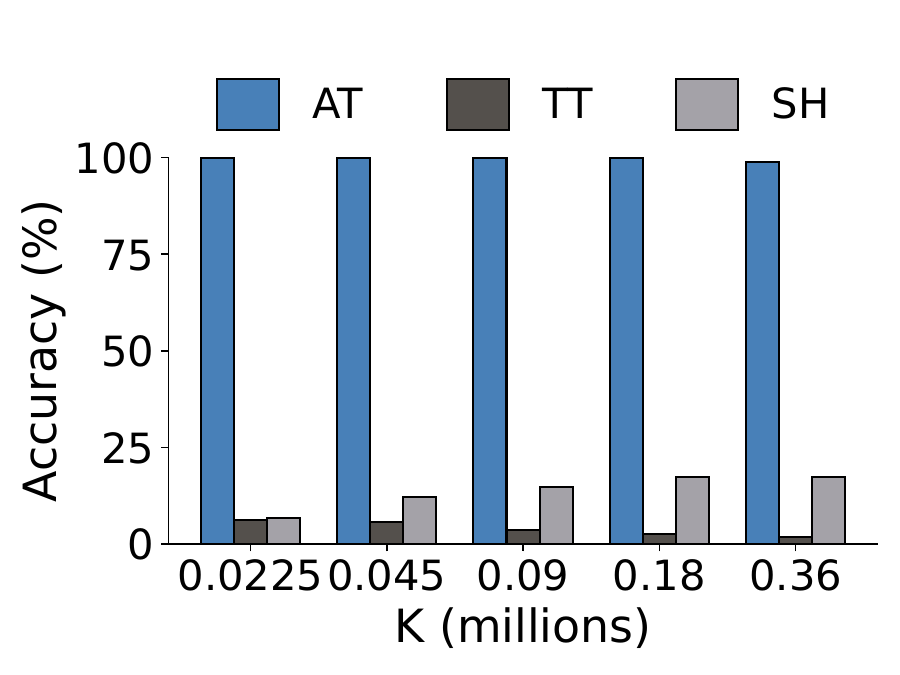}
        \vspace{-6mm}
        \caption{\IoT}
        \label{fig:iot_accuracy_K}
    \end{subfigure}
    \begin{subfigure}{0.19\textwidth}
        \includegraphics[width=\textwidth]{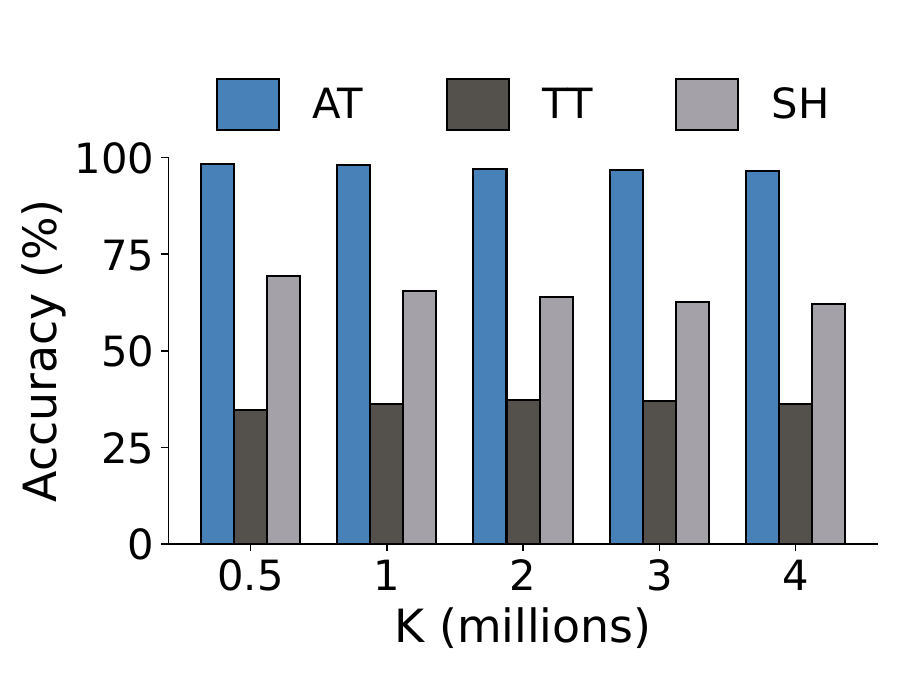}
        \vspace{-6mm}
        \caption{\XML}
        \label{fig:xml_accuracy_K}
    \end{subfigure}
    \begin{subfigure}{0.19\textwidth}
        \includegraphics[width=\textwidth]{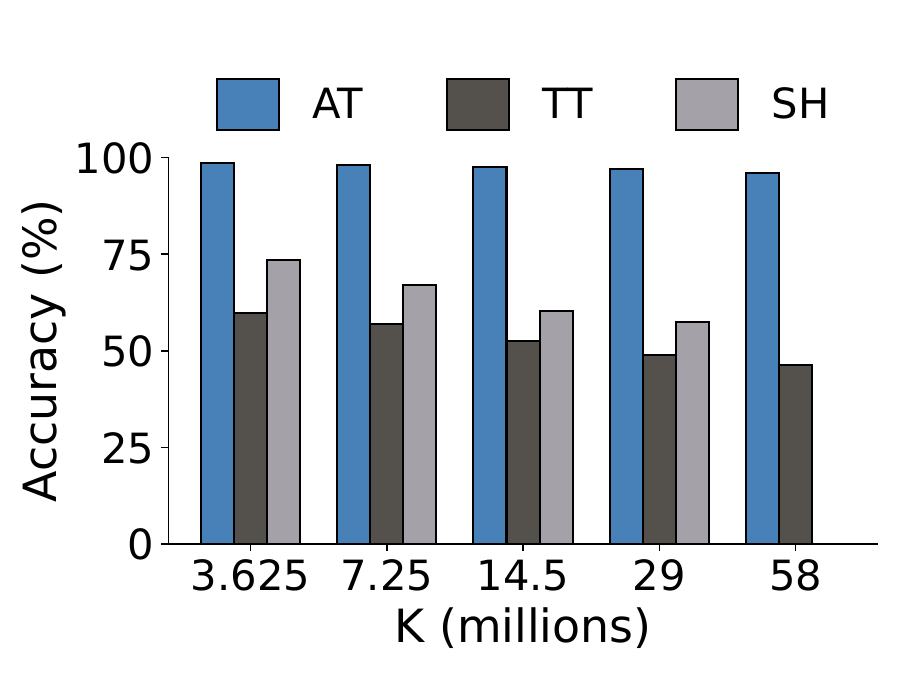}
        \vspace{-6mm}
        \caption{\HUM}
        \label{fig:hum_accuracy_K}
    \end{subfigure}
    \begin{subfigure}{0.19\textwidth}
        \includegraphics[width=\textwidth]{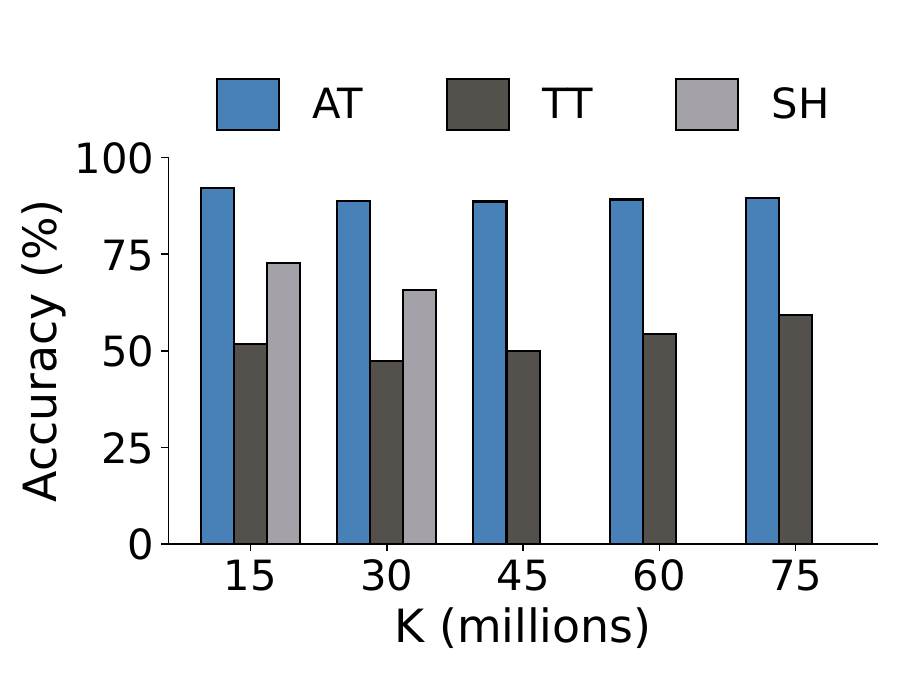}
        \vspace{-6mm}
        \caption{\ECOLI}
        \label{fig:ecoli_accuracy_K}
    \end{subfigure}
    \begin{subfigure}{0.19\textwidth}
        \includegraphics[width=\textwidth]{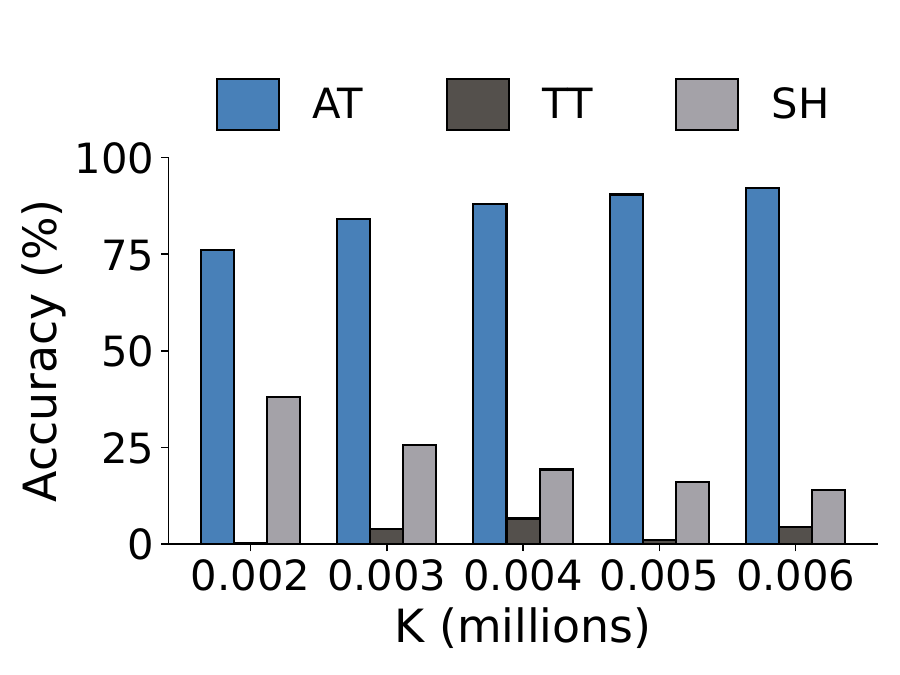}
        \vspace{-6mm}
        \caption{\ADV}
        \label{fig:adv_accuracy_K}
    \end{subfigure}\\
    \begin{subfigure}{0.19\textwidth}
        \includegraphics[trim=6 4 5 30,clip,width=\textwidth]{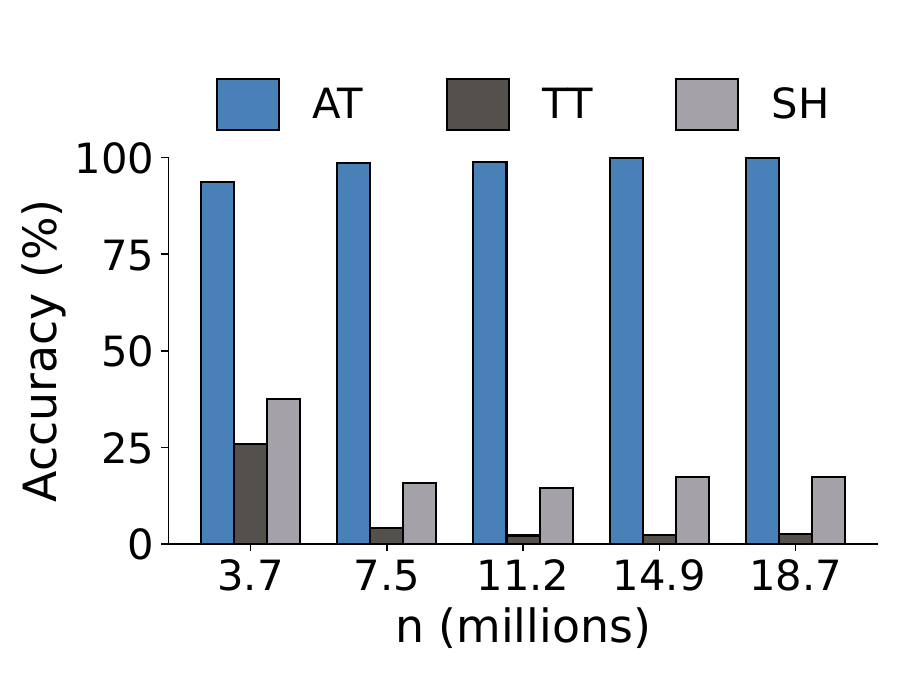}
        \vspace{-6mm}
        \caption{\IoT}
        \label{fig:iot_accuracy_n}
    \end{subfigure}
        \begin{subfigure}{0.19\textwidth}
        \includegraphics[trim=6 4 5 30,clip,width=\textwidth]{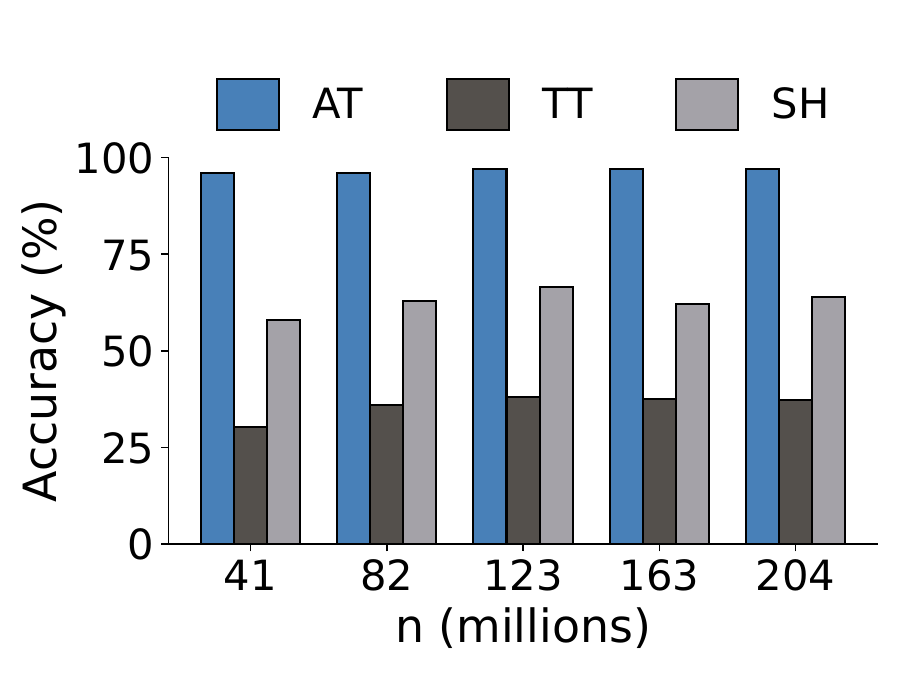}
        \vspace{-6mm}
        \caption{\XML}
        \label{fig:xml_accuracy_n}
    \end{subfigure}
    \begin{subfigure}{0.19\textwidth}
        \includegraphics[trim=6 4 5 30,clip,width=\textwidth]{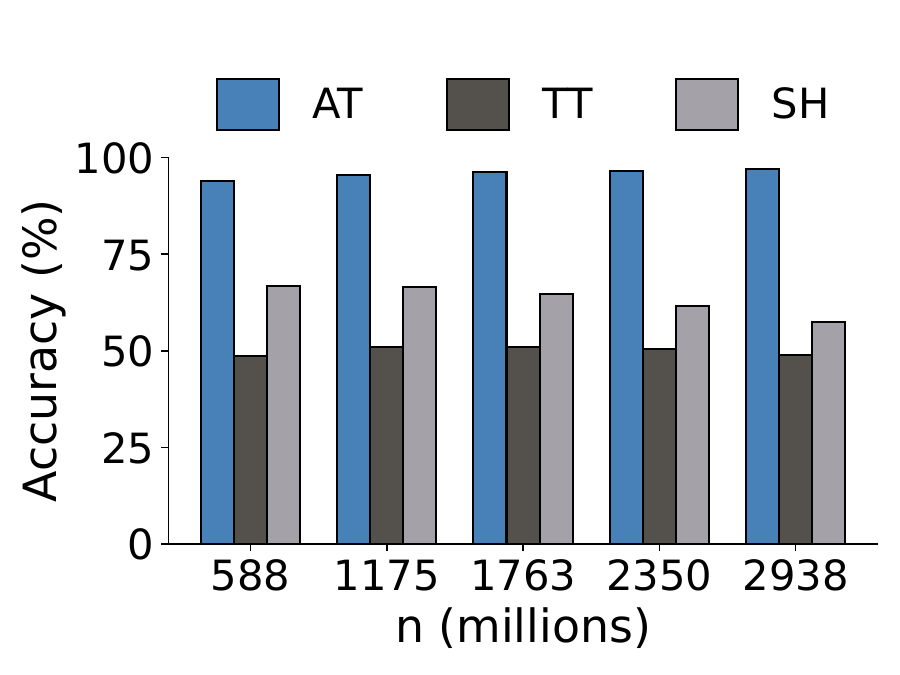}
        \vspace{-6mm}
        \caption{\HUM}
        \label{fig:hum_accuracy_n}
    \end{subfigure}
    \begin{subfigure}{0.19\textwidth}
        \includegraphics[trim=6 4 5 30,clip,width=\textwidth]{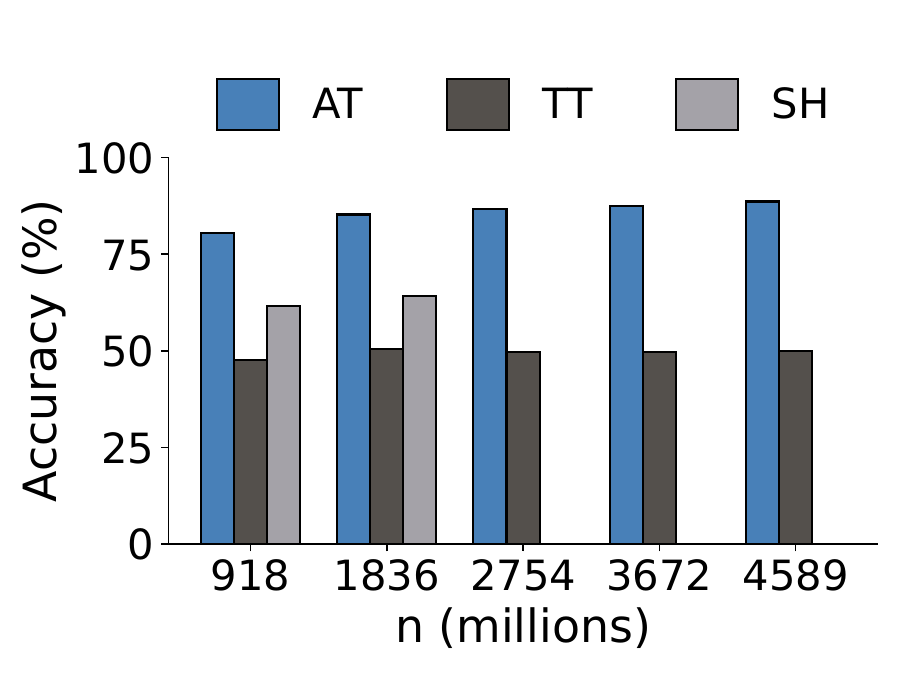}
        \vspace{-6mm}
        \caption{\ECOLI}
        \label{fig:ecoli_accuracy_n}
    \end{subfigure}
    \begin{subfigure}{0.19\textwidth}
        \includegraphics[trim=6 4 5 30,clip,width=\textwidth]{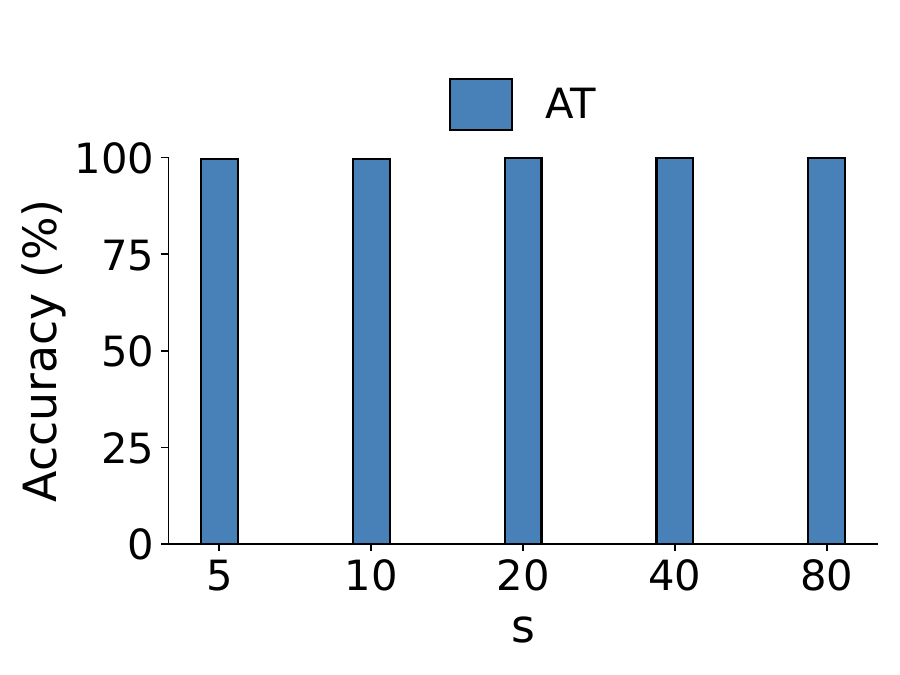}
        \vspace{-6mm}
        \caption{\IoT}
        \label{fig:iot_accuracy_s}
    \end{subfigure}
    \vspace{-2mm}
    \caption{Accuracy vs (a-e) $K$, (f-i) $n$, and (j) $s$ ($s$ affects only $\AT$). We omit \SH when it did not terminate within $5$  days.
    }
\vspace{-4mm}
\end{figure*}

\begin{figure*}[!ht]
    \centering
    \begin{subfigure}{0.19\textwidth}
        \includegraphics[trim=6 4 5 30,clip,width=\textwidth]{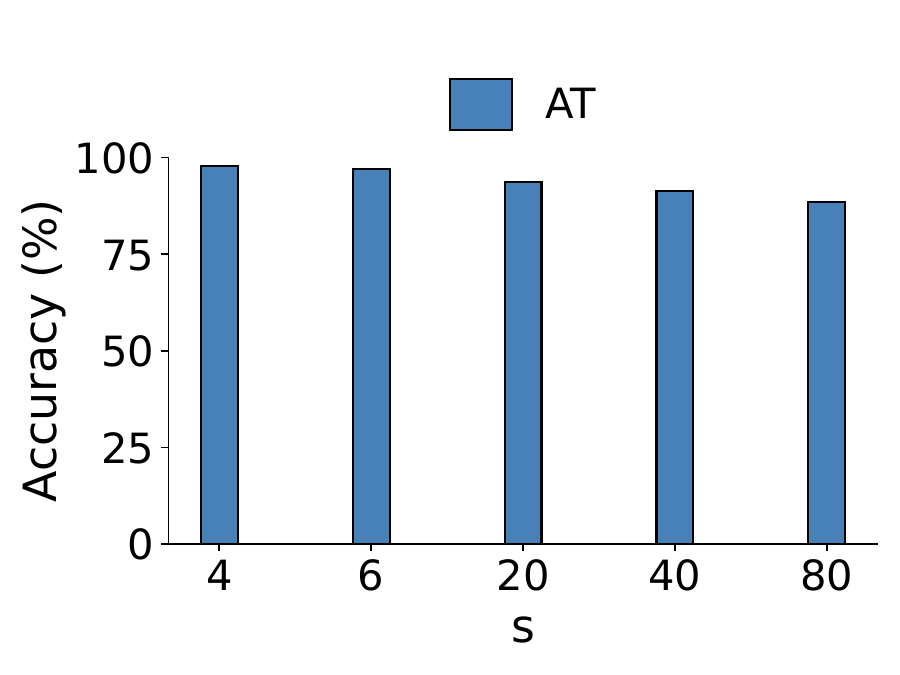}
        \vspace{-6mm}
        \caption{\XML}
        \label{fig:xml_accuracy_s}
    \end{subfigure}
    \begin{subfigure}{0.19\textwidth}
        \includegraphics[trim=6 4 5 30,clip,width=\textwidth]{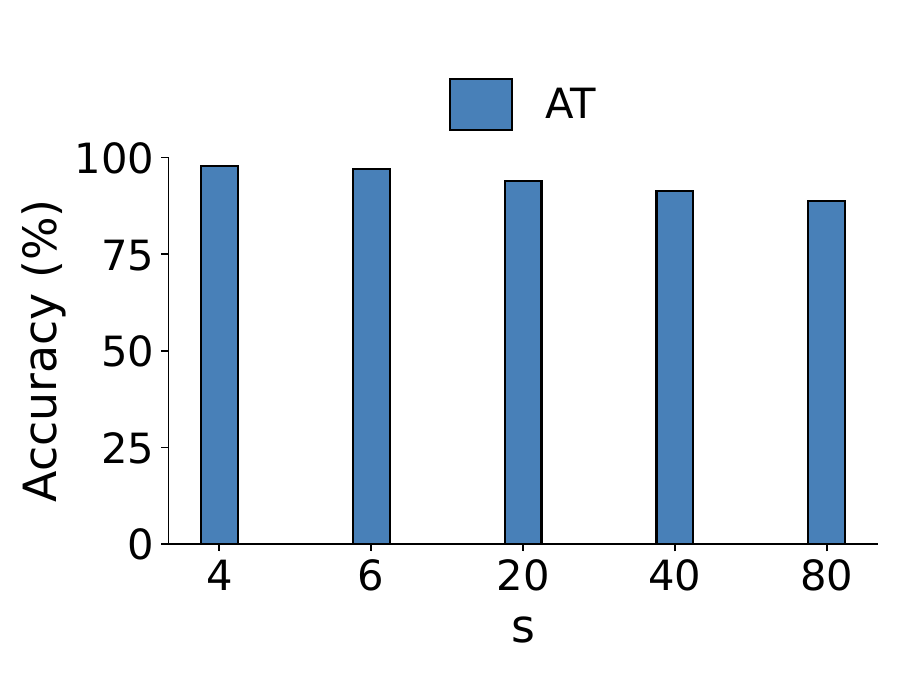}
        \vspace{-6mm}
        \caption{\HUM}
        \label{fig:hum_accuracy_s}
    \end{subfigure}
    \begin{subfigure}{0.19\textwidth}
        \includegraphics[trim=6 4 5 25,clip,width=\textwidth]{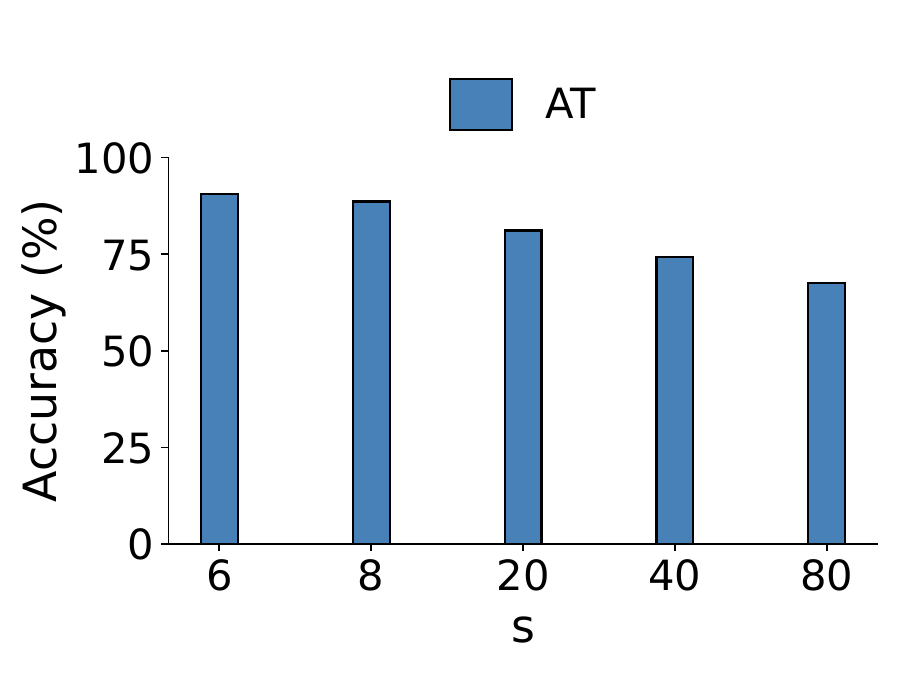}
        \vspace{-6mm}
        \caption{\ECOLI  }
        \label{fig:ecoli_accuracy_s}
    \end{subfigure}
      \begin{subfigure}{0.19\textwidth}
      \hspace{+3mm}
        \includegraphics[trim=0 0 0 0,clip,width=\textwidth]{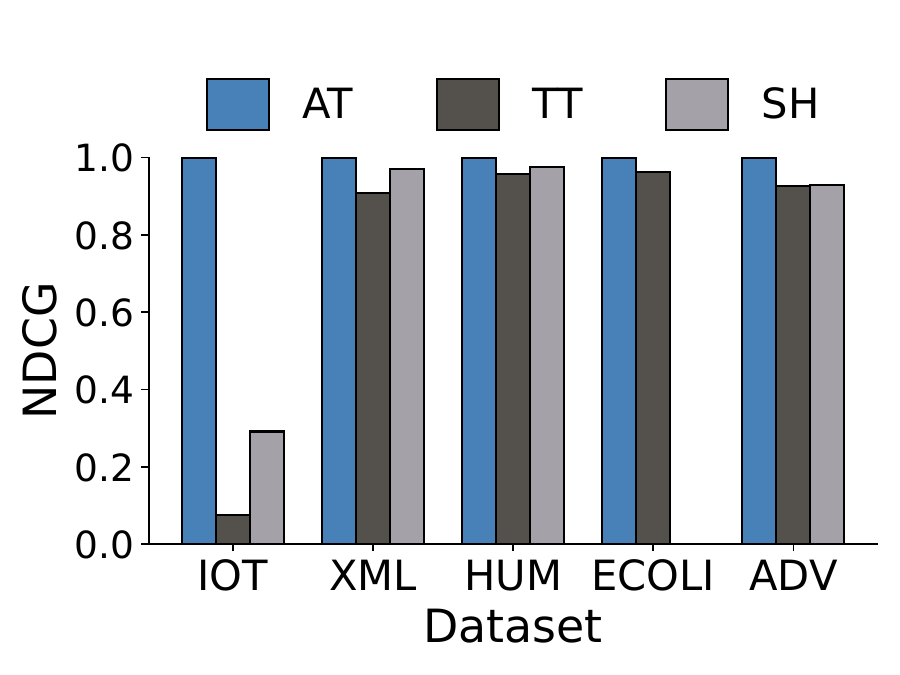}
        \vspace{-6mm}
        \caption{All datasets}
        \label{fig:all_ndcg}
    \end{subfigure}
    \begin{subfigure}{0.19\textwidth}
        \includegraphics[trim=6 4 5 30,clip,width=\textwidth]{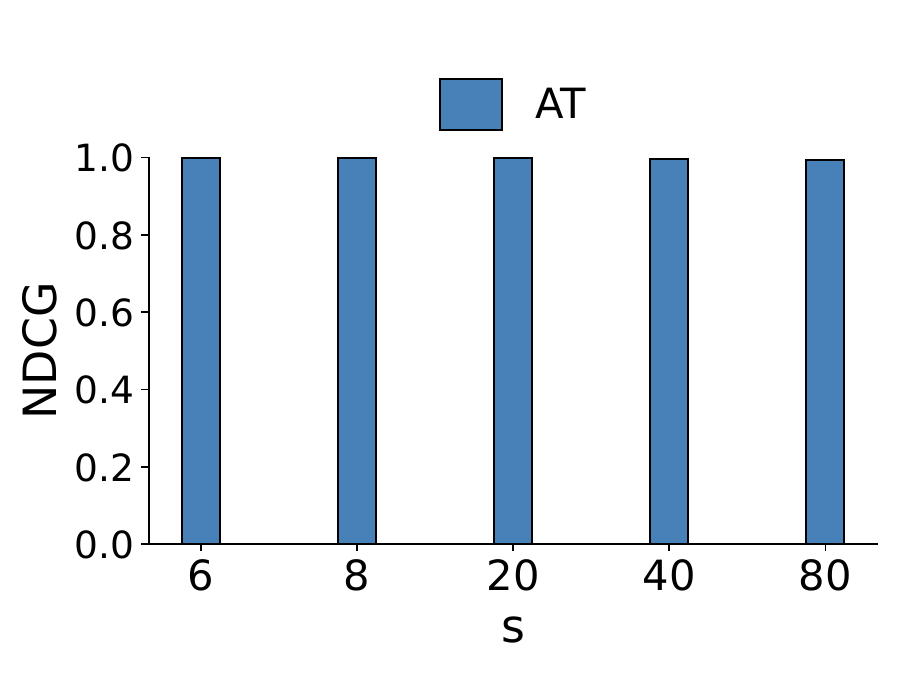}
         \vspace{-6mm}
        \caption{\ECOLI  }
        \label{fig:ecoli_ndcg_s}
    \end{subfigure}
    \vspace{-2mm}
    \caption{Accuracy vs (a-c) $s$. \textsc{NDCG} (d) for all datasets and (e) vs $s$. We omit \SH when it did not terminate within $5$ days.}
\vspace{-4mm}
\end{figure*}

% \begin{figure}
% \vspace{-2mm}
%     \centering
%     \begin{subfigure}{0.19\textwidth}
%         \includegraphics[width=\textwidth]{images/DNA_vss_acc.pdf}
%         \vspace{-4mm}
%         \caption{\HUM}
%         \label{fig:hum_accuracy_s}
%     \end{subfigure}
%     \begin{subfigure}{0.19\textwidth}
%         \includegraphics[width=\textwidth]{images/ECOLI_vss_acc.pdf}
%         \vspace{-4mm}
%         \caption{\ECOLI  }
%         \label{fig:ecoli_accuracy_s}
%     \end{subfigure}
%     \vspace{-2mm}
%     \caption{Accuracy for \AT vs (a, b) $s$.}
% \end{figure}
% \begin{figure}
% \vspace{-2mm}
%     \centering
%     \begin{subfigure}{0.19\textwidth}
%         \includegraphics[width=\textwidth]{images/DNA_vss_acc.pdf}
%         \vspace{-4mm}
%         \caption{\HUM}
%         \label{fig:all_ndcg}
%     \end{subfigure}
%     \begin{subfigure}{0.19\textwidth}
%         \includegraphics[width=\textwidth]{images/NDCG_ECOLI_vss.pdf}
%         \vspace{-4mm}
%         \caption{\ECOLI  }
%         \label{fig:ecoli_ndcg_s}
%     \end{subfigure}
%     \vspace{-2mm}
%     \caption{\textsc{NDCG} (a) for all datasets and (b) vs $s$. We omit \SH when it did not terminate within $5$ days.}
% \end{figure}

\noindent {\bf Environment.}~All experiments were conducted on an AMD EPYC 7282 CPU with 256 GB RAM. All methods were implemented in \texttt{C++}. %and compiled with \texttt{g++} (v.~11.4.0) at optimization level \texttt{-O3}. 
The source code is available at %\url{https://bit.ly/3V5Z0wO}.
\url{https://github.com/chenhuiping/Utility-Oriented-String-Indexing}.

\begin{table}[t]
\caption{Dataset properties and values of parameters. The default values are in bold. M stands for millions.}\label{table:data}
\centering
\resizebox{0.48\textwidth}{!}{%
\begin{tabular}{|c||c|c||c|c|}
\hline
{\bf Dataset} & {\bf Length }                    & {\bf Alphabet}  & {\bf Number of \TOP}                                    & {\bf Number of sampling}                 \\
~ & $n$ &  {\bf size} $\sigma$ & {\bf frequent substrings} $K$ & {\bf rounds} $s$ \\\hline\hline
\ADV~\cite{ad_dataset}    & $2.19\cdot10^5$  & 14        & {[}2K, 6K{]} (\textbf{6K}) &  \textbf{6}\\ \hline
\IoT~\cite{iotdataset}     & $1.9\cdot10^7$ & 63       & {[}0.0225M,0.36M{]} (\textbf{0.18M})         & {[}10, 80{]} (\textbf{20}) \\ \hline
\XML~\cite{xmldataset}     & $2\cdot10^8$    & 95       & {[}0.2M, 3M{]} (\textbf{2M})     & {[}4, 80{]} (\textbf{6}) \\ \hline
\HUM~\cite{fullgenom}     & $2.9\cdot10^9$  & 4        & {[}3.6M, 58M{]} (\textbf{29M})  & {[}4, 80{]} (\textbf{6}) \\ \hline
\ECOLI~\cite{ecolidataset}   & $4.6\cdot10^9$  & 4        & {[}15M, 75M{]} (\textbf{45M}) & {[}4, 80{]} (\textbf{8}) \\ \hline
\end{tabular}}
\end{table}

\subsection{Top-$K$ Frequent Substring Mining}\label{sec:experiments:topkfreq}

\noindent{\bf Methods.}~We compared our \ExactTopK (\ET) and \ApproxTopK (\AT) algorithms to 
\SubstringHeavy (\SH) and \TopKTrie (\TT) 
from Section~\ref{sec:whynot}. The default values for $K$ in \ET and for $K$ and $s$ in \AT are in Table~\ref{table:data}. The range of $s$ values is in $\cO(\log n)$, as discussed in Section~\ref{sec:SSA}.

\noindent{\bf Measures.}~Let $T_K$ be the set of \TOP frequent substrings and $T'_K$ that of the substrings found by an algorithm that estimates $T_K$. We used \emph{Accuracy}, defined as the percentage of substrings in $T'_K$ with the same frequency as those in $T_K$, \emph{Relative Error} (RE), defined as $\frac{\sum_{P\in T_K}| \occ_S(P)|-\sum_{P'\in T'_K}|\occ_S(P')|}{\sum_{P\in T_K}{|\occ_S(P)|}}$, and \emph{Normalized Discounted Cumulative Gain (NDCG)}~\cite{ngcd} using the frequencies of the substrings in $T_K$ in \emph{Ideal DCG} and those of the substrings in $T'_K$ in \emph{DCG}. These quantify the efficiency loss of an algorithm that estimates $T_K$ by $T'_K$ when answering queries with a frequency smaller than those in $T_K$. 

% \noindent{\bf Overview.}~We show that: (1) \AT is remarkably effective, very close to the exact \ET algorithm. For example, the Accuracy of \AT was on average $94.9\%$ over all tested $K$ values. (2) \AT is much more space-efficient than \ET, especially when $K$ is small and $s$ is large. These results are in line with our complexity analyses. (3) \ET is much faster than \AT (recall that it needs $\cO(n+K)$ time instead of $\tilde{\cO}(n +sK)$ time). (4) Both \TT and \SH are \emph{not practical} (e.g., their average accuracy over all tested $K$ values was $25.7\%$ and $44.3\%$, respectively). (5) \TT needs less space and is faster than \ET and \AT, while \SH is slower than \ET and needs similar space as \AT.    

\begin{figure*}[!ht]
    \centering
    \begin{subfigure}{0.19\textwidth}
        \includegraphics[width=\textwidth]{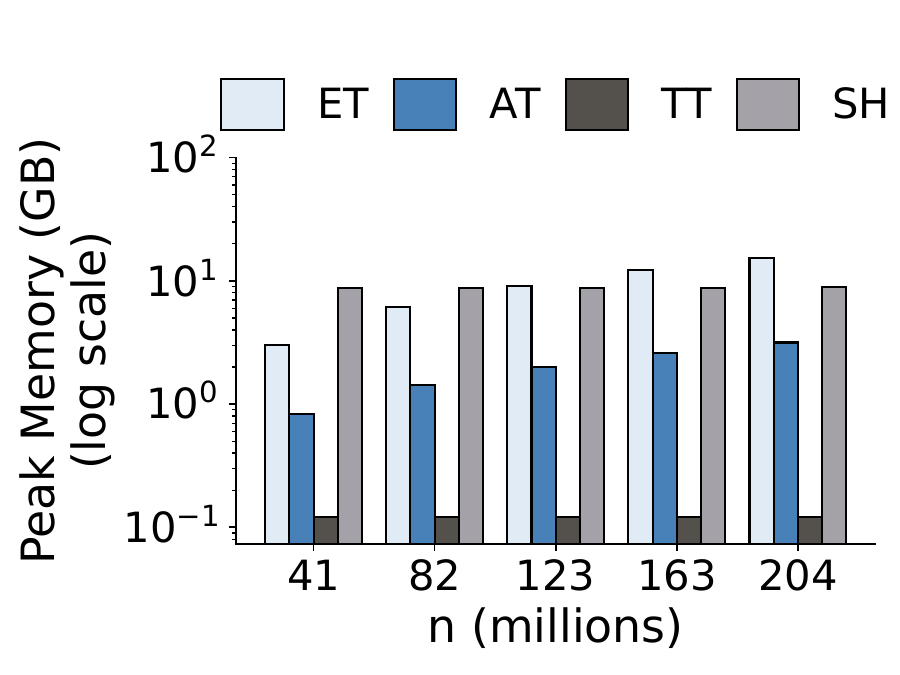}
        \vspace{-6mm}
        \caption{\XML}
        \label{fig:xml_space_n}
    \end{subfigure}
    \begin{subfigure}{0.19\textwidth}
        \includegraphics[width=\textwidth]{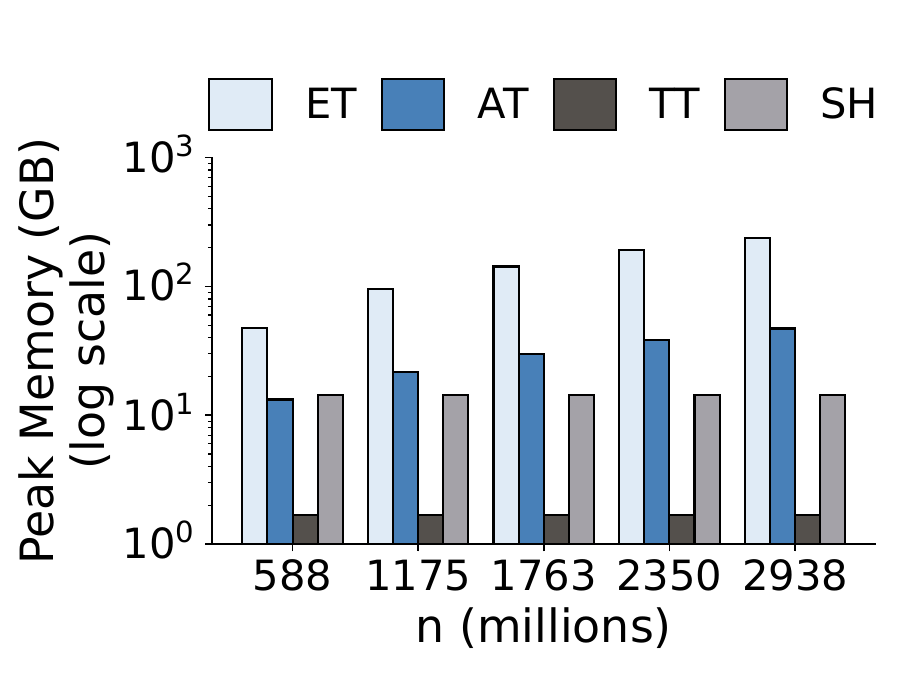}
        \vspace{-6mm}
        \caption{\HUM}
        \label{fig:ecoli_space_n}
    \end{subfigure}
    \begin{subfigure}{0.19\textwidth}
        \includegraphics[width=\textwidth]{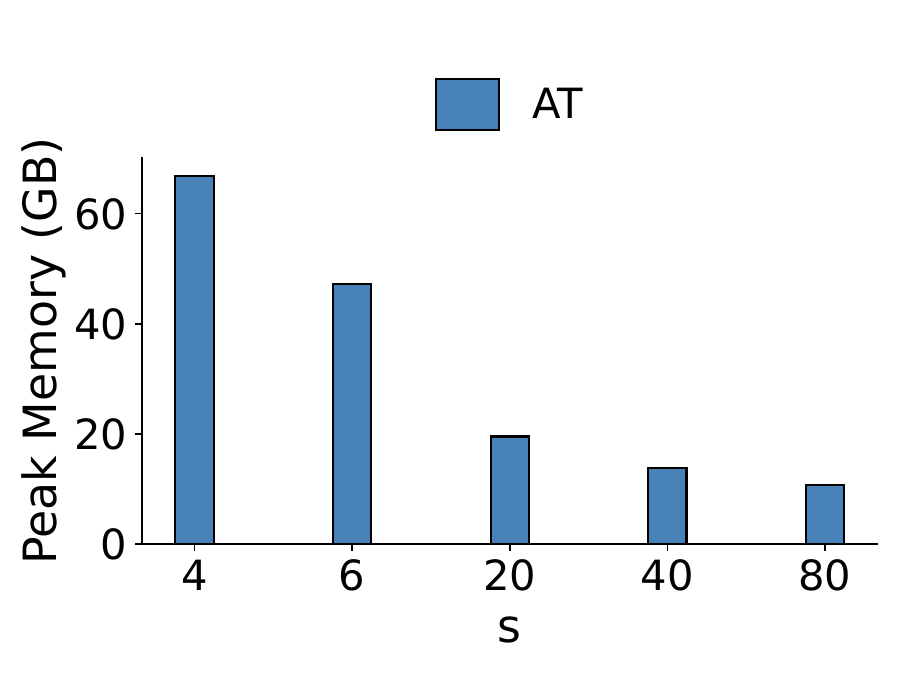}
        \vspace{-6mm}
        \caption{\XML}
        \label{fig:xml_space_s}
    \end{subfigure}
    \begin{subfigure}{0.19\textwidth}
        \includegraphics[width=\textwidth]{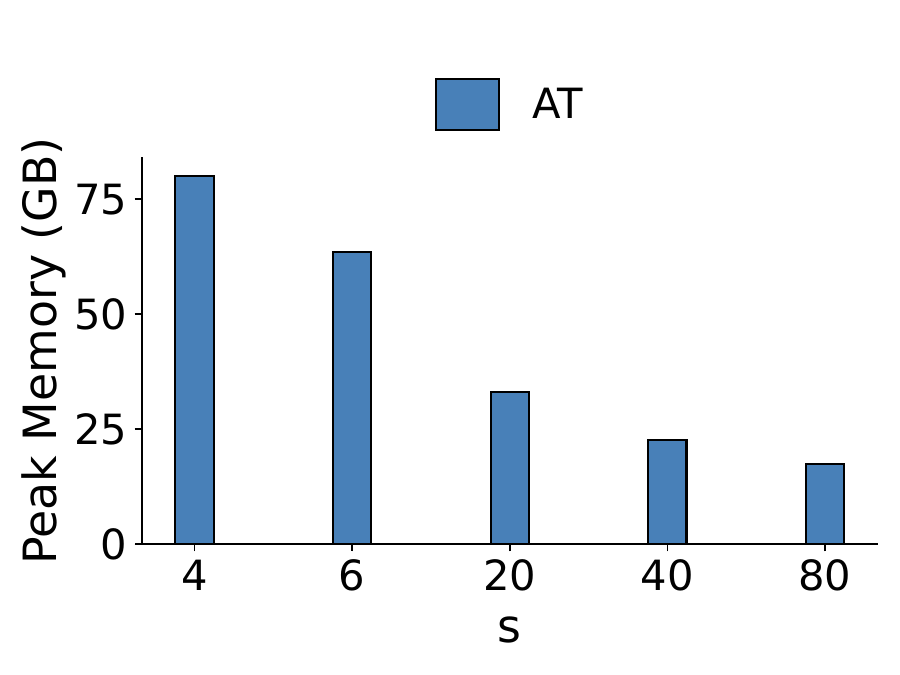}
        \vspace{-6mm}
        \caption{\HUM}
        \label{fig:ecoli_space_s}
    \end{subfigure}
    \begin{subfigure}{0.19\textwidth}
        \includegraphics[width=\textwidth]{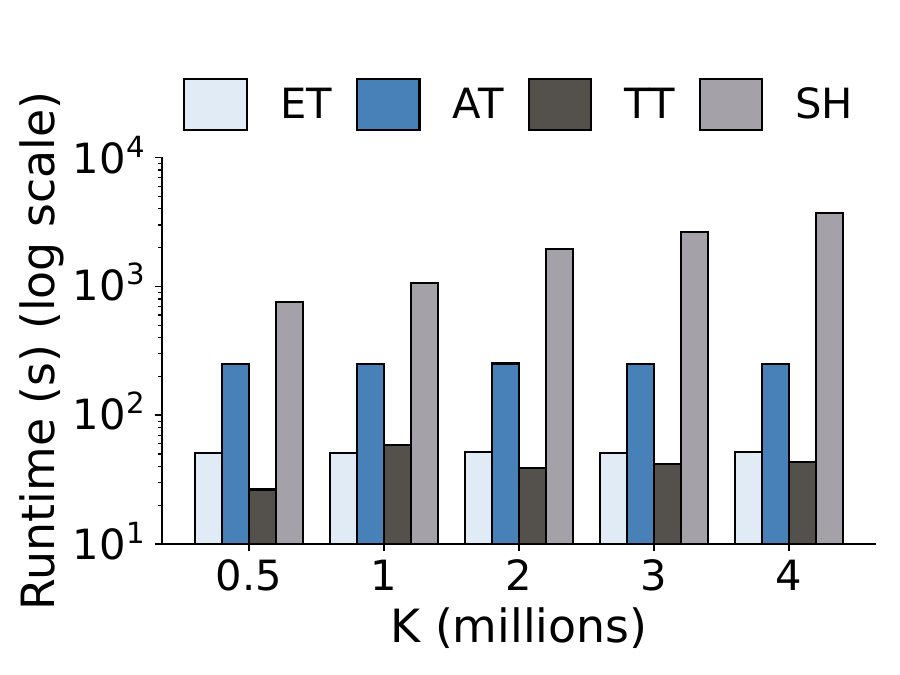}
        \vspace{-6mm}
        \caption{\XML}
        \label{fig:runtime_xml_K}
    \end{subfigure}\\
    \centering
    \begin{subfigure}{0.19\textwidth}
        \includegraphics[width=\textwidth]{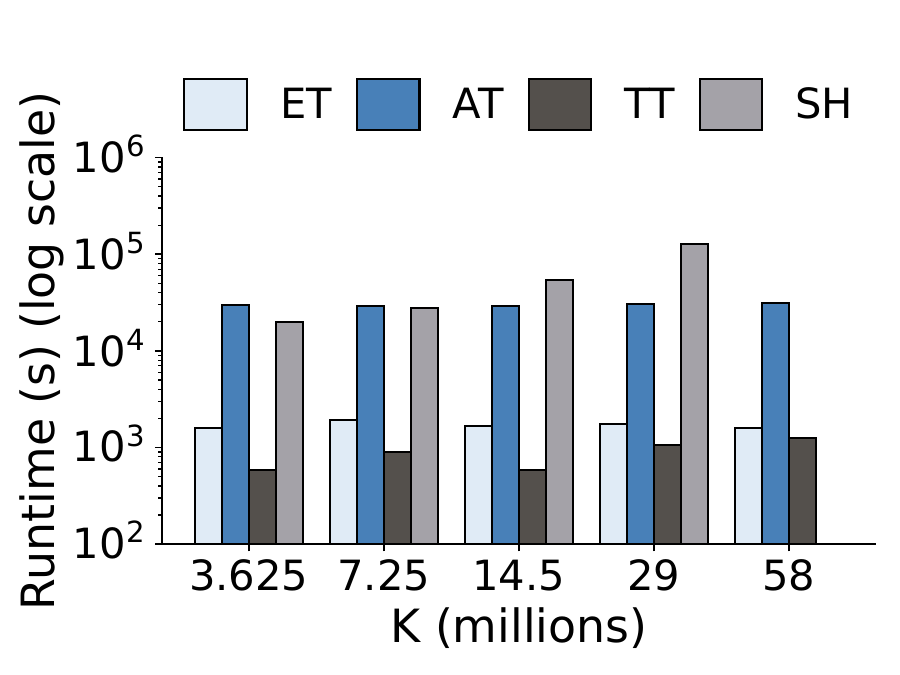}
        \vspace{-6mm}
        \caption{\HUM}
        \label{fig:runtime_ecoli_K}
    \end{subfigure}
    \begin{subfigure}{0.19\textwidth}
        \includegraphics[width=\textwidth]{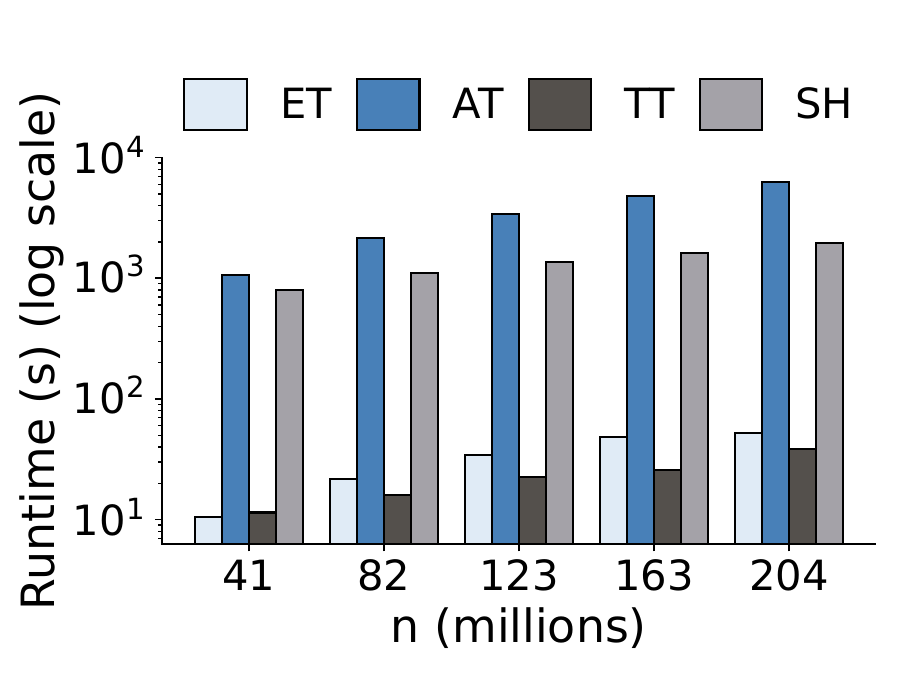}
        \vspace{-6mm}
        \caption{\XML}
        \label{fig:runtime_xml_n}
    \end{subfigure}
    \begin{subfigure}{0.19\textwidth}
        \includegraphics[width=\textwidth]{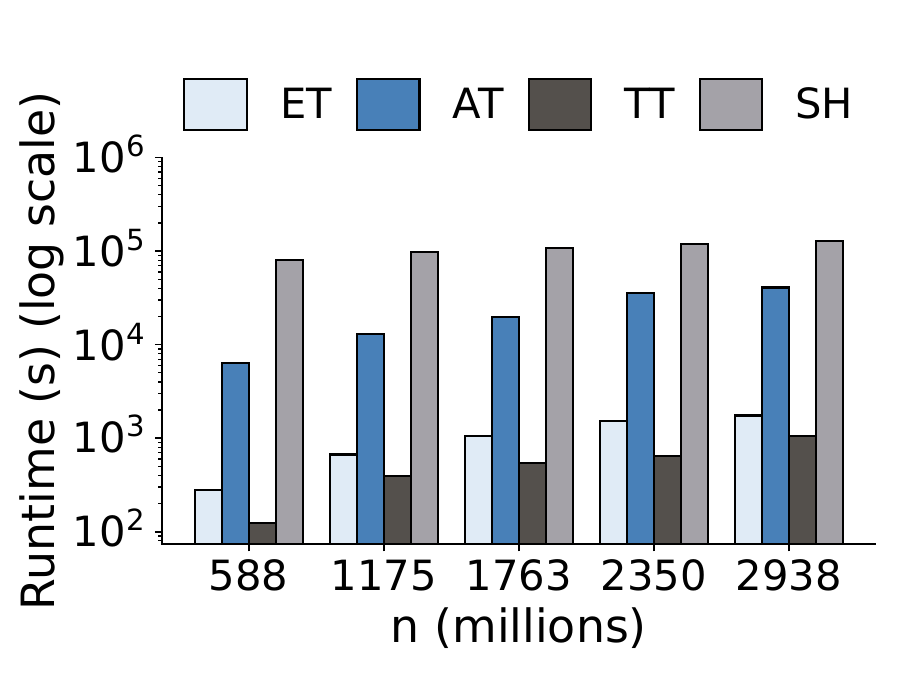}
        \vspace{-6mm}
        \caption{\HUM }
        \label{fig:runtime_ecoli_n}
    \end{subfigure}
    \begin{subfigure}{0.19\textwidth}
        \includegraphics[width=\textwidth]{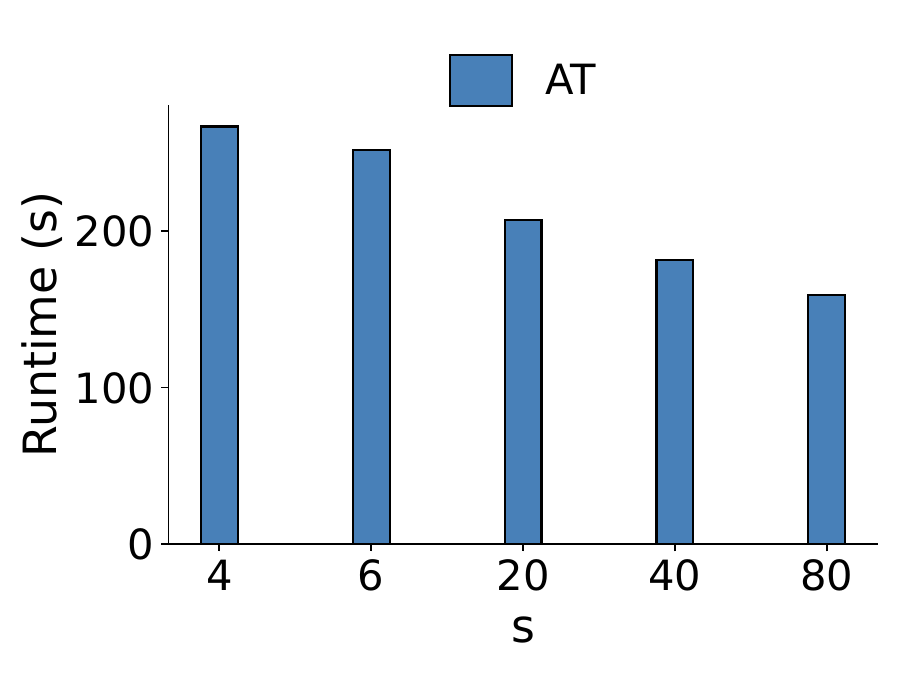}
        \vspace{-6mm}
        \caption{\XML}
        \label{fig:runtime_xml_s}
    \end{subfigure}
    \begin{subfigure}{0.19\textwidth}
        \includegraphics[width=\textwidth]{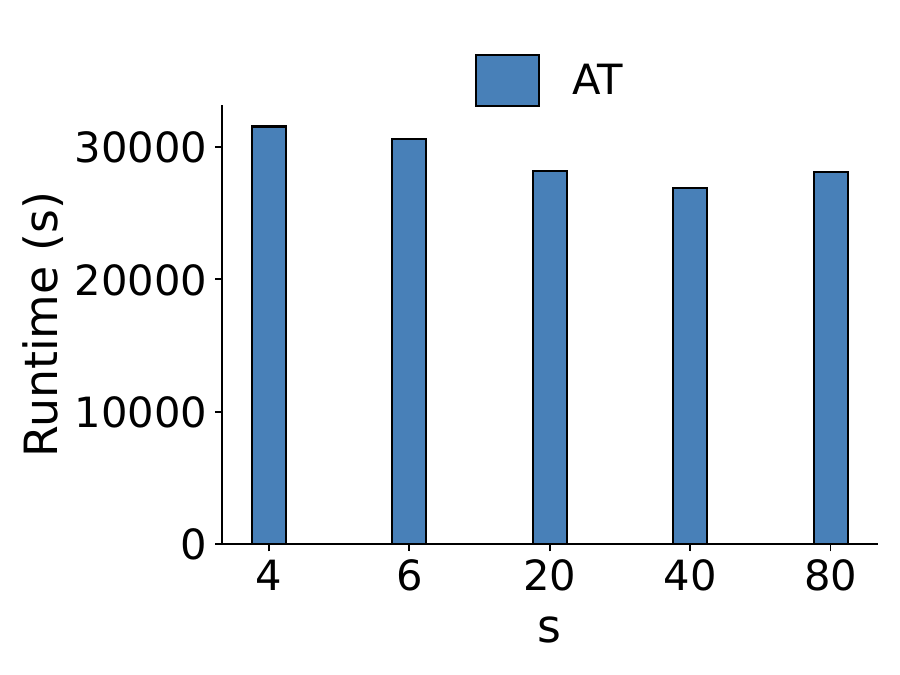}
        \vspace{-6mm}
        \caption{\HUM}
        \label{fig:runtime_ecoli_s}
    \end{subfigure}
    \vspace{-2mm}
    \caption{Space for \ET, \AT, \TT, and \SH vs (a, b) $n$ and  (c, d) $s$ (recall that parameter $s$ affects only $\AT$). Runtime for \ET, \AT, \TT, and \SH (e, f) $K$, (g, h) $n$, and (i, j) $s$. We omit \SH when it did not terminate within $5$ days.}
 \end{figure*}
 \begin{figure*}[!ht]
    \centering
    \begin{subfigure}{0.195\textwidth}       \includegraphics[width=\textwidth]{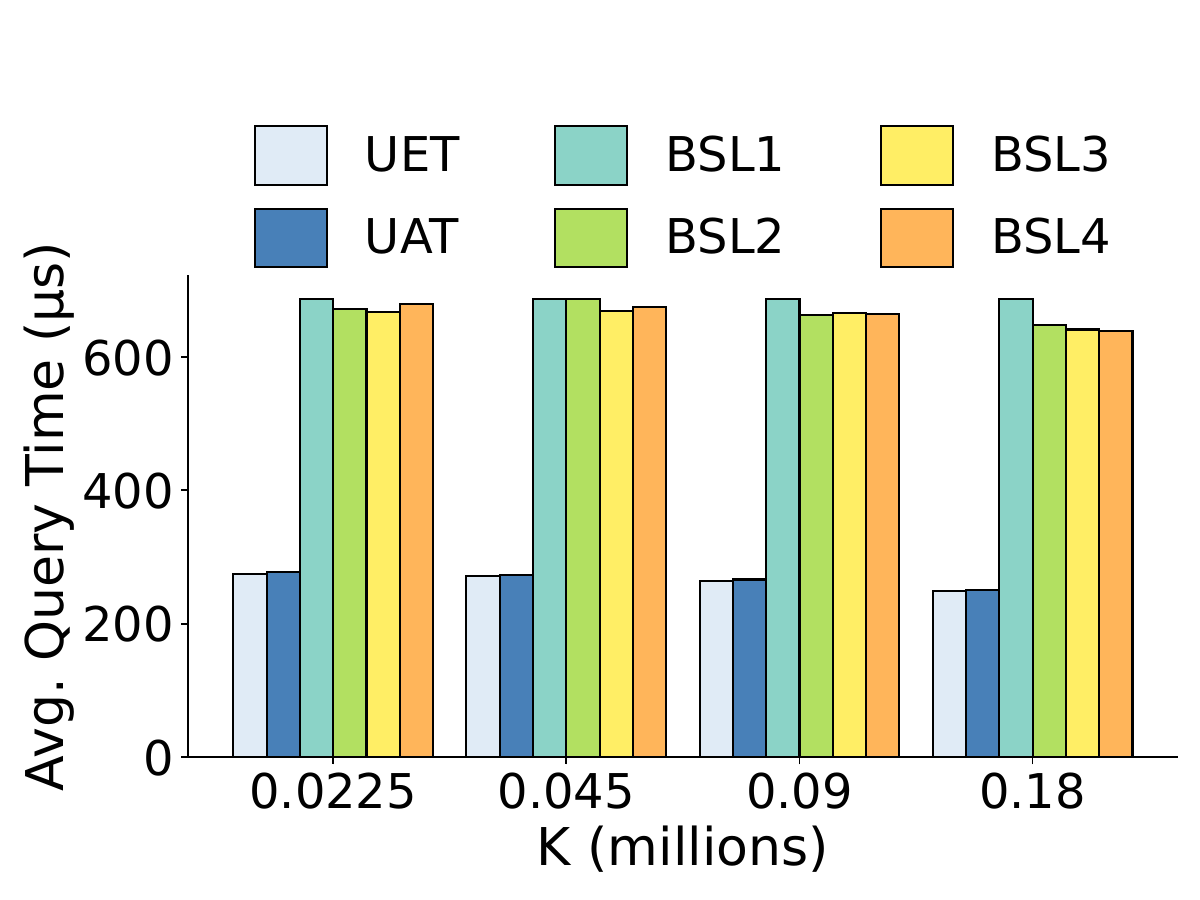}
        \vspace{-6mm}
        \caption{\IoT}
        \label{fig:iot_query_vs_K}
    \end{subfigure}
\begin{subfigure}{0.195\textwidth}        \includegraphics[width=\textwidth]{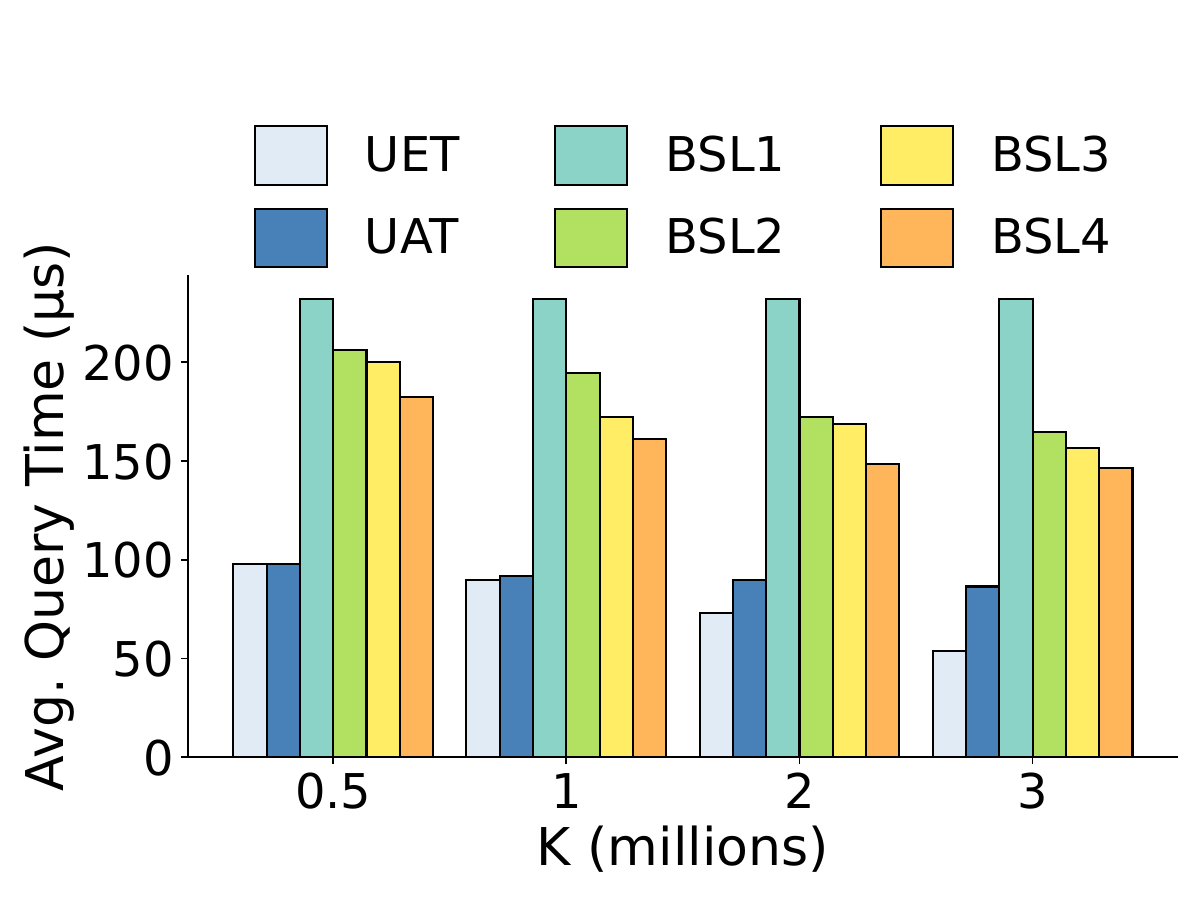}
        \vspace{-6mm}
        \caption{\XML}
        \label{fig:xml_query_vs_K}
    \end{subfigure}
    \begin{subfigure}{0.195\textwidth}
        \includegraphics[width=\textwidth]{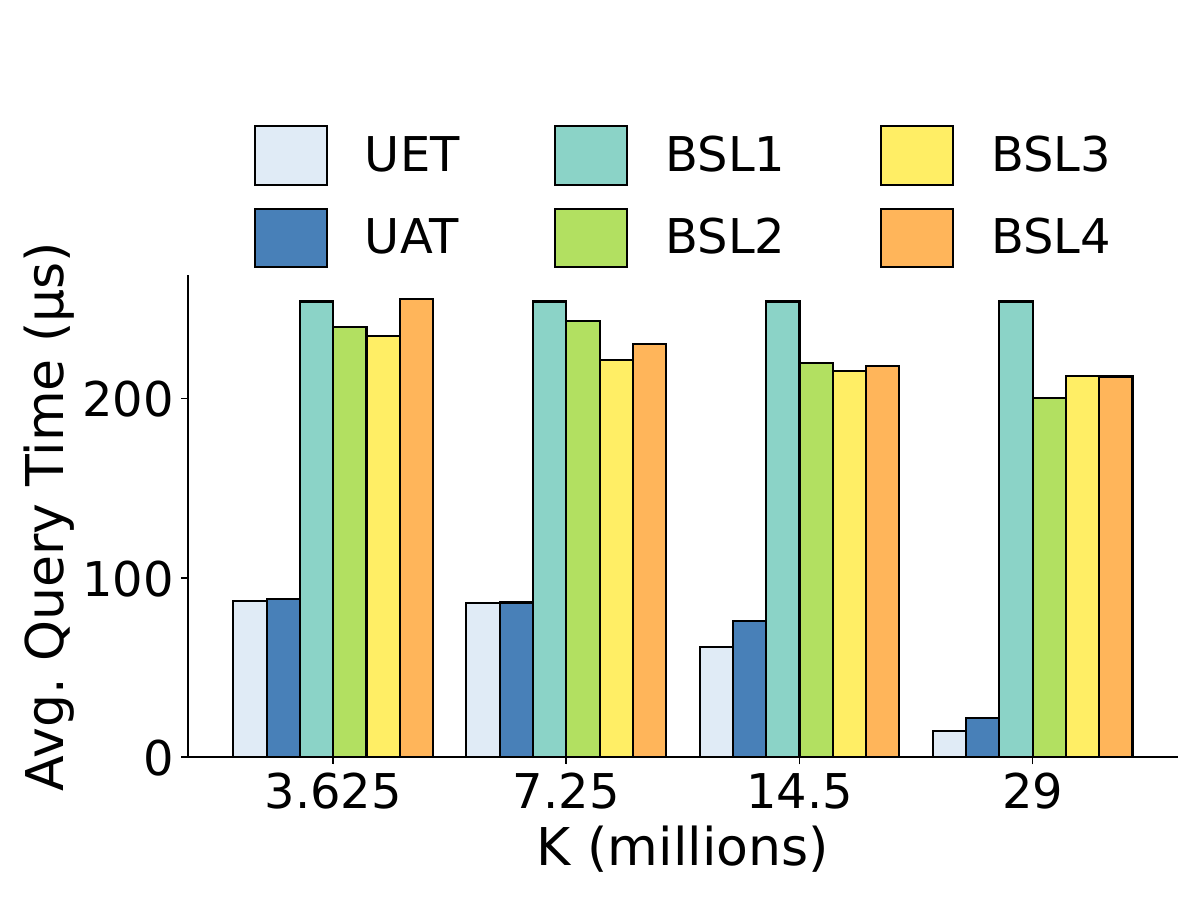}
        \vspace{-6mm}
      \caption{\HUM}
        \label{fig:hum_query_vs_K}
    \end{subfigure}
    \begin{subfigure}{0.195\textwidth}
        \includegraphics[width=\textwidth]{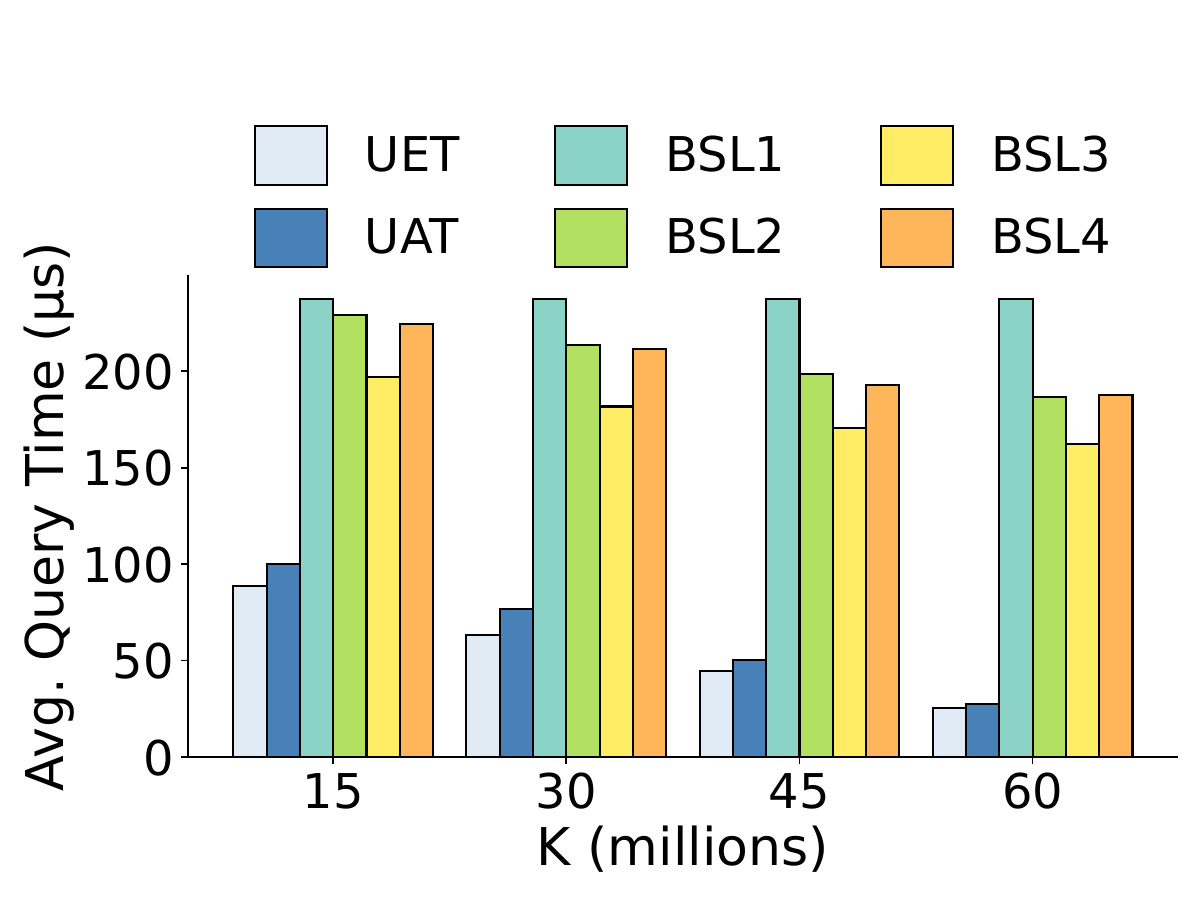}
        \vspace{-6mm}
        \caption{\ECOLI}        
        \label{fig:ecoli_query_vs_K}
    \end{subfigure}
    \begin{subfigure}{0.195\textwidth}
        \includegraphics[width=\textwidth]{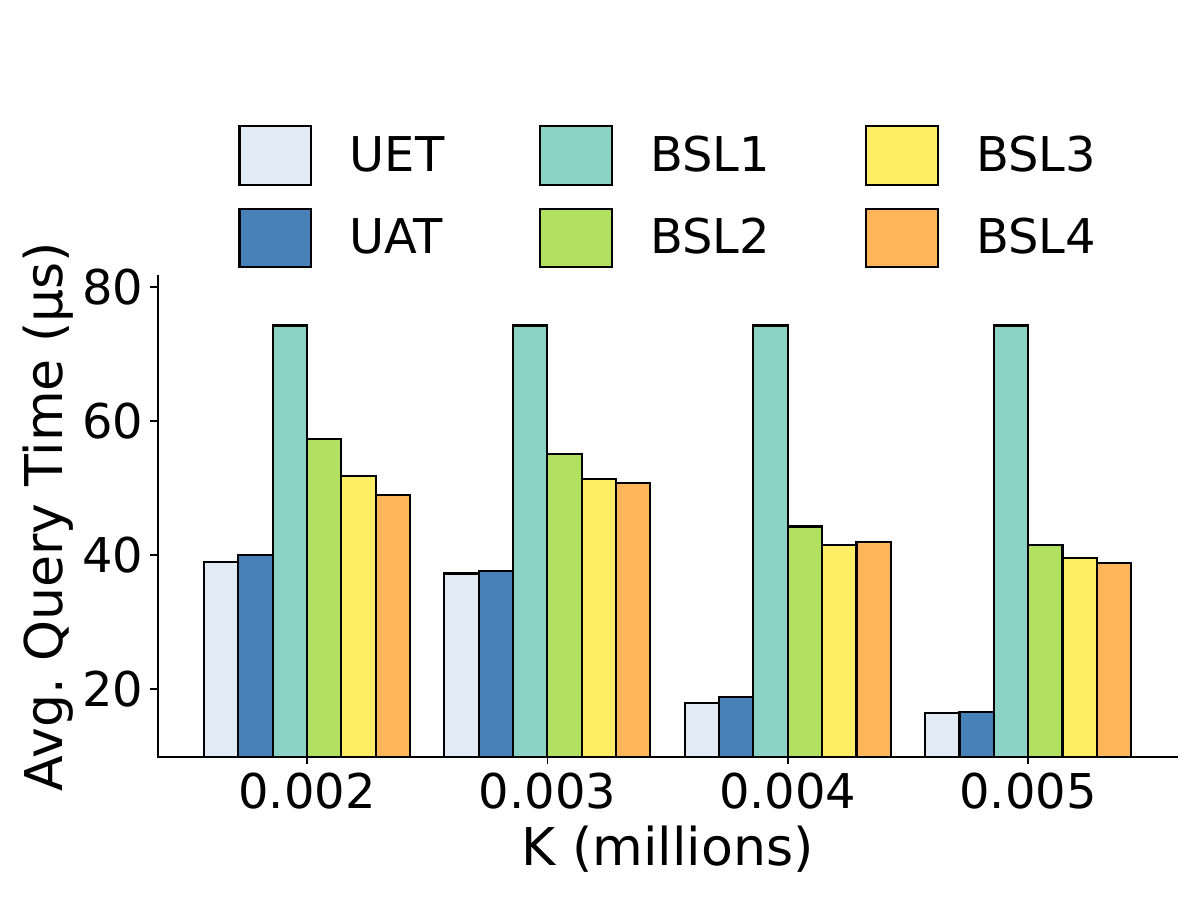}
        \vspace{-6mm}
        \caption{\ADV}        
        \label{fig:adv_query_vs_K}
    \end{subfigure}    
\\\vspace{-0.5mm}
    \begin{subfigure}{0.195\textwidth}
        \includegraphics[trim=6 4 5 30,clip,width=\textwidth]{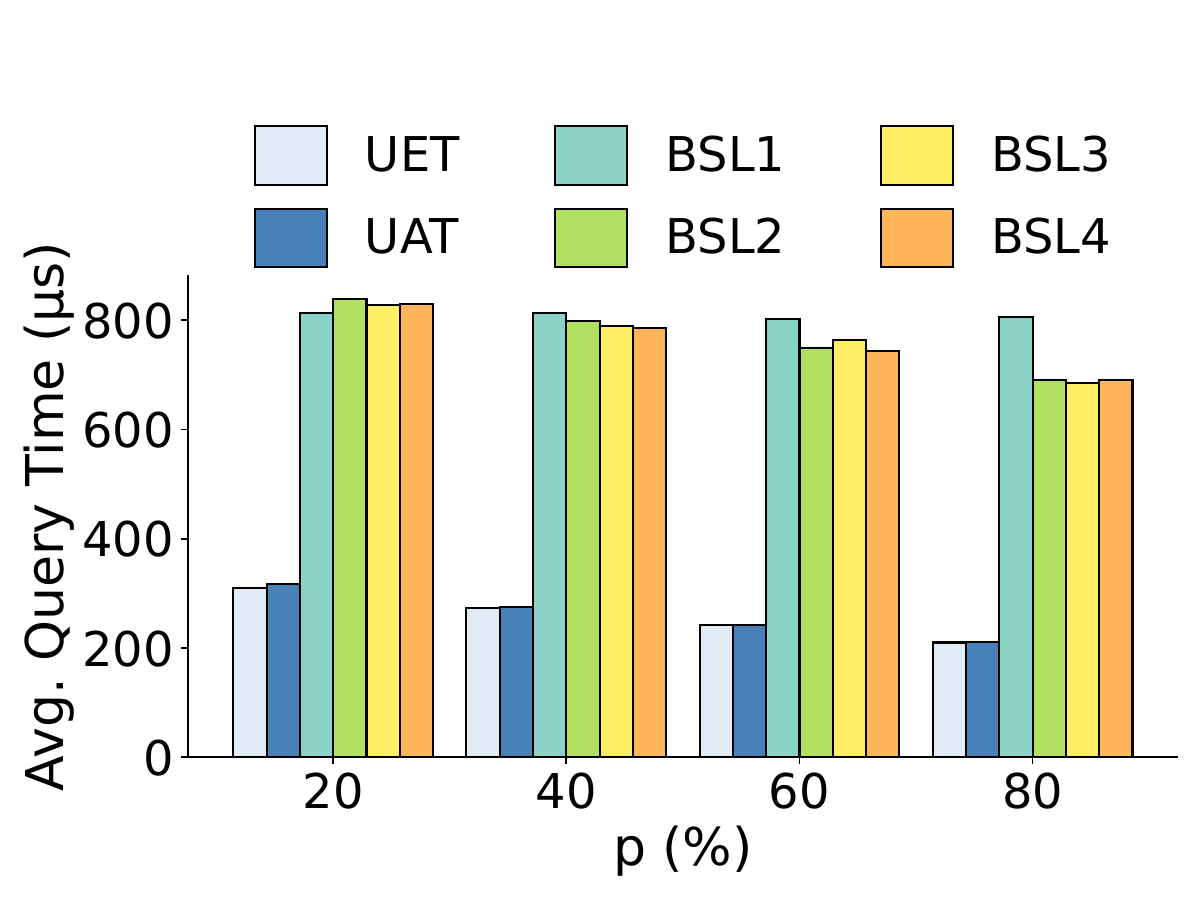}
        \vspace{-6mm}
        \caption{\IoT}
        \label{fig:iot_query_vs_p}
    \end{subfigure}
    \centering
    \begin{subfigure}{0.195\textwidth}
        \includegraphics[trim=6 4 5 30,clip,width=\textwidth]{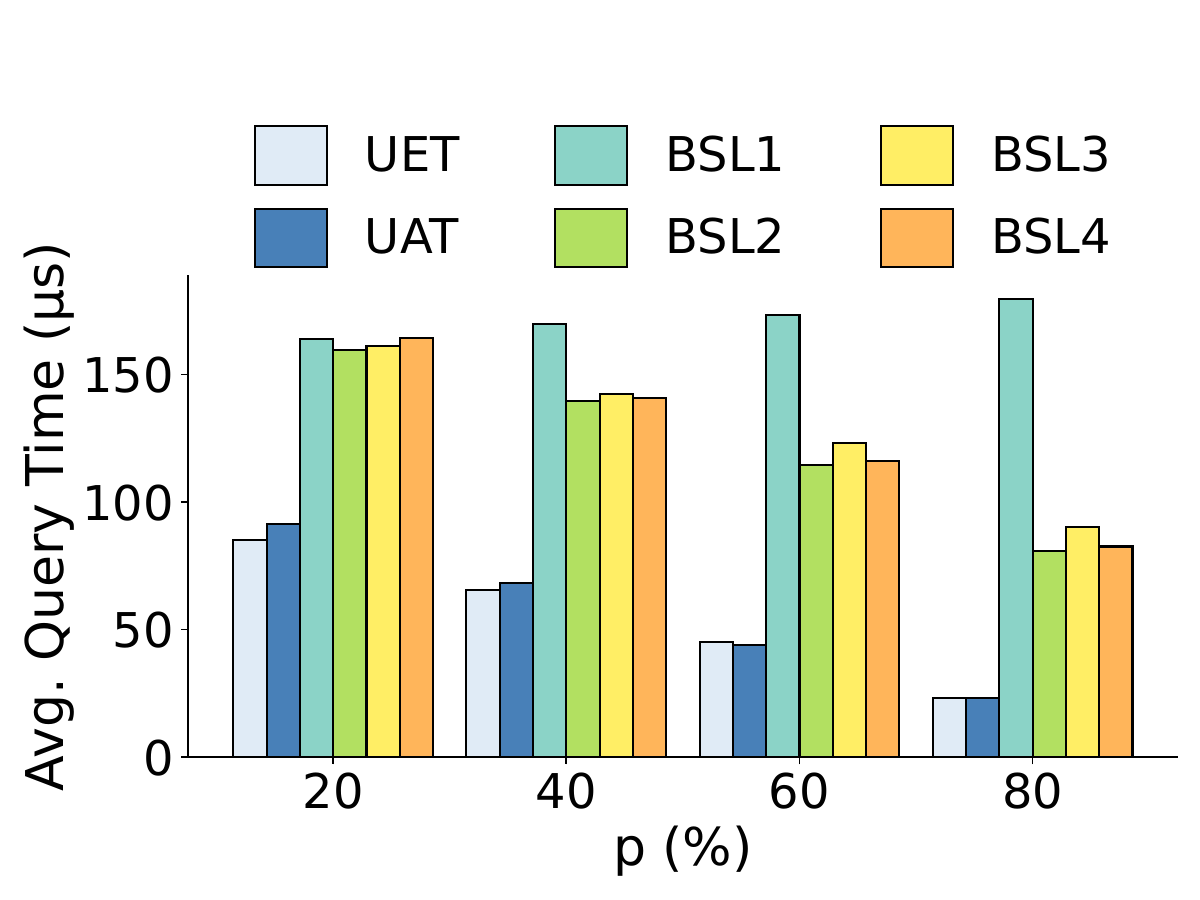}
        \vspace{-6mm}
        \caption{\XML}
        \label{fig:xml_query_vs_p}
    \end{subfigure}
    \begin{subfigure}{0.195\textwidth}
        \includegraphics[trim=6 4 5 30,clip,width=\textwidth]{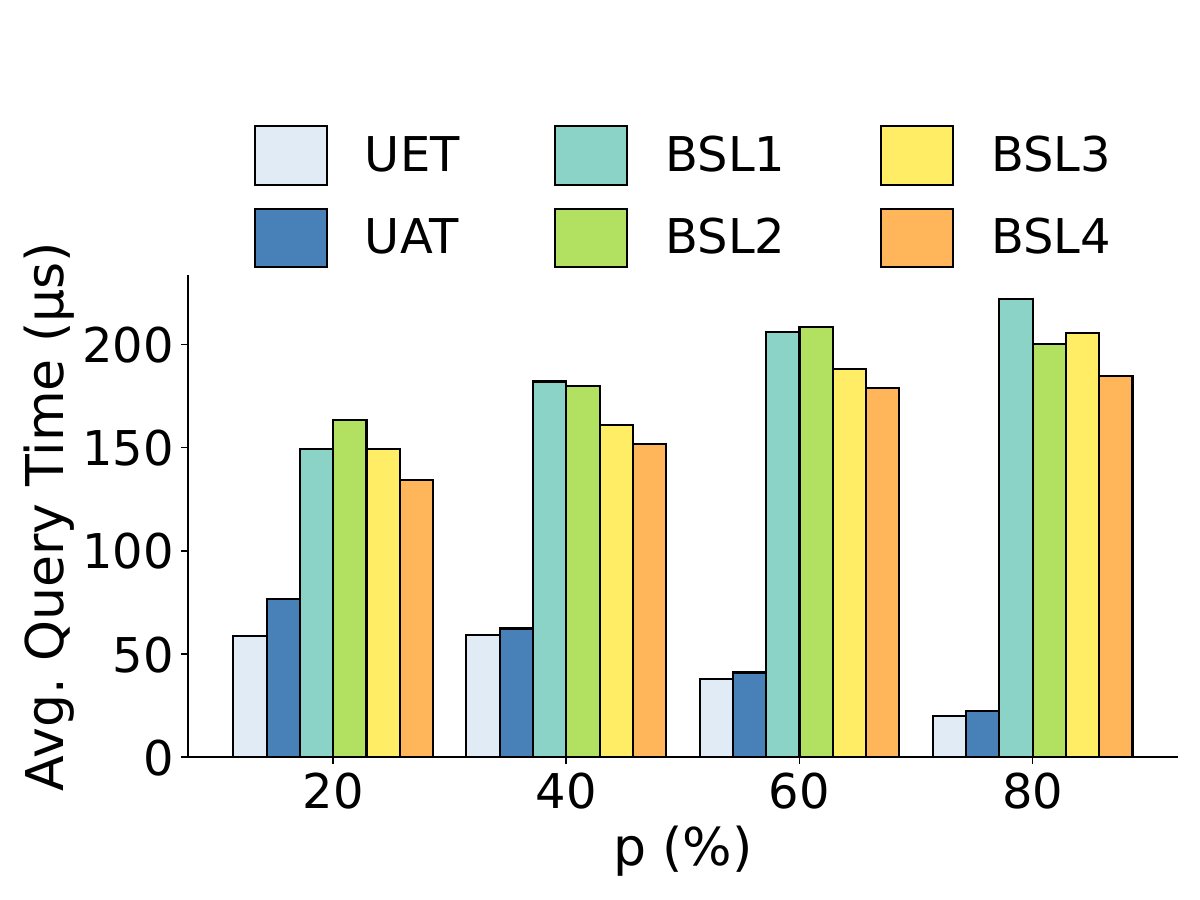}
        \vspace{-6mm}
        \caption{\HUM}
        \label{fig:hum_query_vs_p}
    \end{subfigure}
    \begin{subfigure}{0.195\textwidth}
        \includegraphics[trim=6 4 5 30,clip,width=\textwidth]{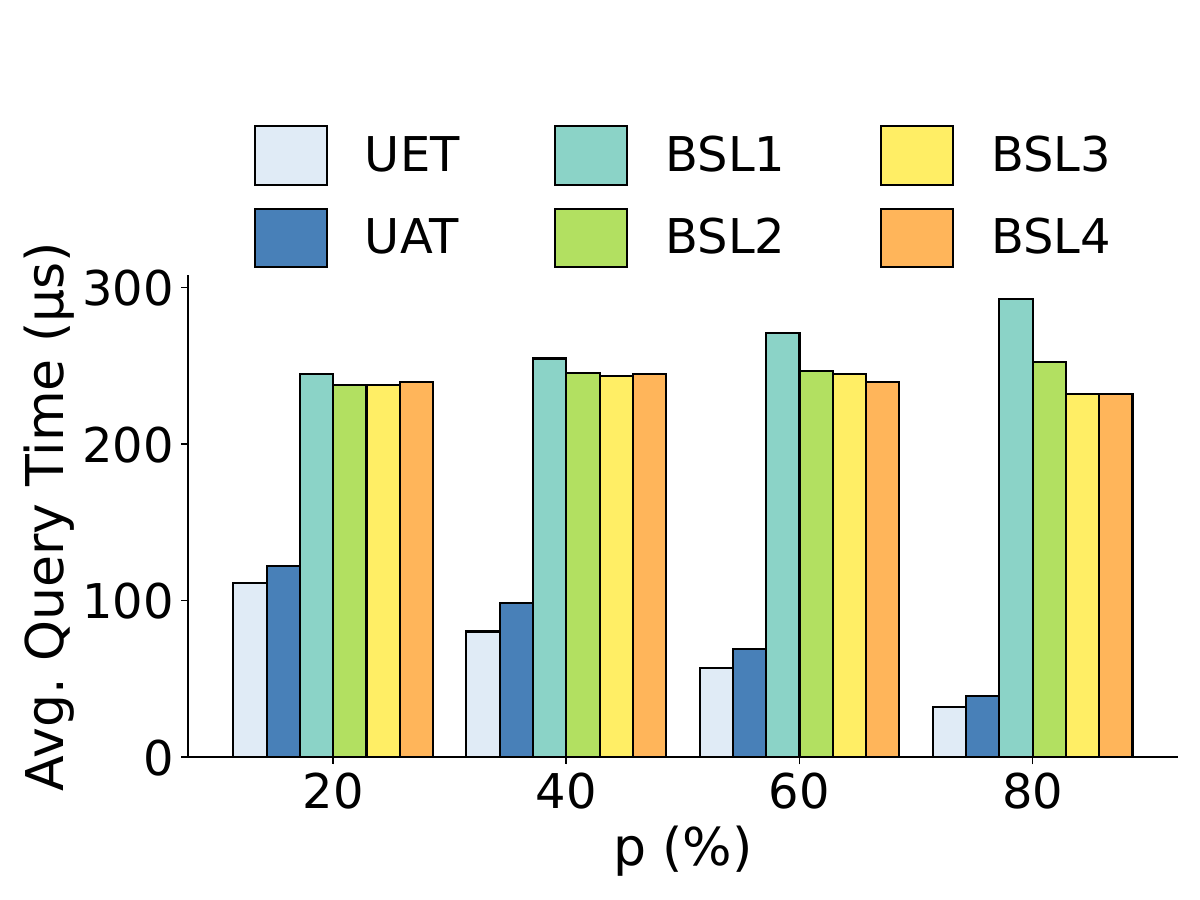}
        \vspace{-6mm}
        \caption{\ECOLI}
        \label{fig:ecoli_query_vs_p}
    \end{subfigure}
    \begin{subfigure}{0.195\textwidth}
        \includegraphics[trim=6 4 5 30,clip,width=\textwidth]{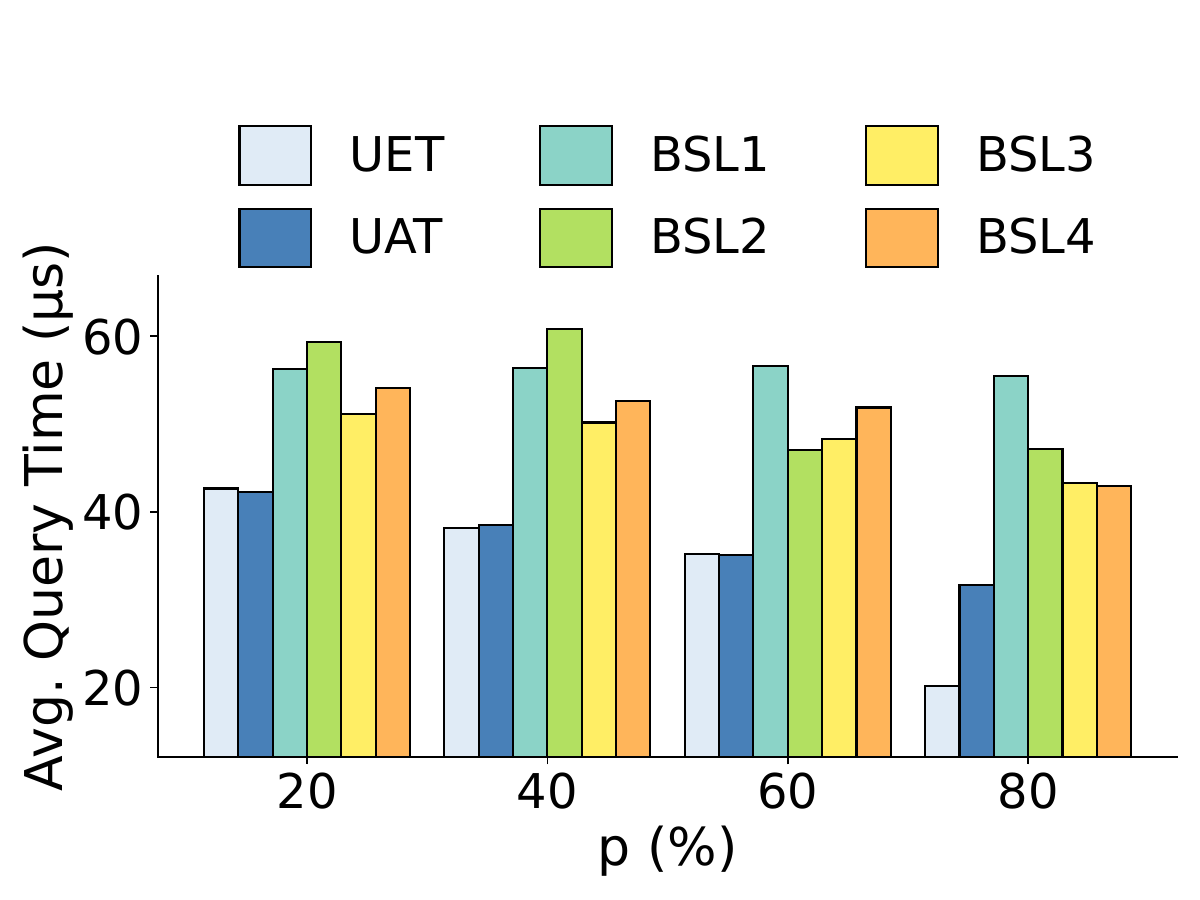}
        \vspace{-6mm}
        \caption{\ADV}
        \label{fig:adv_query_vs_p}
    \end{subfigure}
    \\
    \begin{subfigure}{0.195\textwidth}
        \includegraphics[trim=6 4 5 30,clip,width=\textwidth]{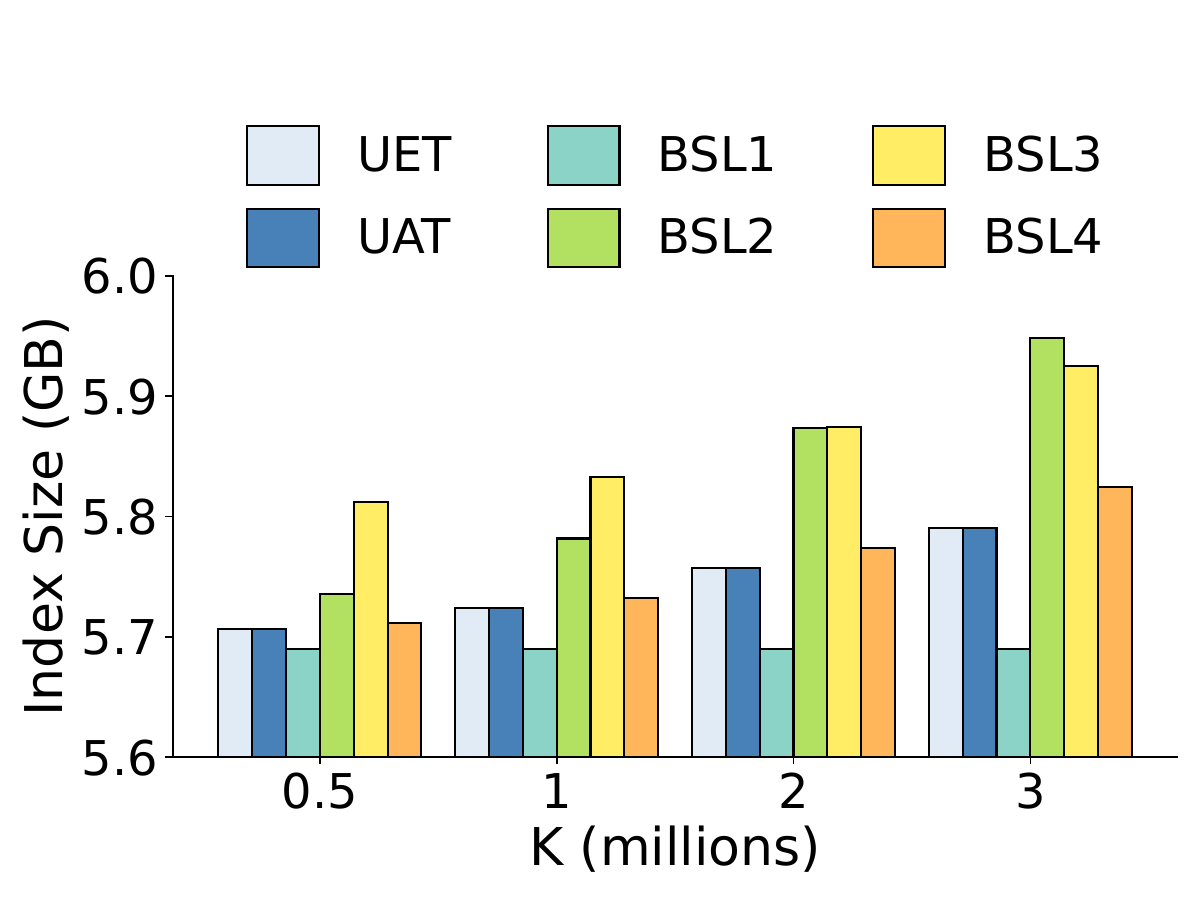}
        \vspace{-6mm}
        \caption{\XML}
        \label{fig:xml_index_size_vs_K}
    \end{subfigure}
    \begin{subfigure}{0.195\textwidth}
        \includegraphics[trim=6 4 5 30,clip,width=\textwidth]{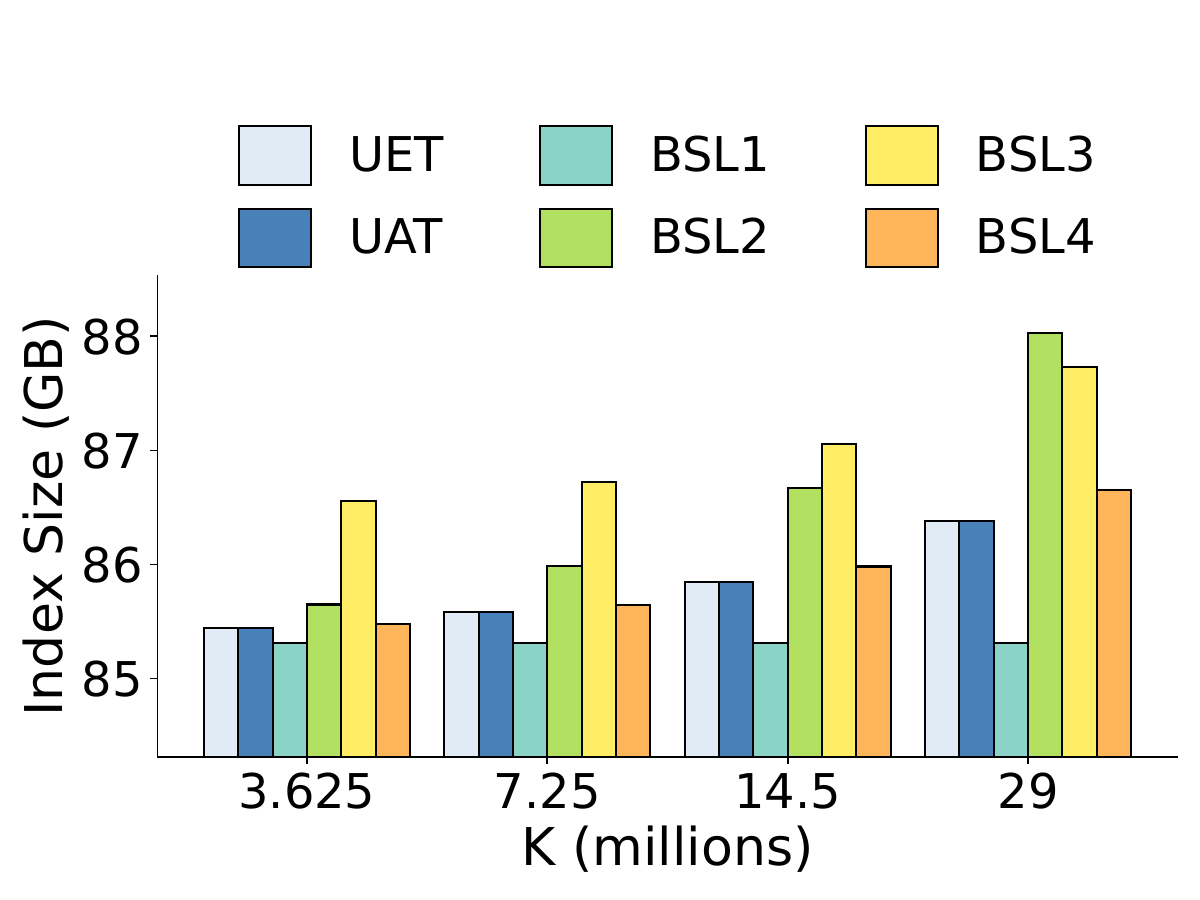}
        \vspace{-6mm}
        \caption{\HUM}
        \label{fig:dna_index_size_vs_K}
    \end{subfigure}
    \begin{subfigure}{0.195\textwidth}
        \includegraphics[trim=6 4 5 30,clip,width=\textwidth]{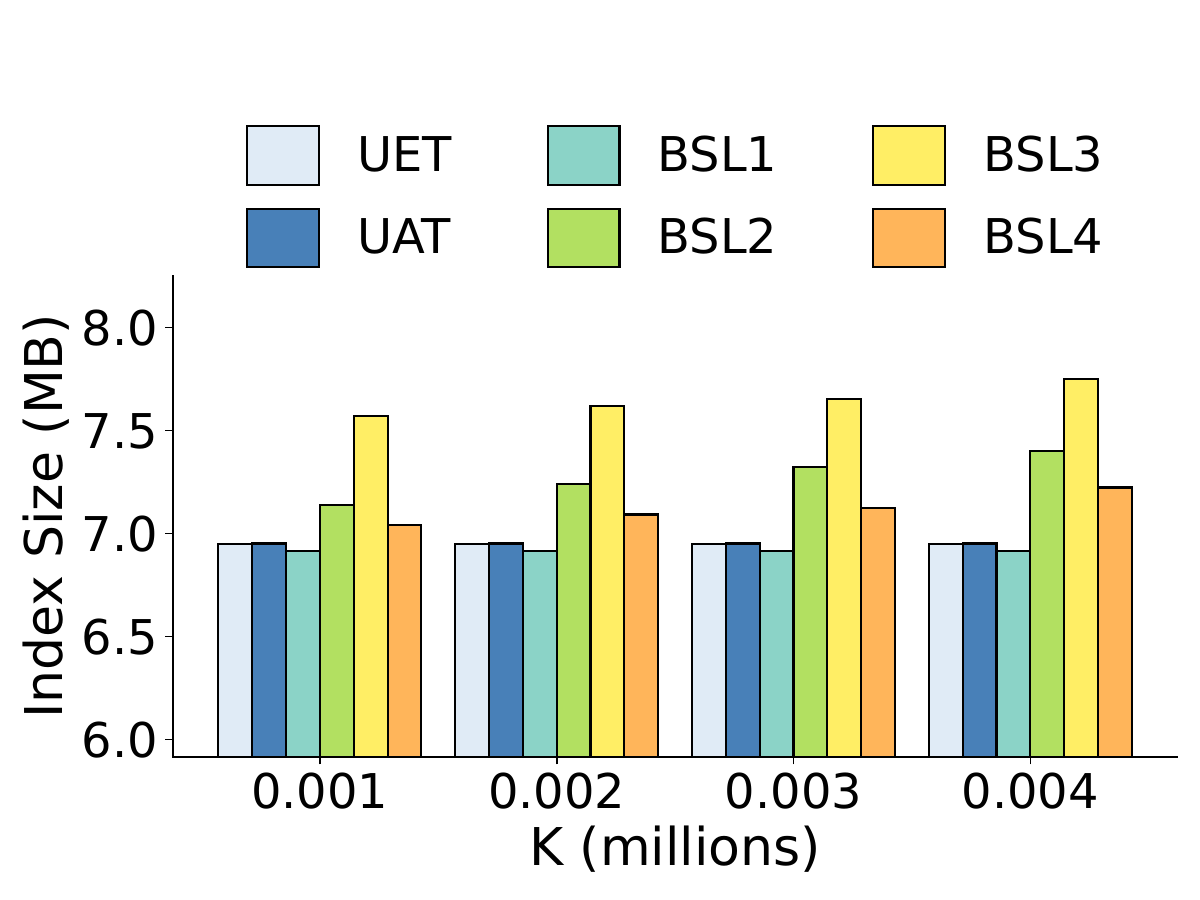}
        \vspace{-6mm}
        \caption{\ADV}
        \label{fig:adv_index_size_vs_K}
    \end{subfigure}
    \begin{subfigure}{0.195\textwidth}
        \includegraphics[trim=6 4 5 30,clip,width=\textwidth]{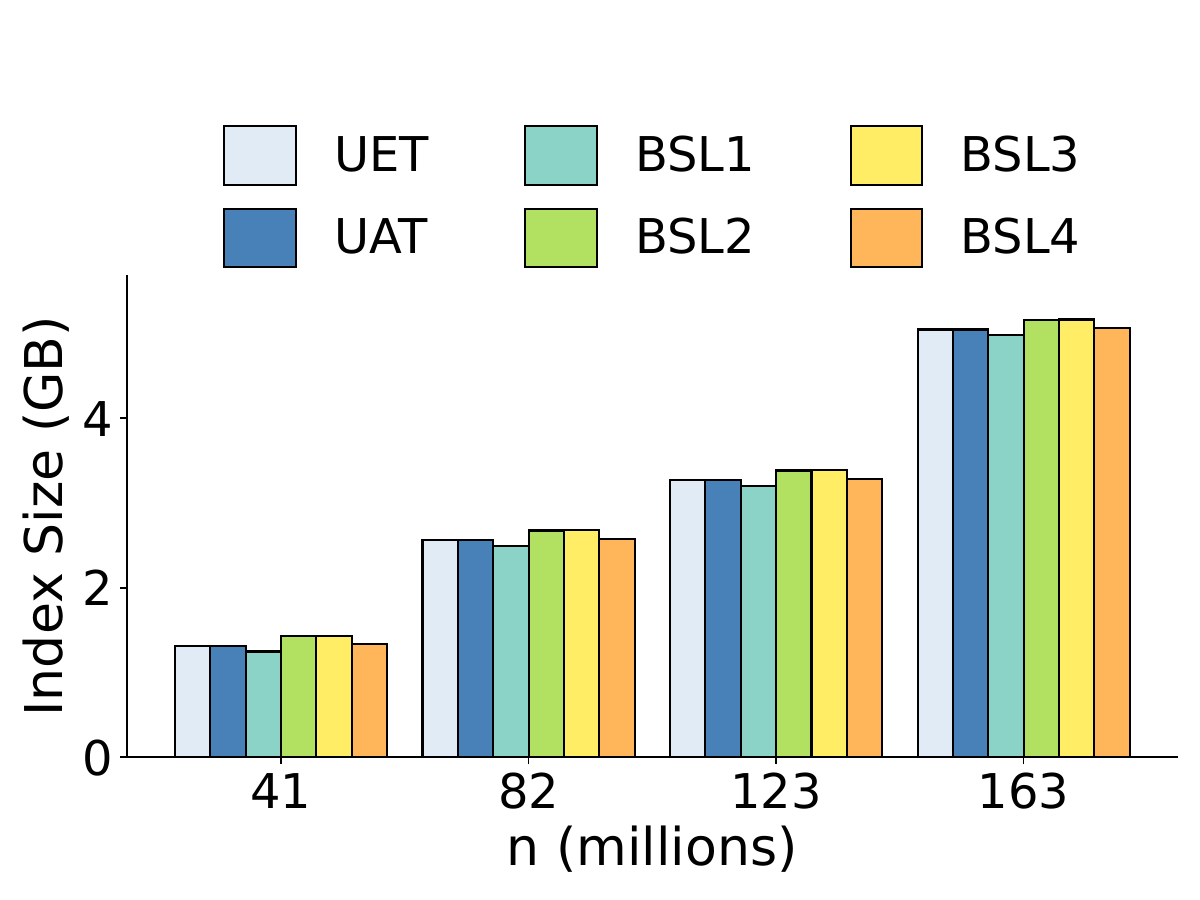}
        \vspace{-6mm}
        \caption{\XML}
        \label{fig:xml_index_size_vs_n}
    \end{subfigure}
    \begin{subfigure}{0.195\textwidth}
        \includegraphics[trim=6 4 5 30,clip,width=\textwidth]{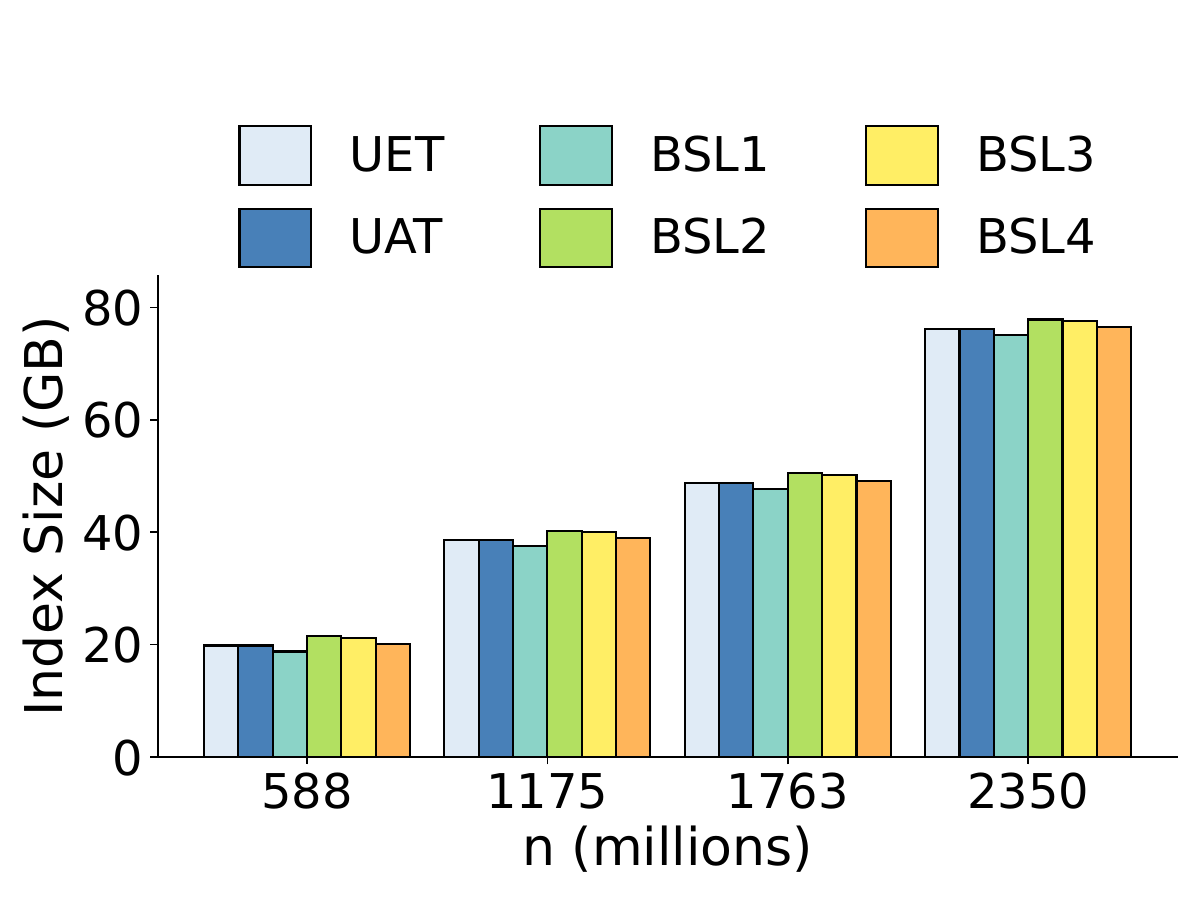}
        \vspace{-6mm}
        \caption{\HUM}
        \label{fig:dna_index_size_vs_n}
    \end{subfigure}\\
    \begin{subfigure}{0.195\textwidth}
        \includegraphics[trim=6 4 5 30,clip,width=\textwidth]{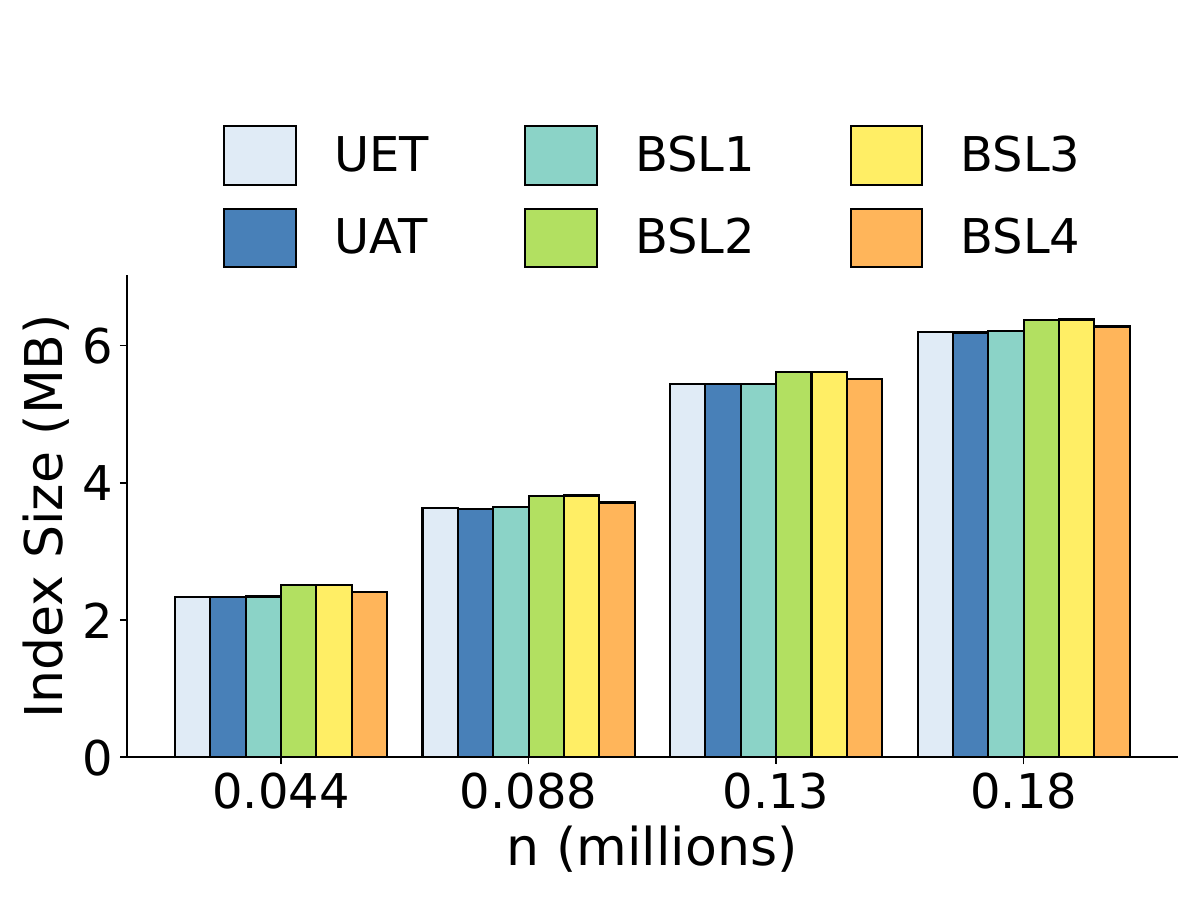}
        \vspace{-6mm}
        \caption{\ADV}
        \label{fig:adv_index_size_vs_n}
    \end{subfigure}
    \begin{subfigure}{0.195\textwidth}
        \includegraphics[trim=6 4 5 30,clip,width=\textwidth]{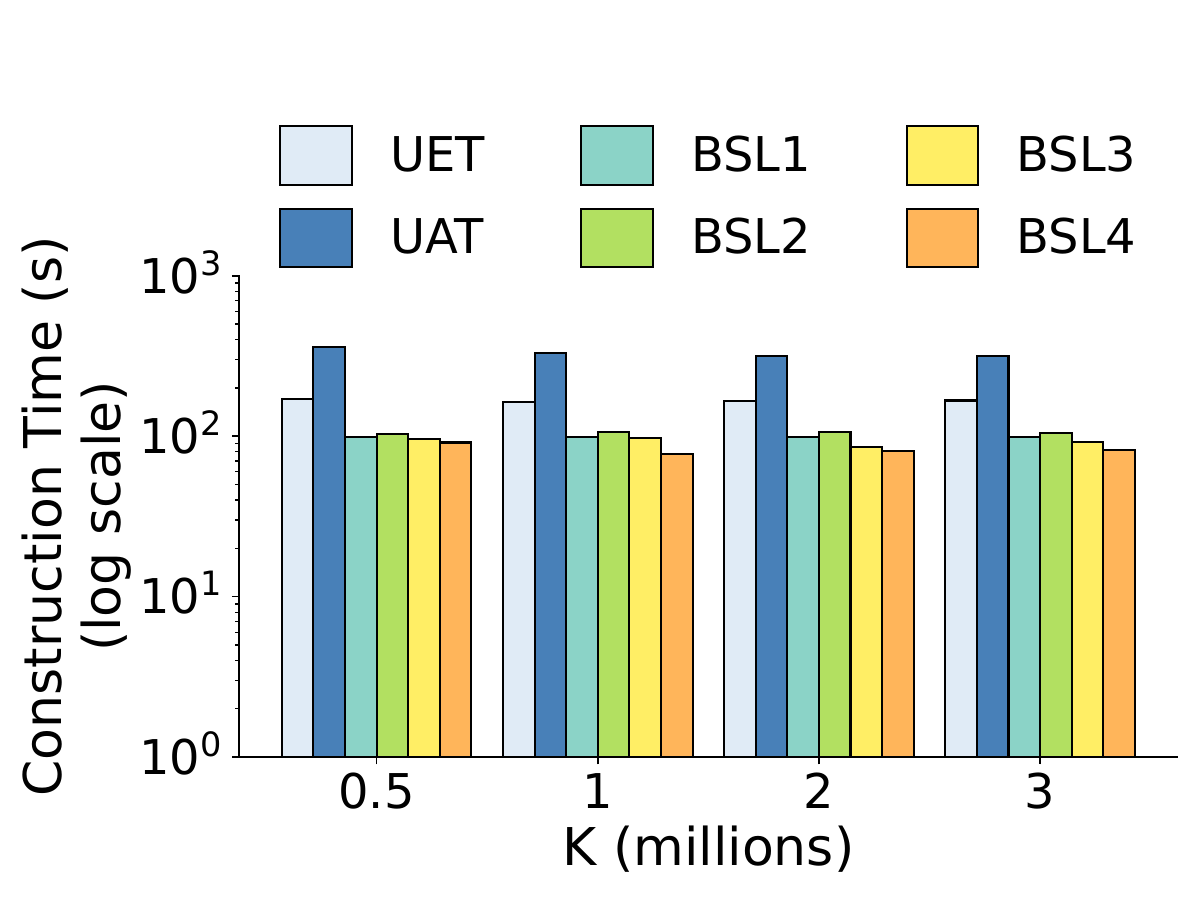}
        \vspace{-6mm}
        \caption{\XML}
        \label{fig:xml_construction_time_vs_K}
    \end{subfigure}
        \begin{subfigure}{0.195\textwidth}
        \includegraphics[trim=6 4 5 30,clip,width=\textwidth]{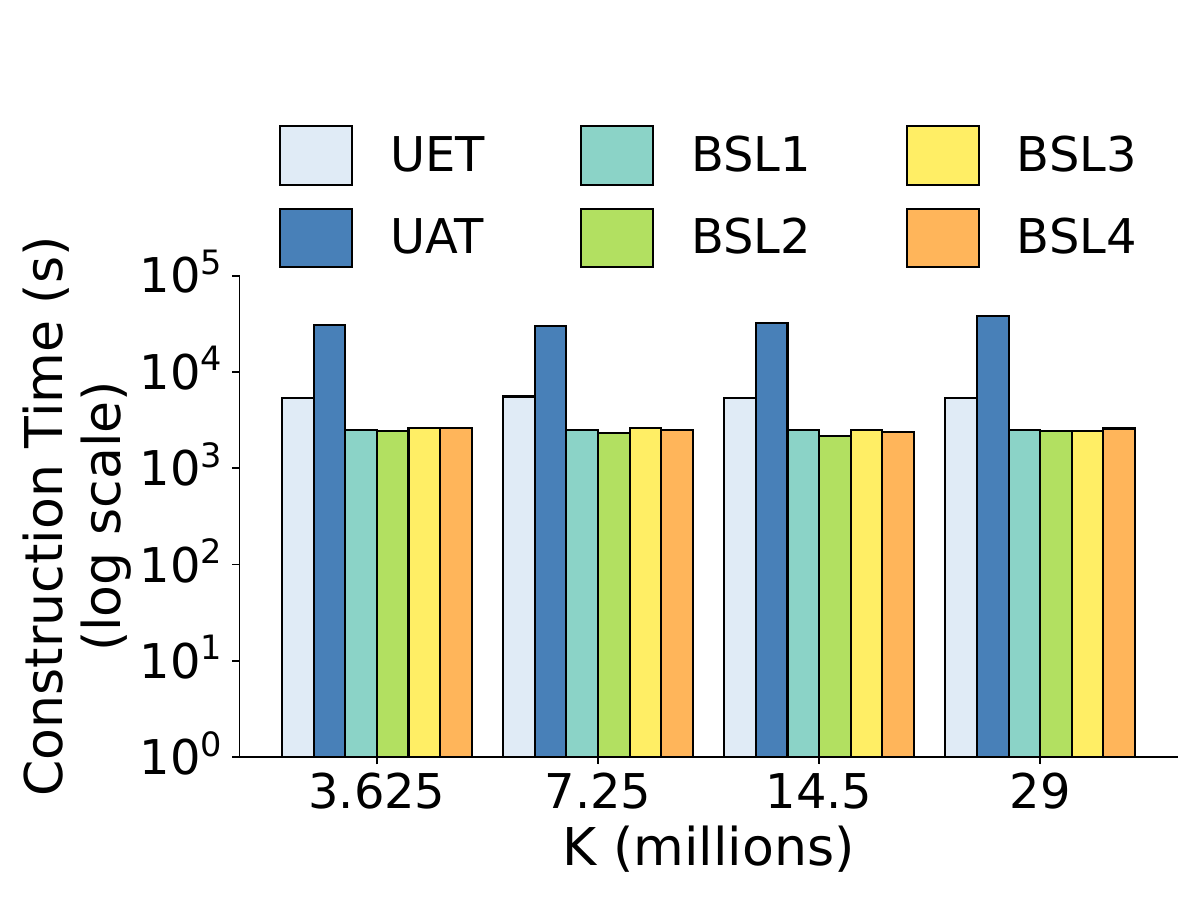}
        \vspace{-6mm}
        \caption{\HUM}
        \label{fig:dna_construction_time_vs_K}
    \end{subfigure}
    \begin{subfigure}{0.195\textwidth}
        \includegraphics[trim=6 4 5 30,clip,width=\textwidth]{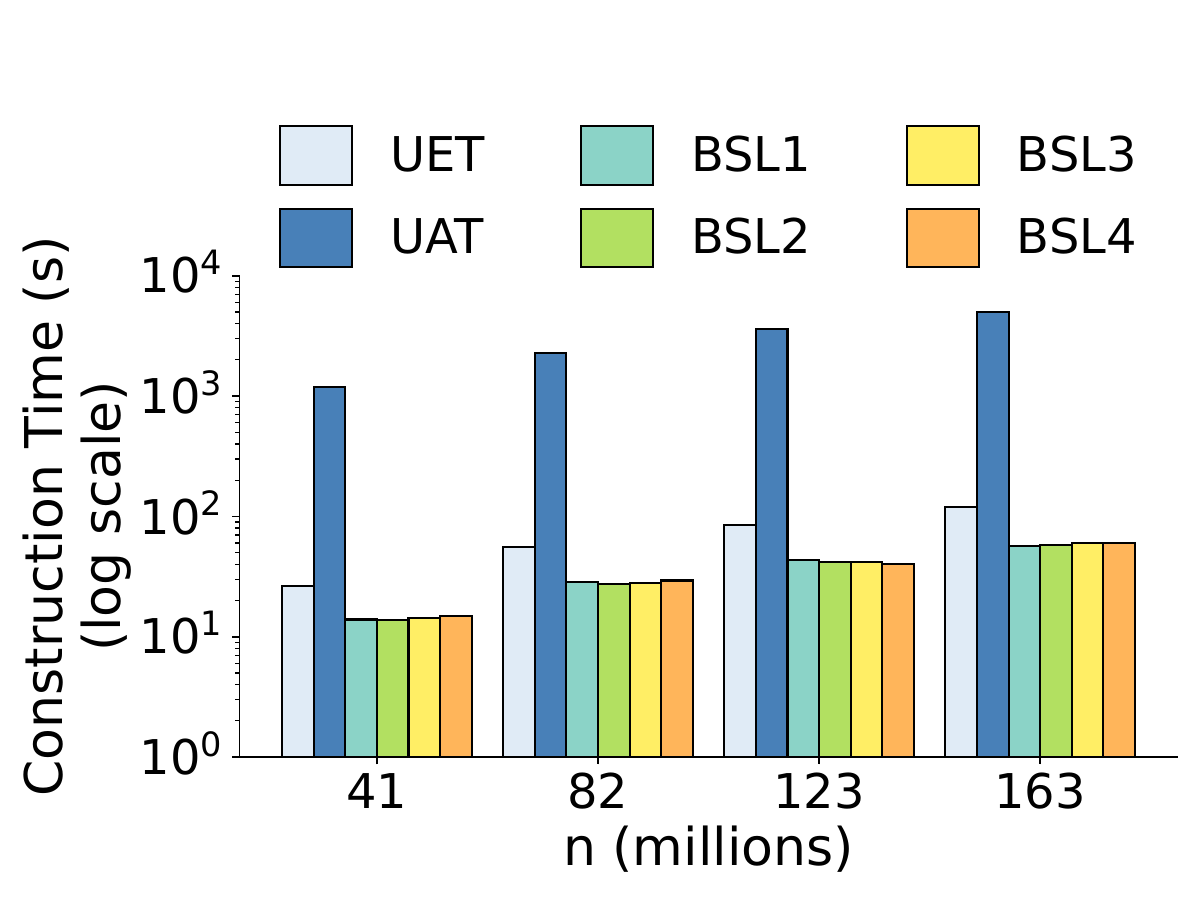}
        \vspace{-6mm}
        \caption{\XML}
        \label{fig:xml_construction_time_vs_n}
    \end{subfigure}
    \begin{subfigure}{0.195\textwidth}
        \includegraphics[trim=6 4 5 30,clip,width=\textwidth]{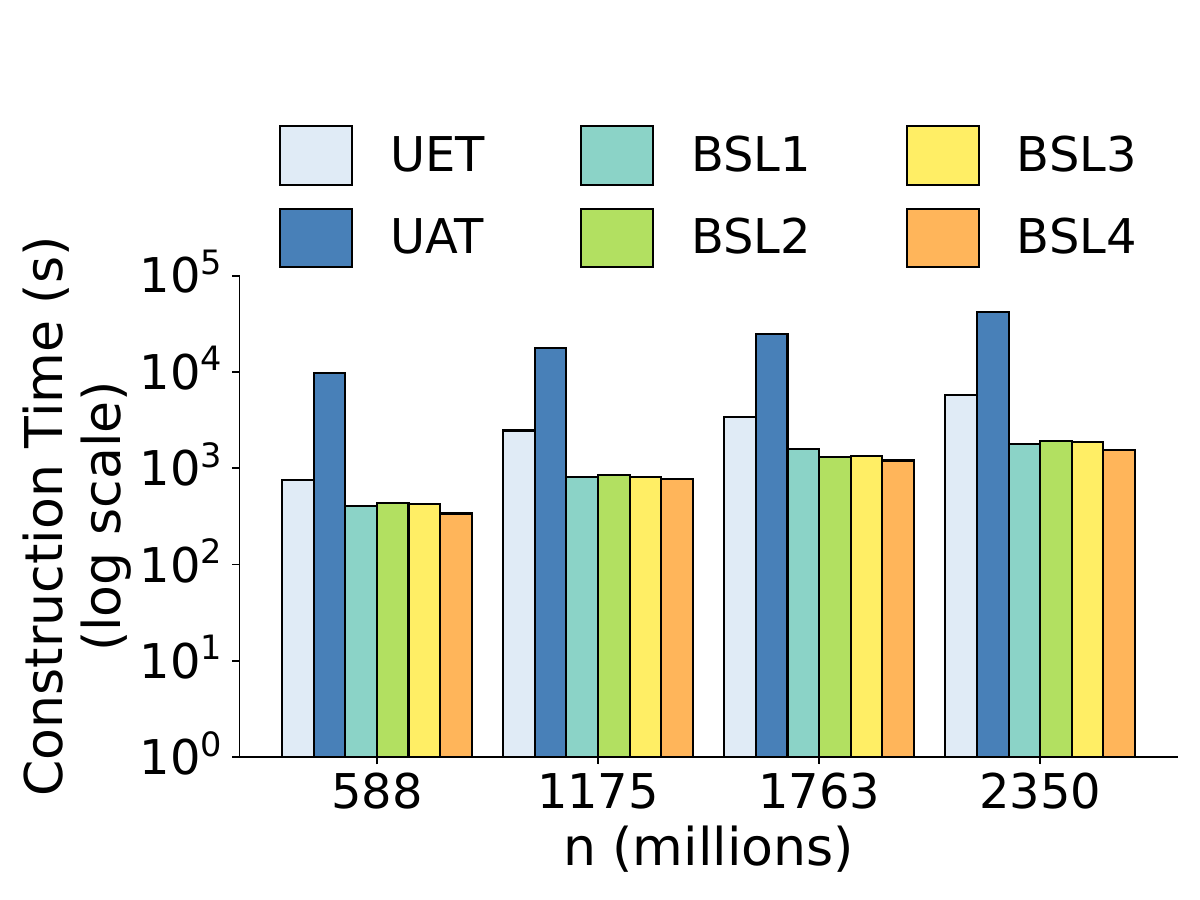}
        \vspace{-6mm}
        \caption{\HUM}
        \label{fig:dna_construction_time_vs_n}
    \end{subfigure}
    \caption{\UET, \UAT, and BSL1, \ldots, BSL4: Average query time vs (a-e) $K$ and (f-j) $p$. Index size vs (k-m) $K$ and (n-p) $n$. Construction  time vs (q, r) $K$ and (s, t) $n$.} 
\end{figure*}

\noindent{\bf Effectiveness.}~Figs.~\ref{fig:iot_accuracy_K} to ~\ref{fig:adv_accuracy_K} show the effectiveness of \AT, \TT and \SH in terms of Accuracy, for varying $K$; we do not report any results for \ET as it is exact. We also omit the results for \SH 
%on \HUM with $K=58$ million and \ECOLI with $K>30$ million, since \SH 
when it did not finish within $5$ days. \AT is remarkably accurate for all $K$ values, 
unlike \TT and \SH. For example, the Accuracy of \AT was $94.9\%$ on average and at least $76.5\%$, while that of \TT (respectively, \SH) is $25.7\%$ (respectively, $44.3\%$) on average and at least $0.15\%$ (respectively, $6.9\%$). \TT and \SH perform the worst on the IoT dataset because they miss long frequent substrings (see Section~\ref{sec:whynot}). For example, the longest string among the exact top-($22500$) frequent substrings has length $11816$, while the longest among the top-($22500$) found by \TT and \SH has length $546$ and $1577$, respectively. 

Figs.~\ref{fig:iot_accuracy_n} to~\ref{fig:ecoli_accuracy_n} show the effectiveness of \AT, \TT, and \SH in terms of Accuracy, for varying $n$; the result for \ADV is analogous (omitted). 
\AT was highly accurate (e.g., its Accuracy was $94.1\%$ on average and at least $80\%$). As expected, the effectiveness of \AT increases with $n$ (e.g., in Fig.~\ref{fig:iot_accuracy_n} it increased from $93.6\%$ to $99.9\%$ as $n$ increased from $3.7\cdot 10^6$ to $18.7\cdot 10^6$). This is because $s$ is the same for all $n$ values and thus more positions of the text are sampled as $n$ increases. \TT and \SH performed much worse than \AT, for the same reasons as before.  
For example, \SH outperformed \TT but its average Accuracy was only $50.6\%$, and it did not terminate within $5$ days in Fig.~\ref{fig:ecoli_accuracy_n} when $n\geq 2754\cdot 10^6$. 

Figs.~\ref{fig:iot_accuracy_s},~\ref{fig:xml_accuracy_s},~\ref{fig:hum_accuracy_s}, and~\ref{fig:ecoli_accuracy_s} show the impact of $s$ in \AT on Accuracy. 
As expected (see Section~\ref{sec:SSA}), a smaller $s$ makes \AT more accurate and $s=\cO(\log n)$ is reasonable ($\log n$ is $25$, $28$, $32$, and $33$ for IoT, XML, HUM, and ECOLI, respectively). 

Fig.~\ref{fig:all_ndcg} shows the NDCG values for all datasets. The results are analogous to those reported for Accuracy. In fact, \AT achieved a result very close to the optimal (i.e., NDCG scores at least $0.9993$), outperforming both \TT and \SH. In the \IoT dataset, the difference from \TT (respectively, \SH) was more than $93\%$ (respectively, $70\%$). Fig.~\ref{fig:ecoli_ndcg_s} shows the impact of $s$ in \textsf{AT} on NDCG; the values decrease with $s$ but very slightly and are at least $0.993$. %sorry I was touching it :-) feel free to modify
        
The results with respect to the RE measure are analogous to those of Accuracy, so we omitted all RE results. 

\noindent{\bf Space.}~We report results for \XML and \HUM; the results for the other datasets are analogous. 
Figs.~\ref{fig:xml_space_n} and~\ref{fig:ecoli_space_n} show the impact of $n$ on space. The space for both \ET and \AT increases linearly with $n$, in line with their space complexities. 
\AT takes $4.5$ times less space than \ET on average, and \TT takes the least space. The space of \TT and \SH does not depend on $n$, in line with their $\cO(K)$ space complexity. Figs.~\ref{fig:xml_space_s} and~\ref{fig:ecoli_space_s} show the impact of $s$ on the space consumption of \AT. As expected by its space complexity, \AT requires less space as $s$ increases, which together with its very high effectiveness highlights the benefit of our sampling approach. 
 
\noindent{\bf Runtime.}~We report results for \XML and \HUM; the results for the other datasets are analogous. Figs.~\ref{fig:runtime_xml_K} and~\ref{fig:runtime_ecoli_K} show the impact of $K$ on the runtime, which is small for all algorithms except \SH. This is because $K$ for \ET and \TT, or $sK$ for \AT in their time complexities is smaller than $n$. \SH takes much more time as $K$ increases because $z$ gets larger (see Section~\ref{sec:whynot}). \TT is the fastest, \ET is slightly slower, \AT is even slower, and \SH is the slowest. Figs.~\ref{fig:runtime_xml_n} and~\ref{fig:runtime_ecoli_n} show that all algorithms scale with $n$ as predicted by their time complexities. Again, \ET is faster than \AT by more than one order of magnitude, \TT is the fastest, and \SH is faster than \AT in \XML but slower in the larger \HUM dataset.  Figs.~\ref{fig:runtime_xml_s} and~\ref{fig:runtime_ecoli_s} show that \AT takes less time as $s$ increases. This is because, in general, constructing \emph{many small} tries of total size $n$ is faster than constructing \emph{few larger} tries of total size $n$. Indeed, the former computation is performed when $s$ is larger. 

\subsection{Useful String Indexing}

\noindent{\bf Methods.}~We refer to the $\USItop$ data structure constructed based on the \ET algorithm (see Section~\ref{sec:ESA}) as \UET and to that based on the \AT algorithm (see Section~\ref{sec:SSA}) as \UAT. We compared these approaches to four nontrivial baselines, as no existing method can be used as a competitor. All baselines employ the suffix array $\textsf{SA}(S)$ for query answering and the \psw array (see  Section~\ref{sec:USI_DS}) for storing the local utility of each prefix of $S$, but they differ in the type of queries that they  may ``cache'' (i.e., answer without using $\textsf{SA}(S)$). 

\begin{enumerate}
    \item{\BLA (No Query Caching).}~This is the baseline from subsection ``Why is USI Challenging?'' in Section~\ref{sec:intro}. It answers all queries with $\textsf{SA}(S)$ and \psw.  
    \item{\BLB (Least Recently Used (LRU)).}~This is similar to \UET, but in the hash table $H$ it stores 
    at most $K$ precomputed global utilities of
the top-$K$ most recently queried substrings, instead of the utilities of the 
\TOP frequent substrings. Like \UET, if the queried pattern is not in $H$, it uses $\textsf{SA}(S)$ and \psw   to compute $U(P)$. 
\item{\BLC (Top-$K$ Seen-so-far).} This baseline is similar to \BLB,  except that it replaces the least \emph{frequently} queried substring in $H$, instead of the least recently queried. The frequencies are maintained using an auxiliary data structure which offers the functionality of a min-heap on substring frequency and of a hash table like $H$ in \BLB. 
\item{\BLD (Space-efficient Top-$K$ Seen-so-far).} \BLD differs from \BLC in that its auxiliary data structure uses the functionality of a count-min-sketch (as in~\cite{HeavyKeeper2018}) instead of that of a hash table for space efficiency. 
\end{enumerate}
\BLA-\textsc{4} differ from \UET and \UAT in that they cache different types of queries and do not have query time guarantees.  

Without the loss of generality, we employed 
the commonly-used~\cite{8845637} ``sum of sums'' global utility function: $U(P)=\sum_{i\in \occ_{S}(P)}u(i, |P|)$, where $u(i, |P|)=\sum_{k\in [i,i+|P|-1]}w[k]$, for $i\in [0, n-|P|]$, and $u=0$ otherwise. 

\vspace{+1mm}
\noindent{\bf Parameters.}~We configured \ET and \AT, used in \UET and \UAT, respectively, using the default $K$ and $s$ values (see Table~\ref{table:data}). We used two types of query workloads per dataset, $\mathcal{W}_1$ and $\mathcal{W}_{2,p}$ (the role of $p$ will be explained next).  
Each workload has $0.7$, $6$, $40$, $70$  and $0.1$,  million query patterns, for the \IoT, \XML, \HUM, \ECOLI and \ADV dataset, respectively; more queries for larger datasets. To construct $\mathcal{W}_1$, we: (1) selected $90\%$ of the query patterns from the top-$\frac{n}{50}$ (respectively, top-$\frac{n}{60}$) frequent substrings of the input dataset when using the \IoT, \XML, \HUM or \ADV  dataset (respectively, the \ECOLI dataset), and (2) selected the remaining query patterns randomly from either the previously selected frequent substrings, or from substrings of the input dataset that have length randomly selected in $[1, 5000]$ for all datasets except \IoT and \ADV. For \IoT, we used $[1, 20000]$ (as its frequent substrings are longer), and for \ADV, we used a range of $[3,200]$ (as it has a small length). To construct $\mathcal{W}_{2,p}$, we selected $p\%$ of the queries randomly from the top-$\frac{n}{100}$ frequent substrings, and the remaining queries as in $\mathcal{W}_1$. We constructed a workload $\mathcal{W}_{2,p}$ for each $p\in\{20, 40, 60, 80\}$ per dataset. In short, our query workloads ensure that there are queries of frequent substrings and/or queries appearing multiple times. 

\noindent{\bf Measures.}~We used all 4 relevant measures of efficiency~\cite{pvldb23}:  (1) query time, (2) index size, (3) construction time, and (4) construction space. For (1) and (3), we used the \texttt{chrono C++} library. For (2), we used the \texttt{mallinfo2 C++} function. For (4), we recorded the maximum resident set size  
using the \texttt{/usr/bin/time -v} command. We do not report construction space results here, as they are essentially the same as those in Section~\ref{sec:experiments:topkfreq}: the top-$K$ frequent substring mining 
determines the construction space of both \UET and \UAT. 

\noindent{\bf Overview.}~We show that \UET and \UAT: (1) have query times up to $15$ times faster than those of the fastest baseline; (2) have size similar to that of the baselines; and (3) take more time to be constructed than the baselines, but scale linearly or near-linearly with the input string length $n$.

\noindent{\bf Query Time.} Figs.~\ref{fig:iot_query_vs_K} to~\ref{fig:adv_query_vs_K} show the average query time for the queries in the workloads of type $\mathcal{W}_1$, for varying $K$. Note that, for all tested $K$ values, both \UET and \UAT are on average $3.1$ and up to  
$15$ times faster than the fastest baseline, \BLC. 
The query time of our data structures decreases with $K$, as more queries are answered efficiently using the hash table; this is more obvious in Fig.~\ref{fig:ecoli_query_vs_K} where more queries are answered. On the contrary, the query time of the baselines stays the same or decreases much more slowly. This directly shows the benefit of efficiently answering query patterns that occur frequently in string $S$, unlike ``caching'' different types of queries as the baselines do. Among our approaches, \UAT is slightly slower but takes smaller space to construct (recall that the construction space is determined by \AT).  

Figs.~\ref{fig:iot_query_vs_p} to~\ref{fig:adv_query_vs_p} show the average query time for the queries in workload $\mathcal{W}_{2,p}$, for varying $p\in \{20, 40, 60, 80\}$. Both \UET and \UAT outperform all baselines (e.g., the best baseline is slower than our slower approach, \UAT, by $199\%$ on average). Also, \UET and \UAT become much faster with $p$, as more queries are answered by their hash table, unlike the baselines.

\noindent{\bf Index Size.}
We report results for XML, \HUM, and \ADV; the results for the other datasets are analogous. Figs.~\ref{fig:xml_index_size_vs_K} to ~\ref{fig:adv_index_size_vs_K} show the index size of all approaches for varying 
$K$. The index sizes are similar (e.g., in Fig.~\ref{fig:dna_index_size_vs_K} they differ by less than $3$GB's or less than $4\%$), since most of the space is occupied by the suffix array \textsf{SA}$(S)$. \BLA has a slightly smaller index size than others, as it does not have a hash table, while that of \BLD is slightly smaller than \BLC due to the use of the sketch in \BLD.  Figs.~\ref{fig:xml_index_size_vs_n} to~\ref{fig:adv_index_size_vs_n} show the index size of all approaches for varying 
$n$. All approaches scale linearly with $n$, as expected by their space complexities, and take roughly the same space for the same reasons as in the last experiment. 

\noindent{\bf Construction Time.}~We report results for \XML and \HUM; the results for the other datasets are analogous. Figs.~\ref{fig:xml_construction_time_vs_K} and~\ref{fig:dna_construction_time_vs_K} show that the baselines need less time to be constructed than \UET and \UAT and that \UET  is constructed faster than \UAT. This is because: (1) the construction for the baselines is much simpler than that of \UET and \UAT, and (2) the construction time of \UAT has an extra term $\tilde{\cO}(n+sK)$ and $sK=\tilde{\cO}(n)$ in our setting. Figs.~\ref{fig:xml_construction_time_vs_n} and~\ref{fig:dna_construction_time_vs_n} show the construction time for varying
$n$. All approaches scale linearly or near-linearly in line with their time complexities. Again, the baselines outperform our approaches and \UET takes less time than \UAT.

\section{Future Work}\label{sec:discussion}

%In this work, we proposed a novel index for \USI.
%, the problem of indexing strings \emph{with utilities} by means of efficient data structures for mining top-$K$ frequent substrings. 
There are three directions for future work. First, it would be worthwhile to employ machine learning to define utility functions based on interestingness measures~\cite{kbs} or to speed up search based on the underlying data distribution~\cite{LISA}. Second, it would be practically useful to investigate how to set the construction parameters $K$ and $\tau$. Our data structure from Section~\ref{sec:ESA} allows us to produce a large number of $(K, \tau)$ values \emph{efficiently}, which could be then used to select a good trade-off~\cite{kossman}. Third, it is interesting to investigate a dynamic version of \USI. We next present a partial solution for the case where \emph{only letter appends are allowed}~\cite{Ukkonen}. 

We assume that we have constructed the index for $S$. 
We maintain two auxiliary dynamic data structures: a heap storing the frequencies of the explicit nodes of $\textsf{ST}(S)$; and a table storing the KR fingerprints of all prefixes of $S$.
%Clearly, the heap can be updated in $\mathcal{O}(\log n)$ time and the table of KR fingerprints in $\mathcal{O}(1)$ time. 
Assume a new letter $\alpha$ is appended to $S$  creating $S'=S\alpha$. We extend \psw by one position, storing the sum of the utility of $\alpha$ and the former last entry of \psw; and we update $\textsf{ST}(S)$ by Ukkonen's algorithm~\cite{Ukkonen} adding
%This takes $\mathcal{O}(1)$ amortized time.
%We may need to add $\mathcal{O}(1)$ new nodes (a new branching node and a leaf node) in the suffix tree of $S'=S\alpha$.
%Two new nodes are added in the suffix tree of $S'=S\alpha$: 
a new branching node and a leaf.
The frequency of the new leaf is trivially $1$, and that of the new branching node is $g\!+\!1$, where $g$ and $1$ are
the frequencies of its two children: the previously existing child and the new leaf.
%So it takes $\mathcal{O}(\log n)$ time to insert these two new frequencies (nodes) in the heap. 
%At this point, we also need to increment by one the frequencies of all ancestors of the new branching node. 
At this point, we insert the frequencies of these two new nodes in the heap and increment the frequencies of all ancestors of the new branching node by one. Incrementing these frequencies is  challenging as \emph{there could be many such ancestors}. We then traverse the heap to list the top-$K$ frequent substrings in $S'$. %Recall that every node can correspond to many substrings, as explained in Section~\ref{sec:ESA}.

%These nodes are $\mathcal{O}(\log n)$ on average, because this is the depth of the suffix tree on average~\cite{DBLP:journals/tit/JacquetS91}, so we need $\mathcal{O}(\log^2 n)$ time to update the heap on average. As for $\psw$, we simply append the utility of $S'$ in $\mathcal{O}(1)$ time. 
%At this point, we can traverse the heap from the most frequent to the least one and enumerate the top-$K$ frequent substrings in $S'$. Recall that every node can correspond to many substrings.}

%For each top-$K$ substring, we compute its KR fingerprint in $\mathcal{O}(1)$ time using the table of KR fingerprints~\cite{tomohiro}. Next we observe that this substring, say $s$ occurring at position $i$ of $S'$, must be a suffix of $S'$ if it has not been in the top-$K$ of $S$. 
We next compute the KR fingerprint of each top-$K$ substring using the  KR fingerprints table~\cite{tomohiro}. Consider a substring $s$, occurring at position $i$ of $S'$, which is in the top-$k$ frequent substrings of $S'$ but not of $S$. We observe that $s$ must be a suffix of $S'$: this is because the frequencies increase monotonically with appending. 
We compute the local utility $u(i,|s|)$ of $s$ 
%in $\mathcal{O}(1)$ time 
using the updated $\psw$ and do the following:
\begin{itemize}
\item If $s$ is in $H$, we add $u(i,|s|)$ to its previous global utility. %in $\mathcal{O}(1)$ time. 
\item If $s$ is not in $H$ (it is new), it must have frequency at most $\tau_K+1$ in $S'$. We access its locus in $\textsf{ST}(S')$, %in constant time because $s$ is a suffix 
compute %in $\mathcal{O}(\tau_K)$ time 
its global utility $U(s)$ in $S'$, and add it to $H$.
Thus \emph{we may spend $\cO(\tau_K)$ time for each of the $\cO(K)$ substrings}.
 \end{itemize}
Finally, we delete any entry in $H$ that does not represent a top-$K$ frequent substring in $S'$. %in $S'$ in $\mathcal{O}(K)$ time. The total amortized time for updates is $\mathcal{\tilde{O}}(\tau_K K)$ on average. 
Querying is not affected as all data structures needed for querying are updated.

Unfortunately, as highlighted above, maintaining the node frequencies in $\textsf{ST}(S')$ and adding the new global utilities in $H$ dynamically can in general be very costly. We thus defer the investigation of this dynamic version of \USI to future work.

\section*{Acknowledgments}

This work was supported by PANGAIA and ALPACA projects funded by the EU under MCSA Grant Agreements 872539 and 956229; by the Next Generation EU PNRR MUR M4 C2 Inv 1.5 project ECS00000017 Tuscany Health Ecosystem Spoke 6 CUP B63C2200068007 and I53C22000780001; by the MUR PRIN 2022 YRB97K PINC; and by the Institute for Interdisciplinary Data Science and Artificial Intelligence Pump Prime Funding at the University of Birmingham.

\bibliographystyle{ieeetr}
\bibliography{biblio}

\end{document}